\documentclass[prd,eqsecnum,twocolumn,nofootinbib,superscriptaddress]{revtex4-2}

\usepackage{amsfonts}
\usepackage{amsmath}
\usepackage{amssymb}
\usepackage{amsthm}
\usepackage{bm}
\usepackage{dcolumn}
\usepackage{epsfig}
\usepackage{graphicx}
\usepackage{graphics}
\usepackage[latin1]{inputenc}
\usepackage{latexsym}
\usepackage{rotating}
\usepackage{hyperref}

\usepackage{xcolor}

\begin{document}

\title{Adiabatic waveforms for extreme mass-ratio inspirals via\\multivoice decomposition in time and frequency}

\author{Scott A.\ Hughes}
\affiliation{Department of Physics and MIT Kavli Institute, Cambridge, MA 02139, United States}

\author{Niels Warburton}
\affiliation{School of Mathematics and Statistics, University College Dublin, Belfield, Dublin 4, Ireland}

\author{Gaurav Khanna}
\affiliation{Physics Department, University of Massachusetts, Dartmouth, MA 02747, United States}
\affiliation{Department of Physics, University of Rhode Island, Kingston, RI 02881, United States}

\author{Alvin J.\ K.\ Chua}
\affiliation{Theoretical Astrophysics Group, California Institute of Technology, Pasadena, CA 91125, United States}

\author{Michael L.\ Katz}
\affiliation{Max-Planck-Institut f\"ur Gravitationsphysik, Albert-Einstein-Institut, Am M\"uhlenberg 1, 14476 Potsdam-Golm, Germany}

\begin{abstract}
We compute adiabatic waveforms for extreme mass-ratio inspirals (EMRIs) by ``stitching'' together a long inspiral waveform from a sequence of waveform snapshots, each of which corresponds to a particular geodesic orbit.  We show that the complicated total waveform can be regarded as a sum of ``voices.''   Each voice evolves in a simple way on long timescales, a property which can be exploited to efficiently produce waveform models that faithfully encode the properties of EMRI systems.  We look at examples for a range of different orbital geometries: spherical orbits, equatorial eccentric orbits, and one example of generic (inclined and eccentric) orbits.  To our knowledge, this is the first calculation of a generic EMRI waveform that uses strong-field radiation reaction.  We examine waveforms in both the time and frequency domains.  Although EMRIs evolve slowly enough that the stationary phase approximation (SPA) to the Fourier transform is valid, the SPA calculation must be done to higher order for some voices, since their instantaneous frequency can change from chirping forward ($\dot f > 0$) to chirping backward ($\dot f < 0$).  The approach we develop can eventually be extended to more complete EMRI waveform models, for example to include effects neglected by the adiabatic approximation such as the conservative self force and spin-curvature coupling.
\end{abstract}

\maketitle

\section{Introduction}

\subsection{Extreme mass-ratio inspirals and self forces}

The large mass ratio limit of the two-body problem is a laboratory for studying strong-field motion in general relativity.  By treating the spacetime of such a binary as an exact black hole solution plus a perturbation due to the less massive orbiting body, it is possible to analyze the binary's dynamics and the gravitational waves it generates using the tools of black hole perturbation theory.  Because the equations of perturbation theory can, in many circumstances, be solved very precisely, large-mass ratio serves as a high-precision limit for understanding the two-body problem in general (see, e.g., \cite{LeTiec:2014lba}).

This limit is also significant thanks to the importance of {\it extreme mass-ratio inspirals} (EMRIs) as gravitational wave sources.  Binaries formed by the capture of stellar mass ($\mu \sim 1-100\,M_\odot$) compact bodies onto relativistic orbits of black holes with $M \sim 10^6\,M_\odot$ in the cores of galaxies will generate low-frequency ($f \sim 0.01$ Hz) gravitational waves, right in the sensitive band of space-based detectors like LISA.  A typical EMRI will execute $\sim 10^4-10^5$ orbits in LISA's band as gravitational-wave backreaction shrinks its binary separation.  Because the small body orbits in the larger black hole's strong field, EMRI waves are highly sensitive to the nature of the large black hole's spacetime.  EMRI events will make it possible to precisely map black hole spacetimes, weighing black holes' masses and spins with exquisite accuracy, and testing the hypothesis that astrophysical black holes are described by the Kerr spacetime \cite{Babak:2017tow}.

To achieve these ambitious science goals for EMRI measurements, we will need waveform models, or templates, that accurately match signals in data over their full duration.  Such templates will provide guidance to algorithms for finding EMRI signals in detector noise, and will be necessary for characterizing astrophysical sources.  Developing such models is one of the goals of the {\it self force} program, which seeks to develop equations describing the motion of objects in specified background spacetimes, including the interaction of that object with its own perturbation to that spacetime --- i.e., including the small body's ``self interaction'' {\cite{Barack:2018yvs}}.  Taking the background spacetime to be that of a black hole, self forces can be developed using tools from black hole perturbation theory, with the mass ratio of the system $\varepsilon \equiv \mu/M$ (where $\mu$ is the mass of the orbiting body, and $M$ the mass of the black hole) serving as a perturbative order counting parameter.

To make this more quantitative, we sketch the general form of such equations of motion in the action-angle formulation used by Hinderer and Flanagan {\cite{Hinderer:2008dm}}.  Let $q_\alpha \doteq (q_t,q_r,q_\theta,q_\phi)$ be a set of angle variables which describe the motion of the smaller body in a convenient coordinate system, let $J_\beta \doteq (J_t, J_r, J_\theta, J_\phi)$ be a set of actions associated with those motions, and let $\lambda$ be a convenient time variable.  The motion of the smaller body is then governed by a set of equations with the form
\begin{eqnarray}
\frac{dq_\alpha}{d\lambda} &=& \omega_\alpha({\bf J}) + \varepsilon\,g^{(1)}_\alpha(q_r,q_\theta;{\bf J}) + \varepsilon^2\,g^{(2)}_\alpha(q_r,q_\theta;{\bf J}) + \ldots
\nonumber\\
\label{eq:dangledt}\\
\frac{dJ_\alpha}{d\lambda} &=& \varepsilon\,G^{(1)}_\alpha(q_r,q_\theta;{\bf J}) + \varepsilon^2\,G^{(2)}_\alpha(q_r,q_\theta;{\bf J}) + \ldots
\label{eq:dJdt}
\end{eqnarray}
The frequency $\omega_\alpha$ describes the rate at which the angles accumulate per unit $\lambda$, neglecting the self interaction.  The $\omega_\alpha({\bf J})$ thus characterize the geodesic motion of the small body in the black hole background.  The terms $g_\alpha^{(n)}$ describe how the small body's trajectory is modified by the self interaction at $O(\varepsilon^n)$; the terms $G_\alpha^{(n)}$ describe how the actions (which are constant along geodesics) are modified.  Note that the self-interaction terms only depend on the angles $q_r$ and $q_\theta$ --- because black hole spacetimes are axisymmetric and stationary, these terms are independent of $q_t$ and $q_\phi$.  Many aspects of this problem are now under control at $O(\varepsilon)$ (see, e.g., \cite{vandeMeent:2017bcc}), and work is rapidly proceeding on the problem at $O(\varepsilon^2)$ \cite{Pound:2019lzj}.

As the issue of the smaller body's motion is brought under control, more attention is now being paid to the gravitational waveforms that arise from this motion.  That is the focus of this paper.

\subsection{The adiabatic approximation and its use}

The forcing terms in Eqs.\ (\ref{eq:dangledt}) and (\ref{eq:dJdt}) can be further decomposed by splitting them into averages and oscillations about their average.  Consider the first-order forcing term $G^{(1)}_\alpha$.  We put
\begin{equation}
G^{(1)}_\alpha(q_r, q_\theta;{\bf J}) = \langle G^{(1)}_\alpha({\bf J}) \rangle + \delta G^{(1)}_\alpha(q_r, q_\theta;{\bf J})\;,
\end{equation}
where the averaged contribution is
\begin{equation}
\langle G^{(1)}_\alpha({\bf J}) \rangle = \frac{1}{(2\pi)^2}\int_0^{2\pi} dq_r\int_0^{2\pi} dq_\theta\, G^{(1)}_\alpha(q_r, q_\theta;{\bf J})\;,
\end{equation}
and we define the oscillations about this average as $\delta G^{(1)}_\alpha(q_r, q_\theta;{\bf J}) \equiv G^{(1)}_\alpha(q_r, q_\theta;{\bf J}) - \langle G^{(1)}_\alpha({\bf J}) \rangle$.  The oscillations vary about zero on a rapid orbital timescale $T_{\rm o} \sim M$; the average evolves on a much slower dissipative inspiral timescale $T_{\rm i} \sim M^2/\mu$, or $T_{\rm i} \sim T_{\rm o}/\varepsilon$.  Because of the large separation of these two timescales for EMRIs, the oscillations nearly average away during an inspiral.  For most orbits, neglecting the impact of the oscillations introduces errors of $O(T_{\rm o}/T_{\rm i}) = O(\varepsilon)$.

The simplest model for inspiral which captures the strong-field dynamics of black hole orbits is known as the {\it adiabatic approximation}.  It amounts to solving the following variants of Eqs.\ (\ref{eq:dangledt}) and (\ref{eq:dJdt}):
\begin{equation}
\frac{dq_\alpha}{d\lambda} = \omega_\alpha({\bf J})\;,\quad
\frac{dJ_\alpha}{d\lambda} = \varepsilon\,\langle G^{(1)}_\alpha({\bf J})\rangle\;.
\label{eq:actionangle_adapp}
\end{equation}
In words, we treat the short-timescale orbital dynamics as geodesic, but use the orbit-averaged impact of the self force on the actions $J_\alpha$.  This amounts to including the ``dissipative'' part of the orbit-averaged self force, since to $O(\varepsilon)$, the action of $\langle G^{(1)}_\alpha\rangle$ is equivalent to computing the rates at which an orbit's energy $E$, axial angular momentum $L_z$, and Carter constant $Q$ change due to the backreaction of gravitational-wave emission.  The adiabatic approximation treats inspiral as ``flow'' through a sequence of geodesics, with the rate of flow determined by the rates of change of $E$, $L_z$, and $Q$ \cite{Hughes:2005qb}.

It is worth noting that the picture we have sketched breaks down near the so-called {\it resonant} orbits \cite{Flanagan:2010cd,Flanagan:2012kg,Yang:2017aht,Bonga:2019ycj}.  These are orbits for which the frequency ratio $\omega_x/\omega_y = n_x/n_y$, where $n_x$ and $n_y$ are small integers.  Resonances have been shown to arise from the gravitational self force itself \cite{Flanagan:2010cd} (for which $x = r$ and $y = \theta$), as well as from tidal perturbations from stars or black holes that are near an astrophysical EMRI system \cite{Yang:2017aht,Bonga:2019ycj} (for which $x = r$ or $\theta$, and $y = \phi$).  Near resonances, some terms change from oscillatory to nearly constant; neglecting their impact introduces errors of $O(\varepsilon^{1/2})$.  As such, they will be quite important, contributing perhaps the leading post-adiabatic contribution to waveform phase evolution.  We neglect the impact of resonances in this analysis, though where appropriate we comment on their importance and how they may be incorporated into future work.

The computational cost associated with even the rather simplified adiabatic waveform model is quite high.  As we outline in Sec.\ \ref{sec:amps_and_evolve}, computing adiabatic backreaction involves solving the Teukolsky equation for many multipoles of the radiation field and harmonics of the orbital motion; tens of thousands of multipoles and harmonics may need to be calculated for tens of thousands of orbits.  To produce EMRI models which capture the most important elements of the mapping between source physics and waveform properties, the community has developed several {\it kludges} as a stop-gap for data analysis and science return studies.  The {\it analytic kludge} of Barack and Cutler \cite{Barack:2003fp} essentially pushes post-Newtonian models beyond their domain of validity.  Their match with fully relativistic models is not good, but they capture the key qualitative features of EMRI physics.  Almost as important, the analytic kludge is fast and is easy to implement; as such, it has been heavily used for many LISA measurement studies.

The {\it numerical kludge} of Babak et al.\ \cite{Babak:2006uv} is closer to the spirit of the adiabatic approximation we describe here, in that it treats the small body's motion as a Kerr geodesic, but uses semi-analytic fits to strong-field radiation emission to describe inspiral.  Wave emission is treated with a crude multipolar approximation based on the small body's coordinate motion.  Despite the crudeness of some of the underlying approximations, the numerical kludge fits relativistic waveform models fairly well.  It is however slower and harder to use than the analytic kludge, and as such has not been used very much.  More recently, Chua and Gair \cite{Chua:2015mua} showed that one can significantly improve matches to relativistic models by using an analytic kludge augmented with knowledge of the exact frequency spectrum of Kerr black hole orbits.

The shortcomings of the kludges illustrate that, ultimately, one needs waveform models that capture the strong-field dynamics of Kerr orbits and that accurately describe strong-field radiation generation and propagation through black hole spacetimes.  Waveforms based on the adiabatic approximation are the simplest ones that accurately include both of these effects.  Though the adiabatic approximation misses important aspects from neglected pieces of the self force, they get enough of the strong-field physics correct that they will be effective and useful tools for understanding the scope of EMRI data analysis challenges, and for accurately assessing the science return that EMRI measurements will enable.

\subsection{This paper}

Although the computational cost of making adiabatic inspiral waveforms is high, the most expensive step of this calculation --- computing a set of complex numbers which encode rates of change of ($E, L_z, Q$), as well as the gravitational waveform's amplitude --- need only be done once.  These numbers can be computed in advance for a range of astrophysically relevant EMRI orbits, and then stored and used to assemble the waveform as needed.  The goal of this paper is to lay out what quantities must be computed, and to describe how to use such precomputed data to build adiabatic waveforms.

Our particular goal is to show how to organize and store the most important and useful data needed to assemble the waveforms in a computationally effective way.  A key element of our approach is to view the complicated EMRI waveform as a sum of simple ``voices.''  Each voice corresponds to a mode $(l, m, k, n)$ representing a particular multipole of the radiation and harmonic of the fundamental orbital frequencies.  The voice-by-voice decomposition was suggested long ago to one of us by L.\ S.\ Finn, and was first presented in Ref.\ \cite{Hughes:2001jr} for the special case of spherical Kerr orbits (i.e., orbits of constant Boyer-Lindquist radius, including those inclined from the equatorial plane).  This paper corrects an important error in Ref.\ \cite{Hughes:2001jr} (which left out a phase that is set by initial conditions), and extends this analysis to fully generic configurations.

The approach that we present has several important features.  First, we find that the data which must be stored to describe the waveform voice-by-voice evolves smoothly over an inspiral.  This suggests that these data can be sampled at a relatively low cadence, and we can then build a high-quality waveform using interpolation.  Such behavior had been seen in earlier work \cite{Hughes:2001jr}; this analysis demonstrates that this behavior is not unique to spherical orbits, but is generic.  We describe the methods we have used for our initial exploration, and that were used in a related study \cite{Chua:2020stf} focusing on the rapid evaluation of waveforms for data analysis. Reference \cite{Chua:2020stf} showed that the chasm between the computational demands of accurate modeling and efficient analysis can be bridged, but much work remains to ``optimize'' these methods --- for example, in determining the best way for the numerical data to be sampled and interpolated.

Second, we show that this framework works well in both the time and frequency domains.  The frequency-domain construction is particularly interesting.  Thanks to their extreme mass ratios, EMRIs evolve slowly enough that the stationary-phase approximation (SPA) to the Fourier transform should provide an excellent representation of the frequency-domain form.  However, for some EMRI voices, the instantaneous frequency grows to a maximum and then decreases.  For such voices, the first time derivative of the frequency vanishes at some point along the inspiral.  The standard SPA is singular at these points.  We show that including information about the second derivative fixes this behavior, allowing us to compute frequency-domain waveforms for all EMRI voices.

Third, we show that several of the waveform's parameters can be included in a simple way which effectively reduces the dimensionality of the waveform parameter space.  These parameters are angles which control the initial position of the orbit in its eccentric radial motion, and its initial polar angle in the range of motion allowed by its orbital inclination.  They are initial conditions on the relativistic analog of the ``true anomaly angles'' used in Newtonian celestial mechanics.  Associated with these initial anomaly angles are phases, originally introduced in Ref.\ \cite{Drasco:2005is}, which correct the complex amplitudes associated with the waveform's voices.  By comparing with output from a time-domain Teukolsky equation solver \cite{Sundararajan:2008zm,Zenginoglu:2011zz}, we show that these phases allow one to match any allowed initial condition with very little computational cost.  This means that we can generate a suite of data using only a single initial choice of the anomaly angles, and then transform to initial conditions corresponding to any other choice.  This significantly reduces the computational cost associated with covering the full EMRI parameter space.

Finally, it should not be difficult to extend this framework to include at least some important effects beyond the adiabatic approximation, many of which are discussed in detail in Refs.\ \cite{Miller:2020bft,Pound:2021qin}.  For example, both conservative self forces and spin-curvature coupling have orbit-averaged effects that are small, but that secularly accumulate over many orbits \cite{Warburton:2017sxk,Osburn:2015duj,Piovano:2020zin,Skoupy:2021iwb}.  Such effects can be included in this framework by allowing the relativistic anomaly angles discussed above (which in the adiabatic limit are constant) to evolve over the inspiral; the phases associated with these angles will evolve as well.   We also expect that the impact of small oscillations (such as arise from both self forces and spin-curvature coupling) can likewise be incorporated, perhaps very efficiently using a near-identity transformation \cite{vandeMeent:2018rms}.

As we were completing the bulk of the calculations which appear here, a similar analysis of adiabatic EMRI waveforms was presented by Fujita and Shibata \cite{Fujita:2020zxe}.  Their analysis focuses to a large extent on the measurability of EMRI waves by LISA, confining their discussion to eccentric and equatorial sources.  Like us, they also take advantage of the fact that one can pre-compute the most expensive data on a grid of orbits, and then assemble the waveform by interpolation.  We are encouraged by the fact that their independently developed framework is substantially similar to what we present here.

\subsection{Organization of this paper}

The remainder of this paper is organized as follows.  Since inspiral in the adiabatic approximation is treated as a sequence of geodesic orbits, we begin by reviewing the properties of Kerr geodesics in Sec.\ \ref{sec:geodesics}.  Nothing in this section is new; it is included primarily to keep the manuscript self contained, and to allow us to carefully define our notation and the meaning of important quantities which are used elsewhere.  Section \ref{sec:amps_and_evolve} reviews how we solve the Teukolsky equation and use its solutions in order to calculate adiabatic backreaction on an orbit.  This allows us to compute how a system evolves from orbit to orbit, as well as the gravitational-wave amplitude produced by that orbit.  This material is likewise not new, but is included for completeness, as well as to introduce and explain all relevant notation.

In Sec.\ \ref{sec:diss_evolve}, we lay out how one ``stitches'' together data describing radiation from geodesics to construct an adiabatic waveform.  This construction essentially amounts to taking the solution to the Teukolsky equation for a geodesic orbit and promoting various factors which are constants on geodesics into factors which vary slowly along an inspiral.  This construction introduces a two-timescale expansion: some quantities vary on the ``fast'' timescale associated with orbital motions, $T_{\rm o} \sim M$; others vary on the ``slow'' timescale associated with the inspiral, $T_{\rm i} \sim M^2/\mu = T_{\rm o}/\varepsilon$.  The adiabatic waveform is only accurate up to corrections of order the system's mass ratio, essentially because it assumes that all time derivatives only include information about the system's ``fast'' time variation.  One way in which a post-adiabatic analysis will improve on these results will be by including information about time derivatives with respect to the slow variation along the inspiral.

Section \ref{sec:fourier} describes how to compute multivoice signals in the frequency domain.  Because EMRI systems are slowly evolving, the stationary phase approximation (SPA) should accurately describe the Fourier transform of EMRI signals.  However, because the evolution of certain voices is not monotonic, the ``standard'' SPA calculation can fail, introducing singular artifacts at moments when a voice's frequency evolution switches sign.  We review the standard SPA Fourier transform and show how by including an additional derivative of the frequency it is straightforward to correct this artifact.  We conclude this section by showing how to combine multiple voices to construct the full frequency-domain EMRI waveform.

In Sec.\ \ref{sec:implement}, we present various important technical details describing how we implement this framework for the results we present in this paper.  We strongly emphasize that there is a great deal of room for improvement on the techniques described here.  We have not, for example, carefully assessed the most effective method for laying out the grid of data on which we store information about adiabatic backreaction and waveform amplitudes, nor have we thoroughly investigated efficient methods for interpolating these data to off-grid points (e.g., \cite{Chua:2020stf}).  These important points will be studied in future work, as we begin assessing how to take this framework and use it to develop EMRI waveforms in support of LISA data analysis and science studies.

In Secs.\ \ref{sec:results_schw} and \ref{sec:results_kerr} we present examples of adiabatic EMRI waveforms.  In both of these sections, we show examples of the complete time-domain waveform produced by summing over many voices, as well as the (much simpler) structure of representative voices which contribute to these waveforms.  Section \ref{sec:results_schw} shows results for inspiral into Schwarzschild black holes, presenting the details of an inspiral with small initial eccentricity ($e_{\rm init} = 0.2$) and another with high initial eccentricity ($e_{\rm init} = 0.7$).  Section \ref{sec:results_kerr} examines several examples for inspiral into Kerr, including a case that is spherical, a case that is equatorial with high eccentricity, and one example that is generic, both eccentric and inclined.  Although generic EMRI waveforms based on various ``kludge'' approximations have been presented before, to our knowledge the generic example shown in Sec.\ \ref{sec:results_kerr} is the first calculation that uses strong-field backreaction and strong-field wave generation for the entire computation.  We find that there is a great deal of similarity between qualitative features of the Kerr and Schwarzschild waveforms.  As such, our presentation of Kerr results is somewhat brief, concentrating on the most important highlights and differences as compared to the Schwarzschild cases.

For all the cases we examine, we demonstrate how one can account for a system's initial conditions by adjusting a set of phase variables which depend on the initial values of the anomaly angles that parameterize the system's radial and polar motions.  We calibrate our calculations in one case (presented in Sec.\ \ref{sec:results_schw}) by comparing to an EMRI waveform computed using a time-domain Teukolsky equation solver \cite{Sundararajan:2008zm,Zenginoglu:2011zz}.  Interestingly, in this case we find a small phase offset that, for most initial conditions, secularly accumulates, causing the waveforms computed with this paper's techniques to dephase from those computed with the time-domain code by up to several radians over the course of an inspiral.  We argue tat this is an artifact of the adiabatic approximation's neglect of terms which vary on the slow timescale, and is not unexpected.  We show in this comparison case that we can empirically compensate for much of the dephasing using an {\it ad hoc} correction that corrects some of the neglected ``slow time'' evolution.  Though this correction is not rigorously justified, its form suggests that the dephasing may arise from a slow-time evolution of the phases variables which are set by the orbit's initial conditions.

Our conclusions are presented in Sec.\ \ref{sec:conclude}.  Along with summarizing the main results of this analysis, we describe plans and directions for future work.  Chief among these plans is to investigate how to optimize the implementation in order to make EMRI waveforms as rapidly as possible, which will be very important in order for these waveforms to be usable for LISA data and science analysis studies, as well as investigating how to include post-adiabatic effects with small modifications of the framework we present here.  We also plan to publicly release the code and data used in this study, and describe the status of our plans as this analysis is being completed.

\section{Kerr geodesics}
\label{sec:geodesics}

We begin by discussing bound Kerr geodesics.  The most important aspects of this content are discussed in depth elsewhere \cite{Misner:1974qy,Schmidt:2002qk,Drasco:2003ky,Fujita:2009bp,vandeMeent:2019cam}; we briefly review this material for this paper to be self contained, as well as to introduce notation and conventions that we use.  Certain lengthy but important formulas are given in Appendix \ref{app:geodesicconstants}.

\subsection{Mino-time formulation of orbital motion}

Consider a point-like body of mass $\mu$ orbiting a Kerr black hole with mass $M$ and spin parameter $a = |{\bf S}|/M$ (where ${\bf S}$ is the black hole's spin angular momentum in units with $G = 1 = c$), and use Boyer-Lindquist coordinates (with the angle $\theta$ measured from the black hole's spin axis) to describe its motion.  We use {\it Mino time} as our time parameter describing these orbits.  An interval of Mino time $d\lambda$ is related to an interval of proper time $d\tau$ by $d\lambda = d\tau/\Sigma$, where $\Sigma = r^2 + \cos^2\theta$, and where $r$ and $\theta$ are the Boyer-Lindquist radial and polar coordinates.  With this parameterization, motion in Boyer-Lindquist coordinates is governed by the equations
\begin{eqnarray}
\left(\frac{dr}{d\lambda}\right)^2 &=& [E(r^2 + a^2) - aL_z]^2 - \Delta[r^2 + (L_z - aE)^2 + Q]
\nonumber\\
&\equiv& R(r)\;,
\label{eq:rdot}\\
\left(\frac{d\theta}{d\lambda}\right)^2 &=& Q - \cot^2\theta L_z^2 - a^2\cos^2\theta(1 - E^2)
\nonumber\\
&\equiv& \Theta(\theta)\;,
\label{eq:thdot}\\
\frac{d\phi}{d\lambda} &=& \csc^2\theta L_z + \frac{2Mr a E}{\Delta} - \frac{a^2 L_z}{\Delta}
\nonumber\\
&\equiv& \Phi_r(r) + \Phi_\theta(\theta)\;,
\label{eq:phdot}\\
\frac{dt}{d\lambda} &=& E\left[\frac{(r^2 + a^2)^2}{\Delta} - a^2\sin^2\theta\right] - \frac{2Mra L_z}{\Delta}
\nonumber\\
&\equiv& T_r(r) + T_\theta(\theta)\;.
\label{eq:tdot}
\end{eqnarray}
We have introduced $\Delta = r^2 - 2Mr + a^2$.  The quantities $E$, $L_z$, and $Q$ are the orbit's energy (per unit $\mu$), axial angular momentum (per unit $\mu$), and Carter constant (per unit $\mu^2$).  These quantities are conserved along any geodesic; choosing them specifies an orbit, up to initial conditions.  Writing $d/d\lambda = \Sigma\, d/d\tau$ puts these equations into more familiar forms typically found in textbooks, such as Eqs.\ (33.32a-d) of Ref.\ \cite{Misner:1974qy}.

Equations (\ref{eq:rdot}) and (\ref{eq:thdot}) tell us that bound Kerr orbits are periodic in $r$ and $\theta$ when parameterized using $\lambda$:
\begin{equation}
r(\lambda) = r(\lambda + n\Lambda_r)\;,\quad
\theta(\lambda) = \theta(\lambda + k\Lambda_\theta)\;,
\end{equation}
where $n$ and $k$ are each integers.  Simple formulas exist for the Mino-time periods $\Lambda_r$ and $\Lambda_\theta$ {\cite{Fujita:2009bp}}; we define the associated frequencies by $\Upsilon_{r,\theta} = 2\pi/\Lambda_{r,\theta}$.

The motions in $t$ and $\phi$ are the sum of secularly accumulating pieces and oscillatory functions:
\begin{eqnarray}
t(\lambda) &=& t_0 + \Gamma\lambda + \Delta t_r[r(\lambda)] + \Delta t_\theta[\theta(\lambda)]\;,
\label{eq:t_of_lambda}
\\
\phi(\lambda) &=& \phi_0 + \Upsilon_\phi\lambda + \Delta\phi_r[r(\lambda)] + \Delta\phi_\theta[\theta(\lambda)]\;.
\label{eq:phi_of_lambda}
\end{eqnarray}
In these equations, $t_0$ and $\phi_0$ describe initial conditions,
\begin{eqnarray}
\Gamma &=& \langle T_r(r)\rangle + \langle T_\theta(\theta) \rangle\;,
\label{eq:Gamma}\\
\Upsilon_\phi &=& \langle \Phi_r(r)\rangle + \langle\Phi_\theta(\theta) \rangle\;,
\label{eq:Upsphi}
\end{eqnarray}
and
\begin{eqnarray}
\Delta t_r[r(\lambda)] &=& T_r[r(\lambda)] - \langle T_r(r)\rangle \equiv \Delta t_r(\lambda)\;,\nonumber\\
\Delta t_\theta[\theta(\lambda)] &=& T_\theta[\theta(\lambda)] - \langle T_\theta(\theta)\rangle \equiv \Delta t_\theta(\lambda)\;;
\label{eq:Deltat}\\
\Delta\phi_r[r(\lambda)] &=& \Phi_r[r(\lambda)] - \langle \Phi_r(r)\rangle \equiv \Delta\phi_r(\lambda)\;,\nonumber\\
\Delta\phi_\theta[\theta(\lambda)] &=& \Phi_\theta[\theta(\lambda)] - \langle \Phi_\theta(\theta)\rangle \equiv \Delta\phi_\theta(\lambda)\;.
\label{eq:Deltaphi}
\end{eqnarray}
The quantity $\Gamma$ describes the mean rate at which observer time $t$ accumulates per unit $\lambda$; the Mino-time frequency $\Upsilon_\phi$ describes the mean rate at which $\phi$ accumulates per unit $\lambda$.  The associated period is $\Lambda_\phi = 2\pi/\Upsilon_\phi$.  Simple formulas likewise exist for $\Gamma$ and $\Upsilon_\phi$ \cite{Fujita:2009bp}.  The averages used in Eqs.\ (\ref{eq:Gamma})--(\ref{eq:Deltaphi}) are given by
\begin{eqnarray}
\langle f_r(r)\rangle &=& \frac{1}{\Lambda_r}\int_0^{\Lambda_r} f_r[r(\lambda)]\,d\lambda\;,
\label{eq:raveraging}\\
\langle f_\theta(\theta)\rangle &=&\frac{1}{\Lambda_\theta}\int_0^{\Lambda_\theta} f_\theta[\theta(\lambda)]\,d\lambda\;.
\label{eq:thetaaveraging}
\end{eqnarray}

The ratio of the Mino-time frequencies to $\Gamma$ gives the observer-time frequencies:
\begin{equation}
\Omega_{r,\theta,\phi} = \frac{\Upsilon_{r,\theta,\phi}}{\Gamma}\;.
\end{equation}
We thus have useful closed-form expressions for all frequencies, conjugate to both Mino time $\lambda$ and observer time $t$, that characterize black hole orbits.

\subsection{Orbit parameterization and initial conditions}
\label{sec:params}

Take the orbit to oscillate over $\theta_{\rm min} \le \theta \le \theta_{\rm max}$, with $\theta_{\rm max} = \pi - \theta_{\rm min}$, and over $r_{\rm min} \le r \le r_{\rm max}$, with
\begin{equation}
r_{\rm min/max} = \frac{p}{1 \pm e}\;.
\end{equation}
Choosing $p$, $e$, and $\theta_{\rm min}$ is equivalent to choosing the integrals of motion $E$, $L_z$, and $Q$.  We have found it particularly convenient to replace $\theta_{\rm min}$ with an inclination angle\footnote{The angle $I$ has been labeled $\theta_{\rm inc}$ in some previous work, such as Ref.\ \cite{Drasco:2005kz}.  We have found this label to be potentially confusing, since the Boyer-Lindquist angle $\theta$ is measured from the black hole's spin axis, whereas the inclination angle is measured from the plane normal to this axis.  We have changed notation to avoid confusion with the coordinates.} $I$, defined by
\begin{equation}
I = \pi/2 - \mbox{sgn}(L_z)\theta_{\rm min}\;.
\label{eq:Idef}
\end{equation}
The angle $I$ varies smoothly from $0$ for equatorial prograde to $\pi$ for equatorial retrograde.  The definition (\ref{eq:Idef}) may seem a bit awkward thanks to the branch associated with the sign of $L_z$.  It is simple to show that
\begin{equation}
\cos\theta_{\rm min} = \sin I\;.
\end{equation}
We have found that $x_I \equiv \cos I$ is a particularly good parameter to describe inclination: $x_I$ varies smoothly from $1$ to $-1$ as orbits vary from prograde equatorial to retrograde equatorial, with $L_z$ having the same sign as $x_I$.  Schmidt {\cite{Schmidt:2002qk}} first showed how to compute $(E, L_z, Q)$ for generic Kerr orbits; a particularly clean representation is provided by van de Meent \cite{vandeMeent:2019cam}.  We summarize his formulas in Appendix \ref{app:geodesicconstants}, tweaking them slightly to use our preferred parameter set $(p, e, x_I)$.

Initial conditions for $t$ and $\phi$ are set by the parameters $t_0$ and $\phi_0$ given in Eqs.\ (\ref{eq:t_of_lambda}) and (\ref{eq:phi_of_lambda}).  To set initial conditions on $r$ and $\theta$, we introduce anomaly angles $\chi_r$ and $\chi_\theta$ to reparameterize those coordinate motions:
\begin{eqnarray}
r &=& \frac{p}{1 + e\cos(\chi_r + \chi_{r0})}\;,
\label{eq:rdef}\\
\cos\theta &=& \sqrt{1 - x_I^2}\cos(\chi_\theta + \chi_{\theta0})\;.
\label{eq:thetadef}
\end{eqnarray}
We put $\chi_\theta = 0$, $\chi_r = 0$, $t = t_0$, and $\phi = \phi_0$ when $\lambda = 0$.  The phase $\chi_{r0}$ then determines the value of $r$ at $\lambda = 0$, and $\chi_{\theta0}$ determines the corresponding value of $\theta$.  When $\chi_{\theta0} = 0$, the orbit has $\theta = \theta_{\rm min}$ when $\lambda = 0$; when $\chi_{r0} = 0$, it has $r = r_{\rm min}$ when $\lambda = 0$.

We define the {\it fiducial geodesic} to be the geodesic that has $\chi_{\theta0} = \chi_{r0} = \phi_0 = 0 = t_0$.  We denote with a ``check-mark'' accent any quantity which is defined along the fiducial geodesic.  For instance, $\check r(\lambda)$ is orbital radius along the fiducial geodesic, $\check\theta(\lambda)$ is the polar angle $\theta$ along the fiducial geodesic.  For non-fiducial geodesics, we define $\lambda = \lambda_{r0}$ to be the smallest positive value of $\lambda$ at which $r = r_{\rm min}$; likewise\footnote{These definitions account for the fact that these motions are periodic: $r(\lambda_{r0} + n\Lambda_r) = r_{\rm min}$ for any integer $n$, and $\theta(\lambda_{\theta0} + k\Lambda_\theta) = \theta_{\rm min}$ for any integer $k$.} $\lambda = \lambda_{\theta0}$ is the smallest positive value of $\lambda$ at which $\theta = \theta_{\rm min}$.  This means that
\begin{equation}
r(\lambda) = \check r(\lambda - \lambda_{r0})\;,
\quad
\theta(\lambda) = \check \theta(\lambda - \lambda_{\theta0})\;,
\end{equation}
There is a one-to-one correspondence between $\lambda_{\theta0}$ and $\chi_{\theta0}$, and between $\lambda_{r0}$ and $\chi_{r0}$.  A useful corollary is
\begin{eqnarray}
\Delta t_r(\lambda) &=& \Delta\check t_r(\lambda - \lambda_{r0}) - \Delta\check t_r(-\lambda_{r0})\;,
\nonumber\\
\Delta t_\theta(\lambda) &=& \Delta\check t_\theta(\lambda - \lambda_{\theta0}) - \Delta\check t_\theta(-\lambda_{\theta0})\;,
\end{eqnarray}
with analogous formulas for $\Delta\phi_r$ and $\Delta\phi_\theta$.

Some of our definitions differ from those used in Ref.\ {\cite{Drasco:2005is}}.  In that reference\footnote{Reference \cite{Drasco:2005is} also used slightly different symbols for the anomaly angles, writing $\chi$ for the polar anomaly, and $\psi$ for radial.  Here, we only use $\psi$ for the Newman-Penrose curvature scalars.}, $\chi_{\theta0}$ and $\chi_{r0}$ were not used.  Instead, $\chi_\theta = 0$ corresponded to $\lambda = \lambda_{\theta0}$, and $\chi_r = 0$ corresponded to $\lambda = \lambda_{r\theta}$.  As we discuss briefly in Secs.\ \ref{sec:diss_evolve} and \ref{sec:conclude}, the angles $\chi_{r0}$ and $\chi_{\theta0}$ will play an important role going beyond adiabatic waveforms.  In the adiabatic approximation, the angles $\chi_{r0}$, $\chi_{\theta0}$, and $\phi_0$ are constant as we move from geodesic to geodesic.  When we include, for example, orbit-averaged conservative self-force effects or orbit-averaged spin-curvature forces, we find secularly accumulating phases associated with each of these motions.  Allowing the angles $\chi_{r0}$, $\chi_{\theta0}$, and $\phi_0$ to evolve during inspiral is a simple and robust way to ``upgrade'' this framework to include these next-order effects.

\section{Adiabatic evolution and waveform amplitudes via the Teukolsky equation}
\label{sec:amps_and_evolve}

The next critical ingredient to constructing an adiabatic inspiral is the backreaction which arises from the orbit-averaged self interaction.  The quantities which encode the backreaction also tell us the amplitude of the inspiral's associated gravitational waveform.  In this section, we briefly summarize how these quantities are calculated using the Teukolsky equation.  As with Sec.\ \ref{sec:geodesics}, the contents of this section are discussed at length elsewhere, but are summarized here to introduce critical quantities and concepts important for later parts of this analysis.

\subsection{Solving the Teukolsky equation}

The Teukolsky equation \cite{Teukolsky:1973ha} describes perturbations to the Weyl curvature of Kerr black holes.  The version that we will use in this analysis focuses on the Newman-Penrose curvature scalar
\begin{equation}
\psi_4 = -C_{\alpha\beta\gamma\delta}n^\alpha\bar m^\beta n^\gamma\bar m^\delta\;,
\end{equation}
where $C_{\alpha\beta\gamma\delta}$ is the Weyl curvature tensor, and $n^\alpha$ and $\bar m^\alpha$ are legs of the Newman-Penrose null tetrad \cite{Newman:1961qr}.  Teukolsky showed that $\psi_4$ is governed by the equation
\begin{widetext}
\begin{eqnarray}
&&
\left[\frac{(r^2 + a^2)^2 }{\Delta} - a^2\sin^2\theta\right]\partial^2_{t}\Psi - 4\left[r + ia\cos\theta - \frac{M(r^2 - a^2)}{\Delta}\right]\partial_t\Psi +\frac{4 M a r}{\Delta}\partial_\phi\partial_t\Psi- \Delta^{2}\partial_r\left(\Delta^{-1}\partial_r\Psi\right) 
\nonumber\\
&&
- \frac{1}{\sin\theta}\partial_\theta \left(\sin\theta\partial_\theta\Psi\right) + \left[\frac{a^2}{\Delta}   -\frac{1}{\sin^2\theta}\right]\partial_\phi^2 \Psi + 4 \left[\frac{a (r - M)}{\Delta} + \frac{i \cos\theta}{\sin^2\theta} \right]\partial_\phi\Psi + \left(4\cot^2\theta + 2\right) \Psi = 4\pi\Sigma{\cal T}\;.
\label{eq:teuk}
\end{eqnarray}
\end{widetext}
The field $\Psi = (r - ia\cos\theta)^4\psi_4$, and ${\cal T}$ is a source term whose precise form is not needed here.  See Ref.\ \cite{Teukolsky:1973ha} for additional details and definitions.

An important point for our analysis is that 
\begin{equation}
\psi_4 = \frac{1}{2}\frac{d^2}{dt^2}\left(h_+ - ih_\times\right)\quad\mbox{as $r\to\infty$}\;,
\label{eq:psi4_to_h}
\end{equation}
so $\psi_4$ far from the source encodes the emitted gravitational waves.  These solutions also encode contributions to the rates of change of $E$, $L_z$, and $Q$ from gravitational-wave backreaction.  This is equivalent to the orbit-averaged self interaction arising from fields which are regular far from the source (see Ref.\ \cite{Quinn:1999kj}, as well as additional discussion on this point in Ref.\ \cite{Flanagan:2012kg}).  As $r \to r_+ \equiv M + \sqrt{M^2 - a^2}$ (the coordinate radius of the event horizon), $\psi_4$ encodes tidal interactions of the orbiting body with the black hole's event horizon.  These solutions encode contributions to the rates of change of $E$, $L_z$, and $Q$ from radiation absorbed by the horizon, which is equivalent to the orbit-averaged self interaction arising from fields which are regular on the event horizon \cite{Quinn:1999kj,Flanagan:2012kg}.  Knowledge of $\psi_4$ in the limits $r \to \infty$ and $r \to r_+$ provides all the data we need to construct adiabatic inspirals.

The frequency-domain approach we use to solve the Teukolsky equation begins by writing $\psi_4$ in a Fourier and multipolar expansion.  Writing
\begin{widetext}
\begin{equation}
\psi_4 = \frac{1}{(r - ia\cos\vartheta)^4}\int_{-\infty}^{\infty}d\omega\sum_{l = 2}^\infty\sum_{m = -l}^l R_{lm}(r;\omega)S_{lm}(\vartheta;a\omega)e^{i[m\varphi - \omega (t - t_0)]}\;,
\label{eq:psi4decomp}
\end{equation}
\end{widetext}
Eq.\ (\ref{eq:teuk}) separates \cite{Teukolsky:1973ha}, with ordinary differential equations governing $R_{lm}(r, \omega)$ and $S_{lm}(\vartheta,a\omega)$.

The field $\psi_4$ is measured at the event $(t, r, \vartheta, \varphi)$.  (Note the distinction between the orbit's polar and axial angles, $\theta$ and $\phi$, and the polar and axial angles at which the field is measured, $\vartheta$ and $\varphi$.)  The function $S_{lm}(\vartheta;a\omega)$ is a spheroidal harmonic (of spin-weight $-2$, left out for brevity); this function and methods for computing it are discussed at length in Appendix A of Ref.\ \cite{Hughes:1999bq}.  For reasons we will explain below, we have also introduced the initial time $t_0$ into our expansion (\ref{eq:psi4decomp}).

The separated radial dependence has simple asymptotic behavior:
\begin{eqnarray}
R_{lm}(r,\omega) &\to& Z^\infty_{lm\omega}r^3 e^{i\omega r_*}\;,\qquad\qquad r\to\infty\;,
\label{eq:Rlmomega_infty}\\
&\to& Z^{\rm H}_{lm\omega}\Delta e^{-i(\omega - m\Omega_H)r_*}\;,\; r\to r_+\;.
\label{eq:Rlmomega_H}
\end{eqnarray}
(We have absorbed a coefficient $C^{\rm trans}_{lm\omega}$ into the definition of $Z^{\infty}_{lm\omega}$, and a coefficient $B^{\rm trans}_{lm\omega}$ into the definition of $Z^{\rm H}_{lm\omega}$; see Refs.\ \cite{Mano:1996vt,Fujita:2004rb,Fujita:2009uz} for further discussion of these quantities.)  These asymptotic forms depend on the ``tortoise coordinate,''
\begin{eqnarray}
r_*(r) &=& r + \frac{Mr_+}{\sqrt{M^2 - a^2}}\ln\left(\frac{r}{r_+} - 1\right)
\nonumber\\
& &- \frac{Mr_-}{\sqrt{M^2 - a^2}}\ln\left(\frac{r}{r_-} - 1\right)\;,
\end{eqnarray}
where $r_{\pm} = M \pm \sqrt{M^2 - a^2}$.  The frequency
\begin{equation}
\Omega_{\rm H} = \frac{a}{2Mr_+}
\end{equation}
is often called the rotation frequency of the horizon.  It describes the frequency at which an observer held infinitesimally outside the event horizon will orbit the black hole as seen by distant observers.  

The quantities $Z^{\infty,{\rm H}}_{lm\omega}$ are computed by integrating homogeneous solutions of the radial Teukolsky equation against the source term of the separated radial Teukolsky equation.  Further detailed discussion can be found in Ref.\ \cite{Drasco:2005kz}, with updates to notation and minor corrections in Ref.\ \cite{OSullivan:2014ywd}.  Of importance for this analysis is that these quantities are computed by evaluating integrals of the form
\begin{equation}
Z^{\infty,{\rm H}}_{lm\omega} = \int_{-\infty}^\infty d\tau\,e^{i\omega [t(\tau) - t_0]} e^{-im\phi(\tau)}I^{\infty,{\rm H}}_{lm\omega}[r(\tau), \theta(\tau)]\;.
\label{eq:Zlmw1}
\end{equation}
The integration variable $\tau$ is proper time along the geodesic, and we subtract off $t_0$ because it is already accounted for in Eq.\ (\ref{eq:psi4decomp}).  The function $I^{\infty,{\rm H}}_{lm\omega}(r,\theta)$ is discussed in Refs.\ \cite{Drasco:2005kz} and \cite{OSullivan:2014ywd}; schematically, it is a Green's function used to solve the radial Teukolsky equation, multiplied by this equation's source term.  Using the properties of Kerr geodesic orbits and using the methods developed in Refs.\ \cite{Fujita:2004rb,Fujita:2009uz,Mano:1996vt}, it is well understood how to build $I^{\infty,{\rm H}}_{lm\omega}[r(\tau),\theta(\tau)]$.

Changing integration variable from proper time $\tau$ to Mino time $\lambda$, and using Eqs.\ (\ref{eq:t_of_lambda}) and (\ref{eq:phi_of_lambda}), this becomes
\begin{equation}
Z^{\infty,{\rm H}}_{lm\omega} = e^{-i m\phi_0}\int_{-\infty}^\infty d\lambda\, e^{i(\omega\Gamma - m\Upsilon_\phi)\lambda} J^{\infty,{\rm H}}_{lm\omega}[r(\lambda),\theta(\lambda)]\;,
\label{eq:Zlmw2}
\end{equation}
where we have introduced
\begin{eqnarray}
J^{\infty,{\rm H}}_{lm\omega}(r,\theta) &=& (r^2 + a^2\cos^2\theta)I^{\infty,{\rm H}}_{lm\omega}(r,\theta) \times
\nonumber\\
& & e^{i\omega\left[\Delta t_r(r) + \Delta t_\theta(\theta)\right]} e^{-im\left[\Delta\phi_r(r) + \Delta\phi_\theta(\theta)\right]}\;.
\nonumber\\
\end{eqnarray}
By virtue of the periodicity of orbit's $r$ and $\theta$ motions with respect to Mino time, the function $J^{\infty,{\rm H}}_{lm\omega}$ can be written as a double Fourier series:
\begin{equation}
J^{\infty,{\rm H}}_{lm\omega} = \sum_{k=-\infty}^\infty\sum_{n = -\infty}^\infty J^{\infty,{\rm H}}_{lmkn} e^{-i(k\Upsilon_\theta + n\Upsilon_r)\lambda}\;,
\label{eq:Jlmwexpand}
\end{equation}
where
\begin{widetext}
\begin{equation}
J^{\infty,{\rm H}}_{lmkn} = \frac{1}{\Lambda_r\Lambda_\theta}\int_0^{\Lambda_r}d\lambda_r\,e^{in\Upsilon_r\lambda_r}
\int_0^{\Lambda_\theta}d\lambda_\theta\,e^{ik\Upsilon_\theta\lambda_\theta}\,J^{\infty,{\rm H}}_{lm\omega}[r(\lambda_r),\theta(\lambda_\theta)]\;.
\label{eq:Jlmkn_def}
\end{equation}
\end{widetext}
Combining Eqs.\ (\ref{eq:Zlmw2}), (\ref{eq:Jlmwexpand}), and (\ref{eq:Jlmkn_def}) plus the relations $\Omega_{r,\theta,\phi} = \Upsilon_{r,\theta,\phi}/\Gamma$, we find that
\begin{equation}
Z^{\infty,{\rm H}}_{lm\omega} = \sum_{k = -\infty}^\infty\sum_{n = -\infty}^\infty Z^{\infty,{\rm H}}_{lmkn}\delta(\omega - \omega_{mkn})\;,
\label{eq:ZlmknZlmw}
\end{equation}
where
\begin{equation}
\omega_{mkn} = m\Omega_\phi + k\Omega_\theta + n\Omega_r
\end{equation}
and
\begin{equation}
Z^{\infty,{\rm H}}_{lmkn} = e^{-i m\phi_0}J^{\infty,{\rm H}}_{lmkn}/\Gamma\;.
\label{eq:Zlmkn_geod}
\end{equation}
These coefficients have the symmetry
\begin{equation}
Z^{\infty,{\rm H}}_{l,-m,-k,-n} = (-1)^{(l+k)}\bar Z^{\infty,{\rm H}}_{lmkn}\;,
\label{eq:Zlmkn_sym}
\end{equation}
where overbar denotes complex conjugation.  Our code respects the symmetry (\ref{eq:Zlmkn_sym}) to double-precision accuracy.  We take advantage of this by computing $Z^{\infty,{\rm H}}_{lmkn}$ for all $l$, all $m$, all $k$, and $n \ge 0$, then using Eq.\ (\ref{eq:Zlmkn_sym}) to infer results for $n < 0$.  This cuts the amount of computation roughly in half.

As we describe in the following two subsections, the coefficients $Z^\infty_{lmkn}$ set the amplitude of the gravitational waves from a specified geodesic orbit, and also encode the contribution to the orbit-averaged self force from fields that are regular at null infinity; the coefficients $Z^{\rm H}_{lmkn}$ encode the contribution to the orbit-averaged self force from fields that are regular on the event horizon.  These sets of coefficients are thus of crucial importance for computing adiabatic inspiral and its gravitational waveform.

\subsection{The gravitational waveform from an orbit}

As shown in Sec.\ 8 of Ref.\ {\cite{Drasco:2005is}}, the phase of $Z^{\infty,{\rm H}}_{lmkn}$ depends on the values of $\lambda_{r0}$ and $\lambda_{\theta0}$, which in turn depend on the initial anomaly angles $\chi_{r0}$ and $\chi_{\theta0}$.  We denote by $\check Z^{\infty,{\rm H}}_{lmkn}$ the value of $Z^{\infty,{\rm H}}_{lmkn}$ for the fiducial geodesic.  For a general geodesic,
\begin{equation}
Z^{\infty,{\rm H}}_{lmkn} = e^{i\xi_{mkn}}\check Z^{\infty,{\rm H}}_{lmkn}\;,
\end{equation}
where the correcting phase is
\begin{eqnarray}
\xi_{mkn} &=& k\Upsilon_\theta\lambda_{\theta0} + n\Upsilon_r\lambda_{r0}
\nonumber\\
&+& m\left[\Delta\check\phi_r(-\lambda_{r0}) + \Delta\check\phi_\theta(-\lambda_{\theta0}) - \phi_0\right]
\nonumber\\
&-&  \omega_{mkn}\left[\Delta\check t_r(-\lambda_{r0}) + \Delta\check t_\theta(-\lambda_{\theta0})\right]\;.
\label{eq:ximkn}
\end{eqnarray}
The form of $\xi_{mkn}$ we use is slightly different from that derived in Ref.\ \cite{Drasco:2005is}.  In particular, we have separated out the dependence on the initial axial phase $\phi_0$, as shown in Eq.\ (\ref{eq:Zlmkn_geod}), and the dependence on $t_0$, which is discussed further in Sec.\ \ref{sec:diss_evolve}.  See Ref.\ {\cite{Drasco:2005is}} for discussion and derivation of the dependence on $\lambda_{\theta0}$ and $\lambda_{r0}$.

The fact that the initial conditions only influence these amplitudes via the phase factor $\xi_{mkn}$ means that we only need to compute and store quantities on the fiducial geodesic.  Using Eq.\ (\ref{eq:ximkn}), we can then easily convert these results to any initial condition.  This vastly cuts down on the amount of computation that must be done to cover the space of physically important EMRI systems.  In addition, notice that $\xi_{mkn}$ can be written
\begin{equation}
\xi_{mkn} = m\xi_{100} + k\xi_{010} + n\xi_{001}\;.
\end{equation}
For each orbit one need only compute ($\xi_{100},\xi_{010},\xi_{001}$) in order to know the phases for all $(l,m,k,n)$.

As we will see below, the values of these phases are irrelevant for the orbit-averaged backreaction\footnote{The phases are {\it not} irrelevant for backreaction as we go beyond the adiabatic limit, and indeed play an important role in determining the strength of backreaction at orbital resonances \cite{Flanagan:2012kg}.}, but they are critical for getting the phase of the gravitational waveform correct given initial conditions.  To write down the gravitational waves, we use Eq.\ (\ref{eq:psi4_to_h}) to relate $h$ to $\psi_4$ far from the source.  Combining Eqs.\ (\ref{eq:psi4decomp}), (\ref{eq:Rlmomega_infty}), and (\ref{eq:ZlmknZlmw}), we further know that
\begin{equation}
\psi_4 = \frac{1}{r}\sum_{lmkn} Z^\infty_{lmkn} S_{lm}(\vartheta, a\omega_{mkn})e^{i[m\varphi - \omega_{mkn}(t - t_0)]}
\label{eq:psi4decomp2}
\end{equation}
as $r \to \infty$.  Here and in what follows, any sum over the set $(l,m,k,n)$ will be assumed to be from $2$ to $\infty$ for $l$, from $-l$ to $l$ for $m$, and from $-\infty$ to $\infty$ for $k$ and $n$.  Let us define
\begin{eqnarray}
h &\equiv& h_+ - i h_\times \equiv \frac{1}{r}\sum_{lmkn}h_{lmkn}
\nonumber\\
&=& \frac{1}{r}\sum_{lmkn} A_{lmkn}S_{lm}(\vartheta;a\omega_{mkn})e^{i[m\varphi - \omega_{mkn}(t - t_0)])}\;.
\nonumber\\
\label{eq:Adef}
\end{eqnarray}
Combining Eqs.\ (\ref{eq:psi4_to_h}), (\ref{eq:psi4decomp2}), and (\ref{eq:Adef}), we see that
\begin{equation}
A_{lmkn} = -\frac{2 Z^\infty_{lmkn}}{\omega_{mkn}^2}\;.
\label{eq:Hlmkndef}
\end{equation}
As with $Z^\infty_{lmkn}$, we define $\check A_{lmkn}$ to be the wave amplitude for the fiducial geodesic, and we have
\begin{equation}
A_{lmkn} = e^{i\xi_{mkn}}\check A_{lmkn}\;.
\end{equation}
The data $\check A_{lmkn}$ interpolate very well and should be stored for generating inspiral waveforms.  It will be convenient for later discussion to further define
\begin{equation}
H_{lmkn} = A_{lmkn}S_{lm}(\vartheta;a\omega_{mkn})\;
\end{equation}
Because spheroidal harmonics slowly change along an inspiral as $\omega_{mkn}$ evolves, we find it useful to examine $H_{lmkn}$ rather than $A_{lmkn}$ when computing wave amplitudes during inspiral.

\subsection{Adiabatic backreaction}
\label{sec:backreaction}

Here we summarize how to use the coefficients $Z^{\infty,{\rm H}}_{lmkn}$ to compute the adiabatic dissipative evolution, or backreaction, on a geodesic.  We assume that a body is on a Kerr geodesic orbit, and so is characterized (up to initial conditions) by the orbital integrals $E$, $L_z$, and $Q$.  Results for $dE/dt$ and $dL_z/dt$ have been known for quite a long time \cite{Teukolsky:1974yv}; these quantities each split into a contribution from fields that are regular at infinity, which can be extracted from knowledge of the distant gravitational radiation, and a contribution from fields that are regular on the black hole's event horizon.  Computing the down-horizon contribution is a little more tricky; one must compute how tidal stresses shear the horizon's generators, increasing its surface area (or its entropy), and then apply the first law of black hole dynamics to infer the change in the hole's mass and spin.  Results for $dQ/dt$ were first derived by Sago et al.\ \cite{Sago:2005fn}, and are found by averaging the action of the dissipative self force on a geodesic.  It also separates into pieces that arise from fields regular at infinity and fields regular on the horizon.

\begin{widetext}

The results which we need for our analysis are:
\begin{eqnarray}
\left(\frac{dE}{dt}\right)^\infty &=& \sum_{lmkn} \frac{|Z^\infty_{lmkn}|^2}{4\pi\omega_{mkn}^2}\;,\qquad\qquad\qquad\qquad\,\,
\left(\frac{dE}{dt}\right)^{\rm H} = \sum_{lmkn} \frac{\alpha_{lmkn}|Z^{\rm H}_{lmkn}|^2}{4\pi\omega_{mkn}^2}\;,
\label{eq:dEdt}\\
\left(\frac{dL_z}{dt}\right)^\infty &=& \sum_{lmkn} \frac{m|Z^\infty_{lmkn}|^2}{4\pi\omega_{mkn}^3}\;,\qquad\qquad\qquad\quad\,
\left(\frac{dL_z}{dt}\right)^{\rm H} = \sum_{lmkn} \frac{\alpha_{lmkn}m|Z^{\rm H}_{lmkn}|^2}{4\pi\omega_{mkn}^3}\;,
\label{eq:dLzdt}\\
\left(\frac{dQ}{dt}\right)^\infty &=& \sum_{lmkn} |Z^\infty_{lmkn}|^2\times\frac{({\cal L}_{mkn} + k\Upsilon_\theta)}{2\pi\omega_{mkn}^3}\;,\quad
\left(\frac{dQ}{dt}\right)^{\rm H} = \sum_{lmkn}\alpha_{lmkn}|Z^{\rm H}_{lmkn}|^2\times\frac{({\cal L}_{mkn} + k\Upsilon_\theta)}{2\pi\omega_{mkn}^3}\;.
\label{eq:dQdt}
\end{eqnarray}
In these equations,
\begin{eqnarray}
{\cal L}_{mkn} &=& m\langle\cot^2\theta\rangle L_z -a^2\omega_{mkn}\langle\cos^2\theta\rangle E\;,
\label{eq:Lmkndef}\\
\alpha_{lmkn} &=& \frac{256(2Mr_+)^5(\omega_{mkn} - m\Omega_{\rm H})[(\omega_{mkn} - m\Omega_{\rm H})^2 + 4\epsilon^2][(\omega_{mkn} - m\Omega_{\rm H})^2 + 16\epsilon^2]\omega_{mkn}^3}{|C_{lmkn}|^2}\;.
\label{eq:alphadef}
\end{eqnarray}
The terms $\langle\cot^2\theta\rangle$ and $\langle\cos^2\theta\rangle$ in Eq.\ (\ref{eq:Lmkndef}) mean $\cot^2\theta$ and $\cos^2\theta$ evaluated at the $\theta$ coordinate along the orbit, and then averaged using Eq.\ (\ref{eq:thetaaveraging}).  The terms $|C_{lmkn}|^2$ and $\epsilon$ in Eq.\ (\ref{eq:alphadef}) are given by
\begin{eqnarray}
|C_{lmkn}|^2 &=& [(\lambda_{lmkn}^2 + 2)^2 + 4am\omega_{mkn} - 4a^2\omega_{mkn}^2](\lambda_{lmkn}^2 + 36am\omega_{mkn} - 36a^2\omega_{mkn}^2)
\nonumber\\
&+& (2\lambda_{lmkn} + 3)(96a^2\omega_{mkn}^2 - 48am\omega_{mkn}) + 144\omega_{mkn}^2(M^2 - a^2)\;,
\label{eq:Clmknsq}\\
\epsilon &=& \frac{\sqrt{M^2 - a^2}}{4Mr_+}\;.
\label{eq:epsilon}
\end{eqnarray}
\end{widetext}
The quantity $\lambda_{lmkn}$ appearing in Eq.\ (\ref{eq:Clmknsq}) is one form of the eigenvalue of the spheroidal harmonic; see Ref.\ \cite{Hughes:1999bq} for discussion of the algorithm we use to compute it, and Appendix C of Ref.\ \cite{OSullivan:2014ywd} for discussion of the various forms of the eigenvalue in the literature (which are simple to convert between).  Note that the factor $\epsilon$ which appears here (and is not used elsewhere in this paper) is distinct from the similar symbol $\varepsilon \equiv \mu/M$, the mass ratio.

In evaluating Eqs.\ (\ref{eq:dEdt}), (\ref{eq:dLzdt}), and (\ref{eq:dQdt}), we must truncate the infinite sums, with cutoffs determined by the needs of the analysis in question.  For the purpose of this paper, we have implemented the following cutoffs:

\begin{itemize}

\item We include all $l$ from 2 to 10.

\item At each $l$, we include all $m$ from $-l$ to $l$.

\item We truncate the $k$ sum when the fractional change to the accumulated sum is smaller than $10^{-5}$ for three consecutive values of $k$.  Holding all other indices fixed, contributions to this sum tend to fall off monotonically and fairly rapidly with increasing $k$, so in practice this condition means that neglected terms change the sum by less than several $\times$ $10^{-7}$.

\item We truncate the $n$ sum when the fractional change to the accumulated sum is smaller than $10^{-6}$ for three consecutive values of $n$.  Especially for $e \gtrsim 0.4$, contributions to this sum do not fall off monotonically with $n$ until some threshold has been passed (see Figs.\ 5 and 6 of Ref.\ \cite{Cutler:1994pb}, and Figs.\ 2 and 3 of Ref.\ \cite{Drasco:2005kz}).  Once past that threshold, convergence is quick, and we find that neglected terms in this case also change the sum by less than several $\times$ $10^{-7}$.

\end{itemize}

We emphasize that these cutoffs have not been selected carefully, but are simply chosen for ease of calculation and to produce results which are ``converged enough'' for the exploratory purposes of this paper.  A more careful analysis and assessment of how to truncate these sums should be done before using these ideas to make ``production quality'' waveforms (e.g., for exploring LISA data analysis questions, or science return with EMRI measurements) in order to make sure that systematic errors in waveform modeling are understood and do not adversely affect one's analysis.

As $E$, $L_z$, and $Q$ change, we require the system to evolve from one geodesic to another.  To do this, we let these orbital integrals change by enforcing a balance law:
\begin{equation}
\left(\frac{d{\cal C}}{dt}\right)^{\rm orbit} = -\left(\frac{d{\cal C}}{dt}\right)^\infty - \left(\frac{d{\cal C}}{dt}\right)^{\rm H}\;,
\end{equation}
for ${\cal C} \in [E,L_z,Q]$.  As these orbital integrals evolve, the orbit's geometry slowly changes in order to keep the system on a geodesic trajectory.  Appendix \ref{app:jacobian} shows how to relate rates of change for $p$, $e$, and $x_I$ to the rates of change of $E$, $L_z$, and $Q$.

\section{Adiabatic inspiral as dissipative evolution along a sequence of geodesics}
\label{sec:diss_evolve}

We now examine solutions to the Teukolsky equation for a slowly evolving source.   Critical to our analysis is the idea of a two-timescale expansion: the waveform phase varies on a ``fast'' orbital timescale $T_{\rm o} \sim M$, and orbit characteristics vary on a ``slow'' inspiral timescale $T_{\rm i} \sim M^2/\mu$.  The two timescales differ by the system's mass ratio: $T_{\rm o}/T_{\rm i} = \mu/M \equiv \varepsilon$.  The waveforms we compute in this way are accurate up to corrections of order $\varepsilon$.

Suppose that we have used the rates of change $dE/dt$, $dL_z/dt$, $dQ/dt$ to compute how a system evolves from geodesic to geodesic.  We parameterize inspiral by a bookkeeper time $t^{\rm i}$ which measures evolution along the inspiral as seen by a distant observer.  We treat the inspiral as a geodesic at each moment $t^{\rm i}$, and call the geodesic at this moment the ``osculating" geodesic \cite{Lincoln:1990ji,Pound:2007th,Gair:2010iv}.   At each such moment, the osculating geodesic's energy $E(t^{\rm i})$, its angular momentum $L_z(t^{\rm i})$, and its Carter constant $Q(t^{\rm i})$ are known.  By our assumption that the inspiral is a geodesic at each moment, we can reparameterize and determine $p(t^{\rm i})$, $e(t^{\rm i})$, and $x_I(t^{\rm i})$.  We can also compute quantities such as the frequencies $\Omega_{r,\theta,\phi}(t^{\rm i})$ and the amplitudes $Z^\infty_{lmkn}(t^{\rm i})$ for each geodesic in this sequence.

To leading order in $\varepsilon$, the curvature scalar $\psi_4$ that arises from this sequence of geodesics is given by
\begin{eqnarray}
\psi_4(t^{\rm i}) &=& \frac{1}{r}\sum_{lmkn}Z^\infty_{lmkn}(t^{\rm i})S_{lm}[\vartheta; a\omega_{mkn}(t^{\rm i})]
\nonumber\\
& &\quad \times e^{i[m\varphi - \Phi_{mkn}(t^{\rm i})]}\;.
\label{eq:psi4_inspiral}
\end{eqnarray}
This is Eq.\ (\ref{eq:psi4decomp2}), but with the amplitude $Z^\infty_{lmkn}$ and the frequency $\omega_{mkn}$ now functions of $t^{\rm i}$.  Notice the dependence on harmonics of the accumulated orbital phase:
\begin{equation}
\Phi_{mkn}(t^{\rm i}) = \int_{t_0}^{t^{\rm i}} \omega_{mkn}(t')\,dt'\;.
\label{eq:Phimkn}
\end{equation}
This reduces to $\omega_{mkn} (t^{\rm i} - t_0)$ in the limit that the orbit does not inspiral.  Equation (\ref{eq:Phimkn}) builds in the dependence on the initial time $t_0$, which is why we leave this parameter out of the factor $\xi_{mkn}$ in Eq.\ (\ref{eq:ximkn}).

To justify this inspiraling solution for $\psi_4$, substitute Eq.\ (\ref{eq:psi4_inspiral}) into the Teukolsky equation, Eq.\ (\ref{eq:teuk}).  Doing so, one finds that it satisfies the Teukolsky equation up to errors of order the orbital timescale over the inspiral timescale, $O(T_{\rm o}/T_{\rm i}) \sim O(\varepsilon)$.  These errors in turn arise from the fact that time derivatives have both fast-time contributions, for which $\partial_t \sim 1/T_{\rm o} \sim \omega_{mkn}$, as well as slow-time contributions, for which $\partial_t \sim 1/T_{\rm i} \sim \varepsilon/T_{\rm o}$.  In the adiabatic approximation, we neglect the slow-time derivatives, expecting that at any moment their contribution will be small as long as the system's mass ratio is large.  Some of the errors arising from this neglect can accumulate secularly, leading to phase errors up to several radians after an inspiral.  Post-adiabatic corrections will change the amplitude and phase of Eq.\ (\ref{eq:psi4_inspiral}) in such a way as to correct the adiabatic approximation's fast time over slow time errors.  See \cite{Miller:2020bft} for further discussion.

In our applications, we are typically more interested in the waveform $h(t^{\rm i})$ than in $\psi_4(t^{\rm i})$.  This is given by
\begin{equation}
h(t^{\rm i}) \equiv \!\sum_{lmkn} h_{lmkn}(t^{\rm i})
= \frac{1}{r}\sum_{lmkn}H_{lmkn}(t^{\rm i}) e^{i[m\varphi - \Phi_{mkn}(t^{\rm i})]}\;,
\label{eq:h_inspiral}
\end{equation}
where
\begin{equation}
H_{lmkn}(t^{\rm i}) = A_{lmkn}(t^{\rm i})S_{lm}[\vartheta;a\omega_{mkn}(t^{\rm i})]
\end{equation}
and
\begin{equation}
A_{lmkn}(t^{\rm i}) = -\frac{2Z^\infty_{lmkn}(t^{\rm i})}{\omega_{mkn}(t^{\rm i})^2}\;.
\end{equation}
Equation (\ref{eq:h_inspiral}) showcases the ``multivoice'' structure of EMRI waveforms: each $h_{lmkn}(t^{\rm i})$ that contributes to $h(t^{\rm i})$ constitutes a single ``voice'' in this waveform.  As we will see in Secs.\ \ref{sec:results_schw} and \ref{sec:results_kerr}, even when the waveform is complicated, each voice tends to be quite simple.

As discussed in Sec.\ \ref{sec:amps_and_evolve}, we can compute the amplitudes $\check A_{lmkn}$ for the fiducial geodesic and then correct using the phase factor $\xi_{mkn}$ in order to get the wave amplitude for our system's initial conditions.  Let us imagine we have computed $\check A_{lmkn}$ on a dense grid of orbits.  Knowing $[p(t^{\rm i}, e(t^{\rm i}), x_I(t^{\rm i})]$, it is then simple to construct the fiducial amplitudes $\check A_{lmkn}(t^{\rm i})$ along an inspiral.

To convert from the fiducial amplitudes to $A_{lmkn}(t^{\rm i})$, we need the phase factor $\xi_{mkn}(t^{\rm i})$ at each moment along the inspiral.  Recall that $\xi_{mkn}$ depends on the angles $\chi_{\theta0}$ and $\chi_{r0}$ which set the polar and radial initial conditions.  An important feature of the adiabatic approximation is that these angles are {\it constant}: $\chi_{\theta0}$ and $\chi_{r0}$ do not change as inspiral proceeds.  However, the mapping between ($\chi_{\theta0}, \chi_{r0}$) and $\xi_{mkn}$ {\it does} change as inspiral proceeds, leading to a slow evolution in this phase factor.  When post-adiabatic physics is included, this story changes: the angles $\chi_{\theta0}$ and $\chi_{r0}$ slowly evolve under the influence of orbit-averaged conservative self forces, and orbit-averaged spin-curvature interactions.  This will change the slow evolution of $\xi_{mkn}$, and is one way in which conservative effects leave an observationally important imprint on EMRI waveforms.

\section{Frequency-domain description}
\label{sec:fourier}

Our description so far has focused on presenting adiabatic EMRI waveforms in the time domain.  We showed that the time-domain waveform can be regarded as a superposition of different ``voices,'' each of which has its own slowly evolving amplitude.  In this section, we will exploit this multivoice structure to compute the Fourier transform of an EMRI waveform, thereby describing these waves in the frequency domain.

Because all quantities in an EMRI evolve slowly (as long as the two-timescale approximation is valid), our expectation is that the {\it stationary phase approximation} (SPA) will provide a high-quality approximation to the Fourier transform.  For some voices, the frequency evolution is not monotonic: some voices rises to a maximum frequency, and then their frequency decreases.  As we discuss below, it is conceptually straightforward to extend the ``standard'' SPA calculation in such a circumstance.  We begin by reviewing the standard calculation, then discuss voices whose frequency evolution is not monotonic.  We conclude this section by describing how to assemble a frequency-domain waveform with many voices.

\subsection{Standard SPA calculation}

Begin by assuming a single voice signal of the form
\begin{equation}
h_+(t) - i h_\times(t) \equiv h(t) = H(t) e^{-i\Phi(t)}\;.
\end{equation}
The Fourier transform of this is given by
\begin{eqnarray}
{\tilde h}(f) &\equiv& \int_{-\infty}^\infty h(t)e^{2\pi i f t}\,dt
\nonumber\\
&=& \int_{-\infty}^\infty H(t) e^{i[2\pi f t - \Phi(t)]}\,dt\;.
\label{eq:singlevoiceFT}
\end{eqnarray}
To compute the stationary phase approximation to the Fourier transform, expand the signal's phase as
\begin{equation}
\Phi(t) = \Phi(t_S) + 2\pi F (t - t_S) + \pi \dot F (t - t_S)^2 + \ldots\;.
\label{eq:phaseexpand1}
\end{equation}
We will define the time $t_S$ momentarily.  In Eq.\ (\ref{eq:phaseexpand1}), we have introduced the signal's instantaneous frequency and the instantaneous frequency derivative at $t = t_S$:
\begin{equation}
F \equiv \frac{1}{2\pi}\frac{d\Phi}{dt}\biggr|_{t_S}\;,\qquad
\dot F = \frac{dF}{dt} \equiv \frac{1}{2\pi}\frac{d^2\Phi}{dt^2}\biggr|_{t_S}\;.
\label{eq:freqinst}
\end{equation}
We will assume that $\dot F$ is small, in a sense to be made precise below.

Using these definitions, we rewrite the Fourier transform integral:
\begin{eqnarray}
{\tilde h}(f) &=& e^{-i\Phi(t_S)}\int_{-\infty}^\infty H(t) e^{2\pi i[ft - F(t - t_S) - (1/2)\dot F (t - t_S)^2]}\,dt
\nonumber\\
&=& e^{i[2\pi f t_S - \Phi(t_S)]}\times\nonumber\\
& &\int_{-\infty}^\infty H(t' + t_S) e^{2\pi i[ft' - Ft' - (1/2)\dot F(t')^2]}\,dt'\;.
\label{eq:fourierintegral}
\end{eqnarray}
On the second line, we changed the integration variable to $t' = t - t_S$.  The integrand of Eq.\ (\ref{eq:fourierintegral}) very rapidly oscillates {\it unless the Fourier frequency $f$ matches the instantaneous frequency $F$.}  When this condition is met, the phase is {\it stationary}: it is approximately constant, varying very slowly due to the contribution from $\dot F$, which is assumed to be small.

We take the time $t_S$ to be the time at which the phase is stationary.  Let us rewrite it $t(f)$, the time at which $F(t) = f$.  Under the assumption that the largest contribution to the integral comes from $t' \simeq 0$, or equivalently from $t \simeq t(f)$, we have
\begin{equation}
{\tilde h}(f) \simeq H[t(f)]e^{i[2\pi f t(f) - \Phi(t(f))]}\int_{-\infty}^\infty e^{-i \pi\dot F[t(f)](t')^2}dt'\;.
\label{eq:SPAintegral}
\end{equation}
This integral can be evaluated with standard methods, and we finally obtain
\begin{equation}
{\tilde h}(f) \simeq \frac{H[t(f)]e^{i[2\pi f t(f) - \Phi(t(f)) - \pi/4]}}{\sqrt{\dot F[t(f)]}}\;.
\end{equation}
Note that $\dot F$ can be positive or negative, and it appears under the square root.  To eliminate ambiguity about the phase of this voice in the frequency domain, we clean this up as follows:
\begin{equation}
{\tilde h}(f) \simeq \frac{H[t(f)]e^{i[2\pi f t(f) - \Phi(t(f)) \mp \pi/4]}}{\sqrt{|\dot F[t(f)]|}}\;.
\label{eq:standardspa}
\end{equation}
We choose the minus sign in the exponential if $\dot F > 0$, and the plus sign for $\dot F < 0$.  This approximation works well when both $F(t)$ and $H(t)$ change slowly,
\begin{equation}
\biggl|\frac{1}{F}\frac{dH}{dt}\biggl| \ll |H|\;,\quad 
\biggl|\frac{1}{F}\frac{dF}{dt}\biggl| \ll |F|\;.
\end{equation}
It also requires that the signal frequency $F$ evolves monotonically --- the sign of $dF/dt$ cannot change.

\subsection{Non-monotonic frequency}

What if our signal has a frequency which does not evolve monotonically?  In particular, what if $F$ rises to a maximum and then decreases, or falls to a minimum and then increases?  When this occurs, two problems arise with the standard SPA analysis.  First, in this circumstance there are multiple solutions to the condition $F(t) = f$.  The signal at frequency $f$ must include contributions from all times from which the signal frequency $F$ equals the Fourier frequency $f$.  Second, $\dot F = 0$ at the moment that the evolution of $F$ switches sign.  The standard SPA Fourier transform is singular at that point.  These issues affect EMRI signals, since the frequency associated with many voices rises to a maximum and then decreases.  In particular, this occurs for EMRI voices which involve harmonics of the radial frequency: $\Omega_r$ reaches a maximum in the strong field, then goes to zero as the inspiral approaches the last stable orbit.

The root cause of the singularity is that the standard SPA assumes $F(t)$ and $\dot F(t)$ completely describe the signal's phase.  If $\dot F$ vanishes at frequency $f$, then the calculation assumes all times contribute to the Fourier integral at $f$, and the integral (\ref{eq:SPAintegral}) diverges.  (This is consistent with the fact that the Fourier transform of a constant frequency signal is a delta function.)  However, for real EMRI signals, $F(t)$ is not constant when $\dot F = 0$; the singularity in the SPA analysis is an artifact of our assumption that $F(t)$ and $\dot F(t)$ completely characterize the signal's phase.  To remove this artifact, we need more information about the signal's phase evolution.  Let us therefore include the next term in the expansion:
\begin{equation}
\Phi(t) = \Phi(t_S) + 2\pi F (t - t_S) + \pi \dot F (t - t_S)^2 + \frac{\pi}{3} \ddot F (t - t_S)^3\ldots\,
\label{eq:phaseexpand2}
\end{equation}
where
\begin{equation}
\ddot F \equiv \frac{1}{2\pi}\frac{d^3\Phi}{dt^3}\biggr|_{t_S}\;.
\end{equation}

Now revisit Eq.\ (\ref{eq:singlevoiceFT}), but use (\ref{eq:phaseexpand2}) to expand the phase.  We again find that $t_S$ is the stationary time, for which $F = f$.  However, there may be multiple roots of the equation $F(t) = f$.  Let us define $t_j(f)$ to be the $j$th time at which $F(t) = f$, and write $\dot F_j \equiv \dot F[t_j(f)]$, $\ddot F_j \equiv \ddot F[t_j(f)]$.  For EMRI waveforms, $N \le 2$.  Each value of $j$ contributes to the Fourier transform, so we have
\begin{eqnarray}
{\tilde h}(f) &\simeq& \sum_{j=1}^NH[t_j(f)]e^{i[2\pi ft_j(f) - \Phi(t_j(f))]}\times
\nonumber\\
& &\qquad\int_{-\infty}^\infty e^{-i \pi[\dot F_j(t')^2 + \ddot F_j(t')^3/3]}dt'\;.
\label{eq:SPAintegral2}
\end{eqnarray}
To perform this integral, put $\alpha = \gamma + 2\pi i \dot F$, with $\gamma$ real and positive.  Define $\beta = 2\pi \ddot F$, and use
\begin{equation}
\int_{-\infty}^\infty e^{-\alpha t^2/2 - i \beta t^3/6}\,dt = \frac{2}{\sqrt{3}}\frac{\alpha}{|\beta|}e^{\alpha^3/3\beta^2} K_{1/3}(\alpha^3/3\beta^2)\;,
\label{eq:Iintegral}
\end{equation}
where $K_n(z)$ is the modified Bessel function of the 2nd kind.  Taking the limit $\gamma \to 0$, we find
\begin{widetext}
\begin{equation}
{\tilde h}(f) \simeq \frac{2}{\sqrt{3}}\sum_{j=1}^NH[t_j(f)]e^{i[2\pi ft_j(f) - \Phi(t_j(f))]}\frac{i\dot F_j}{|\ddot F_j|}e^{-2\pi i\dot F_j^3/3\ddot F_j^2}K_{1/3}(-2\pi i \dot F_j^3/3\ddot F_j^2)\;.
\label{eq:extendedspa1}
\end{equation}
This result defines our ``extended'' SPA.

It is useful to examine two limits of Eq.\ (\ref{eq:extendedspa1}).  To facilitate taking these limits, we define
\begin{equation}
X_j \equiv \frac{2\pi}{3}\frac{\dot F_j^3}{\ddot F_j^2}\;.
\label{eq:Xdef}
\end{equation}
First, we take $\dot F_j$ to be arbitrary, and expand about $\ddot F_j = 0$.  To set this up this, eliminate $\ddot F_j$ in Eq.\ (\ref{eq:extendedspa1}):
\begin{equation}
{\tilde h}(f) \simeq \sqrt{\frac{2}{\pi}}\sum_{j=1}^NH[t_j(f)]e^{i[2\pi ft_j(f) - \Phi(t_j(f))]}\frac{\dot F_j}{\sqrt{\dot F_j^3}}i\sqrt{X_j}e^{-iX_j}K_{1/3}(-iX_j)\;.
\label{eq:extendedspa_v2}
\end{equation}
Expanding about $\ddot F_j$ is equivalent to examining Eq.\ (\ref{eq:extendedspa_v2}) for $X_j \to \pm\infty$.  Using
\begin{eqnarray}
\frac{\dot F_j}{\sqrt{\dot F_j^3}} &=& |\dot F_j|^{-1/2}\qquad \dot F_j > 0\;,
\nonumber\\
&=& i|\dot F_j|^{-1/2} \qquad \dot F_j < 0\;,
\end{eqnarray}
and, for $X$ real,
\begin{equation}
\lim_{X\to \pm\infty} i\sqrt{X} e^{-iX}K_{1/3}(-iX) = \sqrt{\frac{\pi}{2}}e^{-i\pi/4}\left(1 - \frac{5i}{72}\frac{1}{X} + \ldots\right)\;,
\end{equation}
we have
\begin{equation}
{\tilde h}(f) = \sum_{j = 1}^N\frac{H[t_j(f)]e^{i[2\pi ft_j(f) - \Phi(t_j(f)) \mp \pi/4]}}{|\dot F_j|^{1/2}}\left(1 - \frac{5i}{48\pi}\frac{\ddot F_j^2}{\dot F_j^{3}} + \ldots\right)\;.
\label{eq:extendedspa_approx1}
\end{equation}
The minus sign in the exponential is for $\dot F > 0$, the plus sign for $\dot F < 0$.  Equation (\ref{eq:extendedspa_approx1}) is an accurate approximation when $|\ddot F_j|^2 \ll |\dot F_j|^3$.  Notice that we recover the standard SPA result, Eq.\ (\ref{eq:standardspa}), when $\ddot F_j = 0$ and $N=1$.

Next, allow $\ddot F_j$ to be any real value, and expand about $\dot F_j = 0$.  To do this, use Eq.\ (\ref{eq:Xdef}) to replace $\dot F_j$ with powers of $X_j$ and $\ddot F_j$ in Eq.\ (\ref{eq:extendedspa1}),
\begin{equation}
{\tilde h}(f) \simeq \frac{2^{2/3}}{3^{1/6}\pi^{1/3}}\sum_{j=1}^N H[t_j(f)]e^{i[2\pi ft_j(f) - \Phi(t_j(f))]}\frac{\ddot F_j^{2/3}}{|\ddot F_j|}i{X_j}^{1/3}e^{-iX_j}K_{1/3}(-iX_j)\;,
\label{eq:extendedspa_v3}
\end{equation}
and expand about $X_j = 0$.  Using
\begin{equation}
\lim_{X \to 0} i{X}^{1/3}e^{-iX}K_{1/3}(-iX) = \frac{e^{2\pi i/3}\Gamma(\frac{1}{3})}{2^{2/3}}\left[1 - X^{2/3}\frac{e^{2\pi i/3}\Gamma(-\frac{1}{3})}{2^{2/3}\Gamma(\frac{1}{3})}\right]\;,
\end{equation}
we find
\begin{equation}
{\tilde h}(f) \simeq \frac{e^{2\pi i/3}\Gamma(\frac{1}{3})}{3^{1/6}\pi^{1/3}}\sum_{j = 1}^N H[t_j(f)]e^{i[2\pi ft_j(f) - \Phi(t_j(f))]}\frac{\ddot F_j^{2/3}}{|\ddot F_j|}\left[1 - e^{2\pi i/3}\left(\frac{\pi}{3}\right)^{2/3}\frac{\Gamma(-\frac{1}{3})}{\Gamma(\frac{1}{3})}\frac{\dot F_j^2}{\ddot F_j^{4/3}} + \ldots\right]\;.
\label{eq:extendedspa_approx2}
\end{equation}

\end{widetext}

\noindent Equation (\ref{eq:extendedspa_approx2}) is accurate when $|\dot F_j|^2 \ll |\ddot F_j|^{4/3}$.  Notice that this result is well behaved and finite at $\dot F_j = 0$, demonstrating that the extended SPA cures the singularity at points where the rate of change of the instantaneous signal frequency passes through zero.

We note here that the issue of a signal's frequency and frequency derivative both vanishing was examined by Klein, Cornish, and Yunes \cite{Klein:2014bua} in the context of comparable mass binaries with spinning and precessing constituents.  Although superficially similar to the case we discuss here, the root cause of the pathology in their case was rather different.  In addition to having an orbital timescale $T_{\rm o}$ and an inspiral timescale $T_{\rm i}$, the waveforms from precessing binaries vary on precession timescales $T_{\rm p}$ which are typically intermediate to $T_{\rm i}$ and $T_{\rm o}$.  A given binary typically exhibits precession on multiple timescales, depending on the binary's spins.  Such a waveform may also considered ``multivoice,'' due to the evolution of features in different harmonics.  One finds in this case that the stationary points of different voices can coalesce, leading to a pathological SPA estimate for the waveform's Fourier transform.  Each individual voice, however, remains well behaved, in contrast to the case for EMRIs.  As such, their treatment does not need to use additional information about the frequency evolution, as we find is necessary in our analysis.

\begin{figure*}
\includegraphics[width=0.48\textwidth]{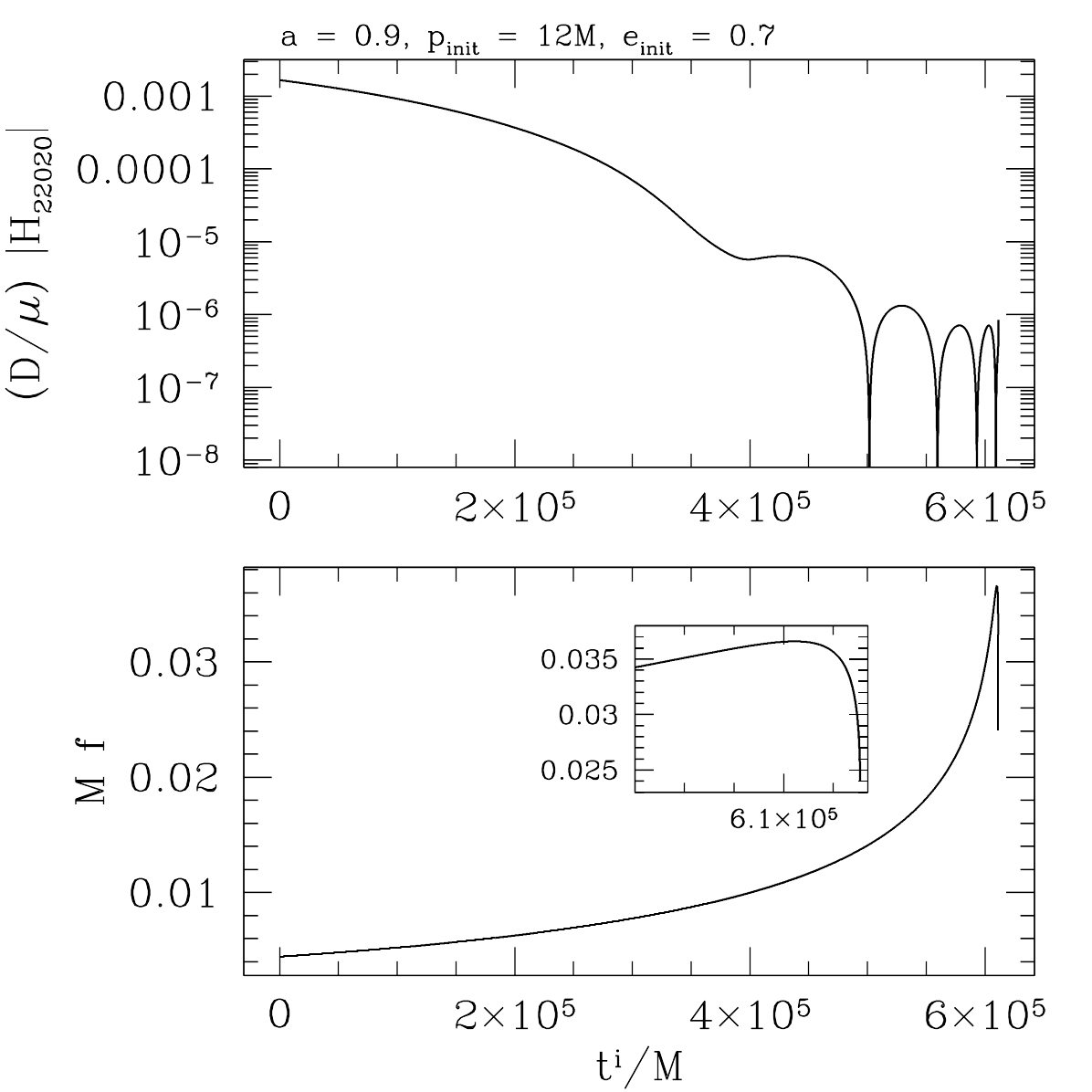}
\includegraphics[width=0.48\textwidth]{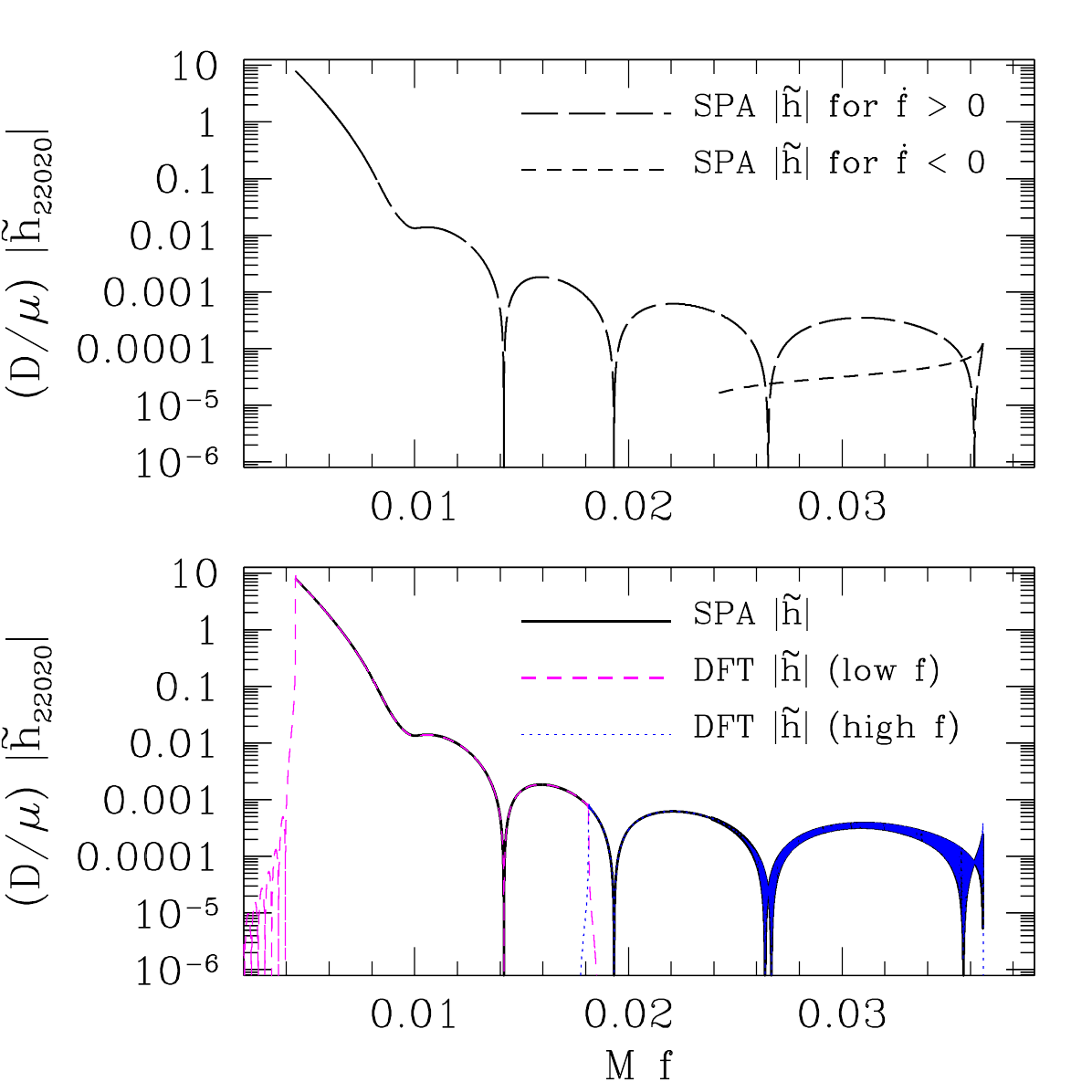}
\caption{An example of the time-domain and frequency-domain structure of one voice: $l = m = 2$, $k = 0$, $n = 20$, for an equatorial Kerr inspiral with $a = 0.9M$, $p_{\rm init} = 12M$, $e_{\rm init} = 0.7$, and mass ratio $\varepsilon = 10^{-3}$.  On the left, we show this voice's amplitude (top) and frequency (bottom) as a function of time over the inspiral.  The voice's frequency increases until it reaches $f_{\rm max} \simeq 0.037/M$; it then reverses and falls to $f_{\rm final} \simeq 0.024/M$ at the end of inspiral.  The spiky features occur when this amplitude passes through zero (note the log scale).  On the right, we show the frequency domain structure computed with the extended SPA.  In the top right, we examine contributions to the SPA along the two branches of $F(t) = f$.  The curve with long dashes is the magnitude of the SPA for the branch with $\dot f > 0$; the curve with short dashes is for the branch with $\dot f < 0$.  Bottom right shows the final SPA Fourier transform obtained by summing contributions from the two branches, and compares this to a discrete Fourier transform (DFT).  Because of the large dynamic range between the low-frequency and high-frequency behavior of this voice, we separately examine the DFT for the early time, low-frequency portion (corresponding to $t^{\rm i} \lesssim 5.5 \times 10^5\,M$; plotted with magenta dashes) and the DFT for the late, high-frequency portion ($t^{\rm i} \gtrsim 5.5 \times 10^5\,M$; plotted with blue dots).  Modulo lobes at the boundaries of the two DFTs (associated with the Tukey window used to taper the time-domain signal near these boundaries), the DFTs coincide perfectly with the SPA.  Notice the interesting behavior in the band $0.024 \lesssim M f \lesssim 0.037$.  This arises from beating between the contributions along the two branches.}
\label{fig:voicedeepdive}
\end{figure*}

Figure \ref{fig:voicedeepdive} illustrates how these elements come together for a voice that showcases many of the features we have discussed here.  We show the $l = m = 2$, $k = 0$, $n = 20$ voice for an equatorial Kerr inspiral with $a = 0.9M$, $p_{\rm init} = 12M$, $e_{\rm init} = 0.7$, and mass ratio $\varepsilon = 10^{-3}$.  The full time-domain waveform and additional voices are discussed for this case in more detail in Sec.\ \ref{sec:kerr_2D}.

On the left-hand side of Fig.\ \ref{fig:voicedeepdive}, we show this voice's time-domain amplitude and the evolution of its frequency.  The voice frequency increases until it reaches a maximum, then rapidly decreases, at least until the inspiral ends several hundred $M$ after reaching this maximum.  The Fourier transform, computed using Eq.\ (\ref{eq:extendedspa1}) and shown in the right-hand panels, has two branches: the branch with $\dot f > 0$, shown as the long-dashed curve in the upper right-hand panel of Fig.\ \ref{fig:voicedeepdive}; and the branch with $\dot f < 0$, shown as the short-dashed curve in this panel.

Combining the two branches yields the solid curve shown in the lower right-hand panel of Fig.\ \ref{fig:voicedeepdive}.  We overlay on this plot a discrete Fourier transform computed using this time-domain voice; because of the large dynamic range in the signal's amplitude, we consider separately a low-frequency DFT (focusing on data for $t^{\rm i} \lesssim 5.5\times10^5\,M$, for which $f \lesssim 0.018M$) and a high-frequency DFT (focusing on data for $t^{\rm i} \gtrsim 5.5\times10^5\,M$).  We use a Tukey window of width $500M$ to taper these segments of the time-domain signal.  Aside from lobes near the boundaries associated with these windows, we see perfect agreement between the DFT and the SPA.  Notice the interesting structure in the band $0.024 \lesssim M f \lesssim 0.037$.  This arises from beating between the contributions along the branches with $\dot f > 0$ and $\dot f < 0$: at each frequency in this band, the signal contributes at two different times, and with two different phases.  Additional examples of voices, in both the time and frequency domains, are shown in Secs.\ \ref{sec:results_schw} and \ref{sec:results_kerr}.

\subsection{Multiple voices}

Generalization to a multivoice signal is straightforward.  Let us write our signal in the time domain
\begin{equation}
h(t) = \sum_{V} H_{V}(t) e^{-i\Phi_{V}(t)}\;,
\end{equation}
where ${V}$ labels the voice, and is shorthand for all the mode indices which describe each voice of the waveform.  For generic EMRIs, ${V} \equiv (l, m, k, n)$.

The calculation proceeds as before, but now the phase is stationary for each voice at some moment.  Define
\begin{equation}
F_{V} = \frac{1}{2\pi}\frac{d\Phi_{V}}{dt}\;,
\end{equation}
and likewise define $\dot F_{V}$, $\ddot F_{V}$.  Assume that the condition $F_V(t) = f$ has $N_{V}$ solutions for voice ${V}$, and define $t_{j,V}(f)$ to be the $j$th such solution.  (As stated in the previous section, $N_{V}$ will be either 1 or 2 for EMRIs.)  The stationary phase Fourier transform is then
\begin{widetext}
\begin{eqnarray}
{\tilde h}(f) &=& \frac{2}{\sqrt{3}}\sum_{V}\sum_{j = 1}^{N_{V}}H_{V}[t_{j,V}(f)]e^{i[2\pi ft_{j,V}(f) - \Phi_V(t_{j,V}(f))]}\frac{i\dot F_{j,V}}{|\ddot F_{j,V}|}\exp\left[-\frac{2\pi i}{3}\frac{\dot F_{j,V}^3}{\ddot F_{j,V}^2}\right]K_{1/3}\left[-\frac{2\pi i}{3}\frac{\dot F_{j,V}^3}{\ddot F_{j,V}^2}\right]
\nonumber\\
&\equiv& \sum_V {\tilde h}_{lmkn}(f)\;.
\label{eq:extendedspa_manyvoices}
\end{eqnarray}
\end{widetext}
We have introduced $\dot F_{j,V} \equiv \dot F_V[t_{j,V}(f)]$ and $\ddot F_{j,V} \equiv \ddot F_V[t_{j,V}(f)]$.  Depending on the relative values of various powers of $\dot F_{j,V}$ and $\ddot F_{j,V}$, one can expand the ${V}$th voice as in Eqs.\ (\ref{eq:extendedspa_approx1}) or (\ref{eq:extendedspa_approx2}).

\section{Implementation}
\label{sec:implement}

In this section, we describe various technical details by which we implement this formalism for computing EMRI waveforms.  To make an adiabatic inspiral and its associated waveform, we lay out a grid of orbits, parameterized by each orbit's $(p, e, x_I)$.  We store all the data at each grid point needed to construct the inspiral and the waveform.  We then interpolate to estimate the values of each datum at locations away from the grid points.  In this section, we describe this data grid and the data which are stored on it, and details of how we interpolate data off grid.  We emphasize that there is surely a great deal of room to improve on the techniques we present here; indeed, we used different algorithms to design our data grid and to perform interpolation in a closely-related companion analysis, Ref.\ \cite{Chua:2020stf}.  For this initial study, the grids and interpolation techniques we use are chosen for ease of use.  In later work, we plan to investigate how best to optimize the grids and interpolation methods for speed and accuracy of waveform calculation.

\subsection{EMRI data grids}
\label{sec:grid}

We store our data on a grid that is rectangular in $p - p_{\rm LSO}$, $e$, and $x_I$, where $p_{\rm LSO}$ parameterizes the last stable orbit (LSO).  The value of $p_{\rm LSO}$ is easy to calculate as a function of $e$ and $x_I$ \cite{Stein:2019buj}, making it simple to set up a grid in this space.  We set our innermost grid point to $p_{\rm min} = p_{\rm LSO} + 0.02M$, slightly outside of the LSO.  The radial frequency $\Omega_r \to 0$ as $p \to p_{\rm LSO}$, which means that Fourier expansions in $\Omega_r$ tend to be badly behaved as the LSO is approached; this can be regarded as a precursor to the small body's plunge into the black hole \cite{Sundararajan:2008bw,Apte:2019txp} at the end of inspiral.  Very little inspiral remains when the small body has reached our choice of $p_{\rm min}$, so we are confident that the error incurred by truncating at $p_{\rm min}$ (rather than closer to $p_{\rm LSO}$) is negligible.  That said, it should be emphasized that this choice of $p_{\rm min}$ has not been carefully evaluated.  It will be useful to systematically examine how to select the grid's inner edge in future ``production quality'' work.

Many important quantities vary rapidly near the LSO.  Of particular importance is the phase of $\check A_{lmkn}$, which tends to rotate rapidly as $p \to p_{\rm LSO}$.  It is crucial to resolve this behavior in order to compute accurate waveforms.  To account for this behavior, we use a grid whose density increases near $p_{\rm LSO}$.  For this paper, our grid is uniformly spaced in
\begin{equation}
u \equiv \frac{1}{\sqrt{p - 0.9p_{\rm LSO}}}\;.
\end{equation}
Using this spacing, we have laid down 40 points between $p_{\rm min} = p_{\rm LSO} + 0.02M$ and $p_{\rm max} = p_{\rm min} + 10M$.  Different choices certainly could be used; for example, a grid reaching to larger $p$ and with a different algorithm for increasing density near $p_{\rm LSO}$ was used in Ref.\ \cite{Chua:2020stf}.  The choice of $p$ spacing is an example of an issue that should be more carefully investigated, and perhaps empirically designed depending on what works best given computing resources and accuracy needs for one's application.

For any $X$ stored on our grid, $dX/de \to 0$ as $e \to 0$.  Empirically, we find it is important to have dense grid coverage for small $e$ in order for this behavior to be accurately captured.  For this paper, we have used grids that run over $0 \le e \le 0.8$ in steps of $\Delta e = 0.1$.  Work in progress \cite{TidalHeating_inprep} suggests that this spacing may introduce small systematic errors in computing the inspiral rate as a function of initial eccentricity; using higher density across at least the small $e$ part of this range appears to effectively address this.  More detailed discussion of this point will be presented in later work \cite{TidalHeating_inprep}.

For fixed $p - p_{\rm LSO}$ and fixed $e$, all our stored data tends to be very smooth and indeed nearly linear as a function of $x_I$.  Our grid covers the range $-1 \le x_I \le 1$, with 16 points spaced by $\Delta x_I = 2/15 \simeq 0.1333$.

We store the following data at each grid point:

\begin{itemize}

\item The rates of change $(dE/dt)^{\infty,{\rm H}}$, $(dL_z/dt)^{\infty,{\rm H}}$, and $(dQ/dt)^{\infty,{\rm H}}$ obtained by summing over many modes until convergence has been reached, using the convergence criteria described in Sec.\ \ref{sec:backreaction}.

\item The rates of change of the orbital elements $(dp/dt)^{\infty,{\rm H}}$, $(de/dt)^{\infty,{\rm H}}$, and $(dx_I/dt)^{\infty,{\rm H}}$ obtained by using the Jacobian described in App.\ \ref{app:jacobian} with these fluxes.

\item The fiducial amplitude $\check A_{lmkn}$ for all modes used to compute $(dE/dt)^\infty$, $(dL_z/dt)^\infty$, and $(dQ/dt)^\infty$.

\end{itemize}
So far, we have constructed such data sets for spherical ($e = 0$) and equatorial orbits ($x_I = \pm 1$) for $a/M \in [0, 0.1, 0.2, \ldots, 0.9, 0.95, 0.99]$, as well as for generic orbits for spin $a = 0.7M$ covering $0 \le e \le 0.4$.  These data are produced using the code GREMLIN, a frequency-domain Teukolsky solver primarily developed by author Hughes, with significant input from collaborators.  The core methods of this code are described in Refs.\ \cite{Hughes:1999bq, Drasco:2005kz}, updated to use methods developed in Refs.\ \cite{Fujita:2004rb,Fujita:2009uz} for solving the homogeneous Teukolsky equation; see \cite{Throwe:2010} for further discussion.

Two-dimensional orbits (eccentric and equatorial, or spherical and inclined) typically require several hundred to several thousand modes in order to converge as described in Sec.\ \ref{sec:backreaction}, reaching $\sim 10^4$ modes in the strongest fields.  The number of modes needed for convergence of generic orbits is an order of magnitude or two larger.  Each spherical orbit mode requires about 0.1 seconds of CPU time for small values of $l$, increasing to roughly 0.25 seconds for modes with $l = 10$.  Computing eccentric orbit modes is more time consuming, since each mode involves an integral over the radial domain covered by its orbit.  This cost also varies significantly by radial mode number $n$.  For small $l$, modes for small eccentricity ($e \lesssim 0.2$) take on average 1 CPU second or less; medium eccentricity modes ($e \approx 0.5$) average about 5--10 CPU seconds; and large eccentricity modes ($e = 0.8$) average 30--40 CPU seconds each.  These averages are skewed significantly by larger values of $n$, for which the integrand of Eq.\ (\ref{eq:Jlmkn_def}) rapidly oscillates, and the integral tends to be small compared to the magnitude of the integrand.  At $l = 10$, these times increase: small eccentricity modes take on average up to 20 CPU seconds; medium eccentricity modes average about 40--50 CPU seconds; and large eccentricity modes require on average 150 seconds.

The CPU cost per mode is ameliorated by the fact that each orbit $(p,e,x_I)$ and each mode $(l,m,k,n)$ is independent of all others.  As such, this problem is embarrassingly parallelizable, and data sets can be effectively generated on distributed computing clusters.  The data sets described above are publicly available through the Black Hole Perturbation Toolkit \cite{BHPToolkit}.  Plans to extend these sets, develop further examples, and release the GREMLIN code, are described in Sec.\ \ref{sec:conclude}.

\subsection{Interpolating and integrating across the grid}
\label{sec:interpolate}

To find data away from the grid points, in this analysis we use cubic spline interpolation in the three directions.  Because our grid is rectangular in $(p - p_{\rm LSO}, e, x_I)$, this can be implemented effectively, and is adequate for demonstrating how to build adiabatic waveforms and illustrating the results.  Cubic spline interpolation for the individual mode amplitudes will not scale well to ``production-level'' code, in terms of computational efficiency and memory considerations. In Ref.\ \cite{Chua:2020stf}, a set of the present authors used reduced-order methods with machine learning techniques to construct a global fit to the set of mode amplitudes, finding outstanding efficiency gains in an initial study of Schwarzschild EMRI waveforms.  Future work will apply these techniques to the more generic conditions we examine here.

Other data needed to construct the waveform (for example, the geodesic frequencies $\Omega_{r,\theta,\phi}$ and the phases $\xi_{mkn}$) are calculated at each osculating geodesic as inspiral proceeds.  Substantial computing speed could be gained by storing and interpolating such data; indeed, all such data tends to evolve smoothly and fairly slowly over an inspiral, so it is likely that effective interpolation could be implemented.  We leave an investigation of what to store and interpolate versus what to compute for future work and future optimization. 

To make an adiabatic inspiral, we construct a sequence of osculating geodesics, parameterized by $[p(t^{\rm i}), e(t^{\rm i}), x_I(t^{\rm i})]$ given initial conditions $[p_{\rm init}, e_{\rm init}, x_{I,{\rm init}}]$.  To do this, we interpolate to find $dp/dt$, $de/dt$, and $dx_I/dt$ at each inspiral time $t^{\rm i}$, then use a fixed-stepsize 4th-order Runge--Kutta integrator to construct the inspiral.  This simple method is an obvious point for improvement in future work; indeed, in the related analysis \cite{Chua:2020stf}, we used a variable-stepsize 8th-order Runge--Kutta integrator.

Note that both $(dp/dt)$ and $(de/dt)$ are singular as the LSO is approached.  This is because the Jacobian (App.\ \ref{app:jacobian}) relating these quantities has zero determinant at the LSO.  This singularity can adversely affect the accuracy of interpolating these quantities.  Multiplying both $(dp/dt)$ and $(de/dt)$ by $(p - p_{\rm LSO})^2$, which exactly describes the singularity in the zero eccentricity limit and approximately describes it in general, yields smooth data which interpolate very well.  Another approach is to interpolate the fluxes $dE/dt$, $dL_z/dt$, and $dQ/dt$, normalized by post-Newtonian expansions of these quantities in order to ``divide out'' their most rapid variations across orbits.  This would construct a sequence of osculating geodesic parameterized as $[E(t^{\rm i}, L_z(t^{\rm i}), Q(t^{\rm i})]$.  Determining the most efficient way to parameterize these orbits in order to optimize waveform construction for speed and accuracy are very natural directions for future work.

\section{Results I: Example Schwarzschild EMRI waveforms and their voices}
\label{sec:results_schw}

We now present examples of EMRI waveforms and their voices in the time and frequency domains.  Our goal is not to exhaustively catalog EMRI waveforms, but just to present examples which showcase the behavior that we find, and how this behavior tends to correlate with source properties.  We begin here with results for Schwarzschild; the next section shows Kerr results.

%

Thanks to spherical symmetry, Schwarzschild orbits are confined to a plane, which we define as equatorial.  We can thus set $Q = 0$ and focus on voices with $k = 0$ (i.e., neglecting harmonics of the $\theta$ motion).  We examine two cases: one starts at $(p_{\rm init}, e_{\rm init}) = (12M,0.2)$ and inspirals to $e_{\rm final} \simeq 0.107$; the other starts at $(p_{\rm init}, e_{\rm init}) = (12M, 0.7)$ and inspirals to $e_{\rm final} \simeq 0.374$.  In both cases, $p_{\rm final}$ follows from the Schwarzschild last stable orbit: $p_{\rm LSO} = (6+2e)M$.  We use mass ratio $\varepsilon = 10^{-3}$ in all the cases we examine, both here and in the following section.  Results for other extreme mass ratios can be inferred by scaling durations and accumulated phases with $1/\varepsilon$.

\subsection{Waveforms in the time domain}
\label{sec:schw_h_TD}

\begin{figure*}[t]
\includegraphics[width=0.48\textwidth]{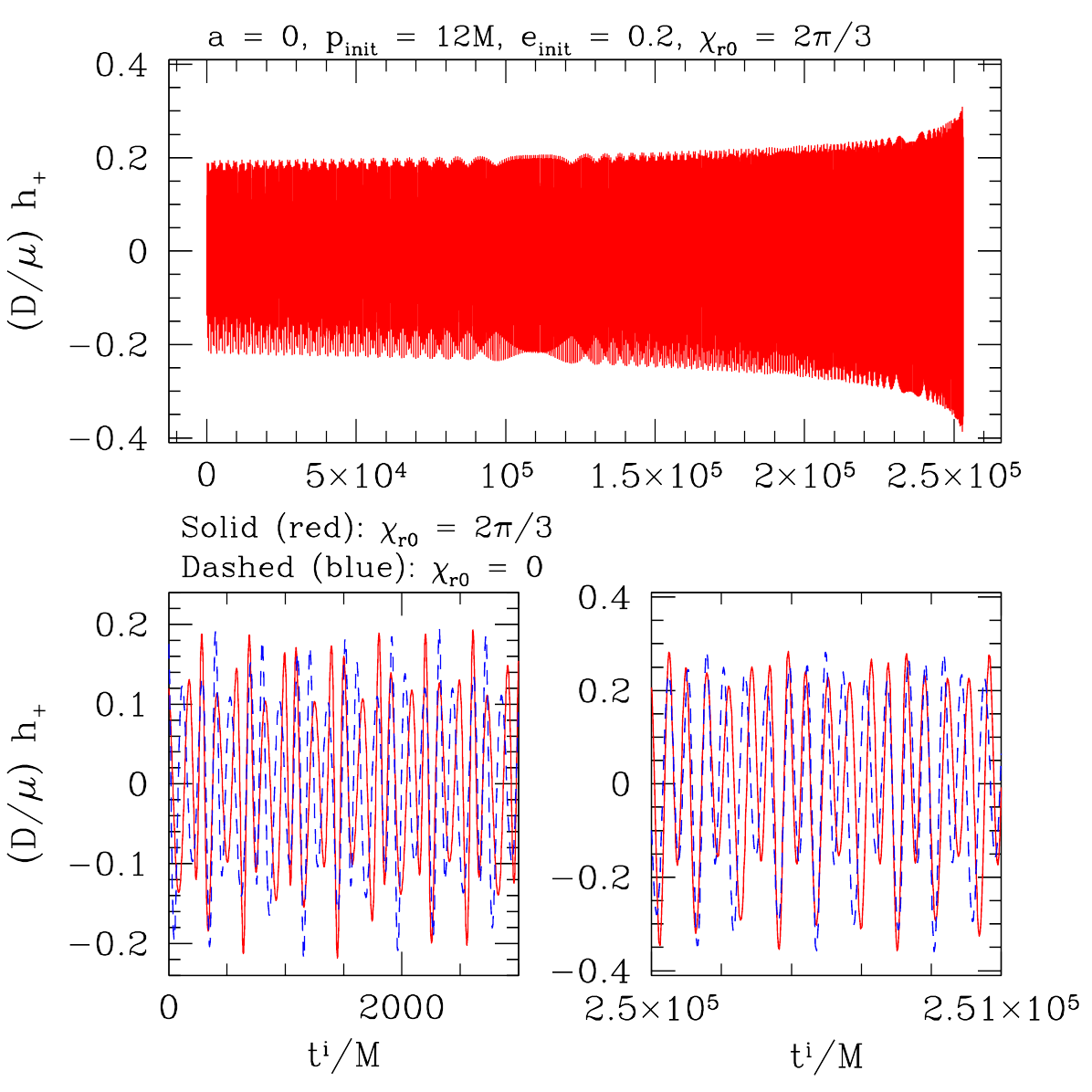}
\includegraphics[width=0.48\textwidth]{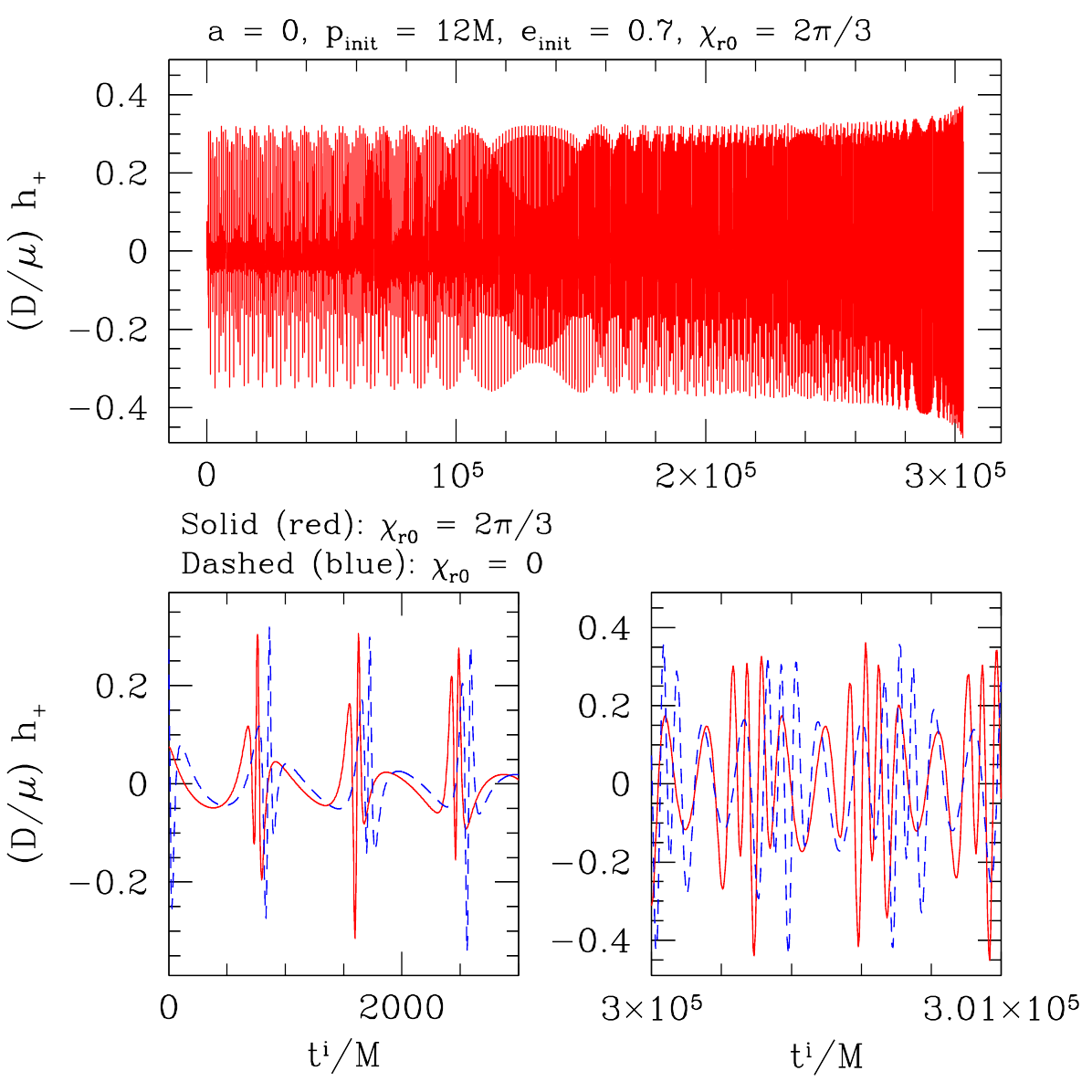}
\caption{The waveform $h_+$ for inspiral into a Schwarzschild black hole with $(p_{\rm init}, e_{\rm init}, x_{I,{\rm init}}) = (12M, 0.2, 1)$ (left-hand panels) and $(p_{\rm init}, e_{\rm init}, x_{I,{\rm init}}) = (12M, 0.7, 1)$ (right-hand panels).  The waves are observed in the plane of the orbit, so $h_\times = 0$, and the mass ratio is $\varepsilon = 10^{-3}$ in all cases.  The top panels show the complete inspiral waveform over this domain, with the initial radial anomaly angle $\chi_{r0} = 2\pi/3$.  The bottom panels compare $h_+$ including the phase corrections $\xi_{m0n}$ (solid [red] curves) with $h_+$ neglecting these corrections (i.e., incorrectly using the fiducial amplitudes $\check A_{lm0n}$; dashed [blue] curves).  In both cases, we compare early times (the first $3000M$ of inspiral) and late times (an interval of $1000M$ near the end of inspiral). Neglecting $\xi_{m0n}$ leads to significant differences in the waveforms.}
\label{fig:a0.0_hp}
\end{figure*}

The top two panels of Fig.\ \ref{fig:a0.0_hp} show $h_+$ for the two Schwarzschild inspirals we examine, both with initial anomaly angle $\chi_{r0} = 2\pi/3$.  The waveform is shown in the system's equatorial plane, so $h_\times = 0$.  For the case with $e_{\rm init} = 0.2$, we plot contributions from all modes with $l \in [2, 3, 4]$, $m \in [-l, \ldots, l]$, $n \in [0, \ldots, 10]$ (as well as modes simply related by symmetry); for the case with $e_{\rm init} = 0.7$, we use the same $l$ and $m$ range, but go over $n \in [0, \ldots, 40]$.  As in our discussion of the convergence of adiabatic backreaction in \ref{sec:backreaction}, we emphasize that we have not carefully analyzed how many modes to include.  This range produces visually converged waveforms --- additional modes do not change the waveform enough to impact the figures.  This is adequate for this paper.

The lower panels of Fig.\ {\ref{fig:a0.0_hp}} show the influence of the phase $\xi_{m0n}$ on the waveform, zooming in on early and late times.  The red curves include all $\xi_{m0n}$ corrections, and the blue curves neglect them, showing inspirals made using the fiducial amplitudes $\check A_{lm0n}$.  Both early and late in the inspiral, the phase correction has a noticeable influence.  This is not surprising, since the impact of $\xi_{m0n}$ is to adjust the system's initial conditions --- different choices of $\xi_{m0n}$ correspond to physically different inspirals.  The influence on the large eccentricity case is particularly strong.

\begin{figure}[htp]
\includegraphics[width=0.48\textwidth]{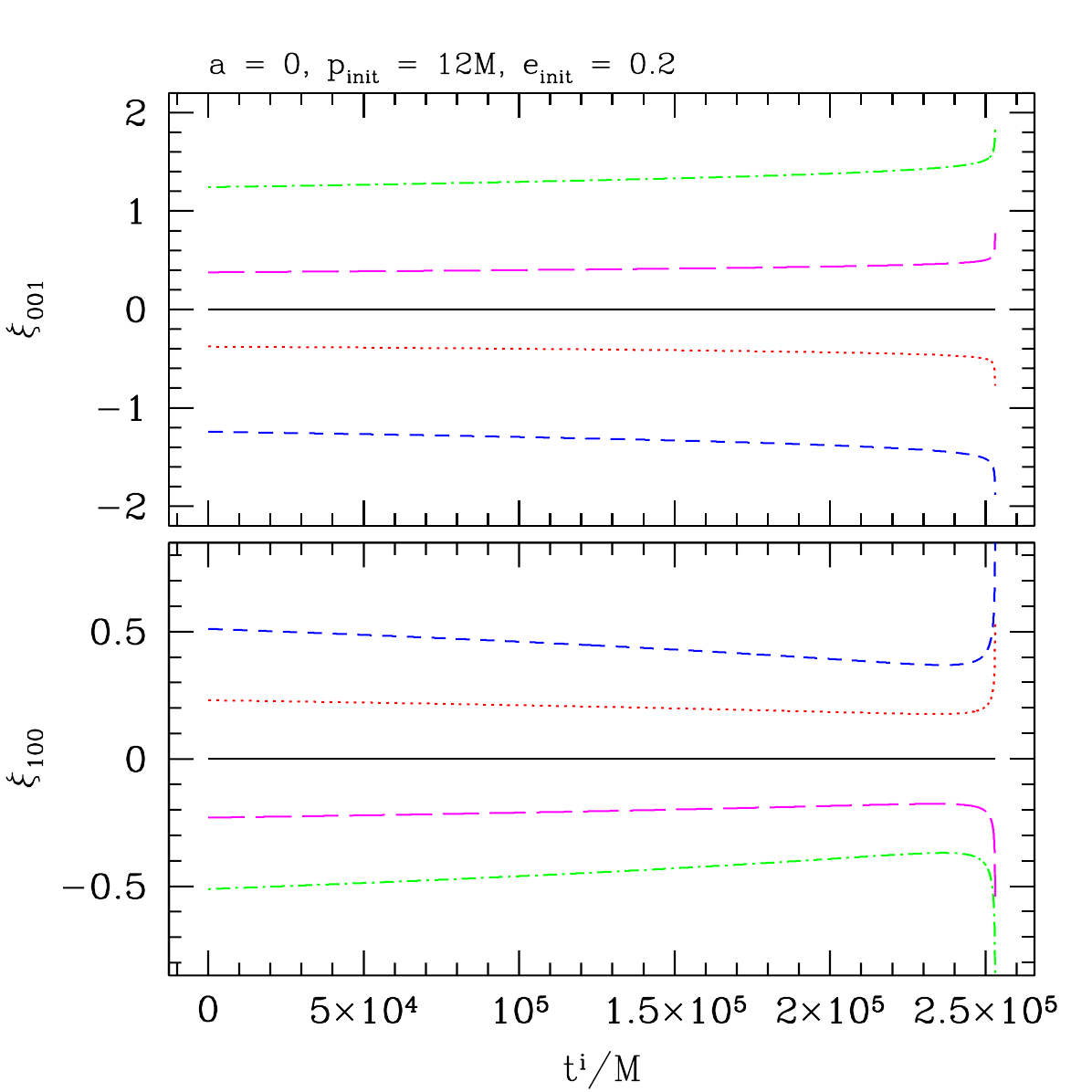}
\caption{The phase correction $\xi_{100}$ (bottom panel) and $\xi_{001}$ (top panel) for inspiral with $p_{\rm init} = 12M$, $e_{\rm init} = 0.2$ at mass ratio $\varepsilon = 10^{-3}$.  In both panels, the solid (black) line corresponds to $\chi_{r0} = 0$, dotted (red) curves show $\chi_{r0} = \pi/6$, short-dashed (blue) shows $\chi_{r0} = \pi/2$, long-dashed (magenta) shows $\chi_{r0} = 11\pi/6$, and dot-dashed (green) shows $\chi_{r0} = 3\pi/2$.  Notice that these curves show only gentle variation until nearly the end, with many changing rapidly as the inspiral approaches the last stable orbit.}
\label{fig:a0.0_smallecc_xi}
\end{figure}

\begin{figure}[htp]
\includegraphics[width=0.48\textwidth]{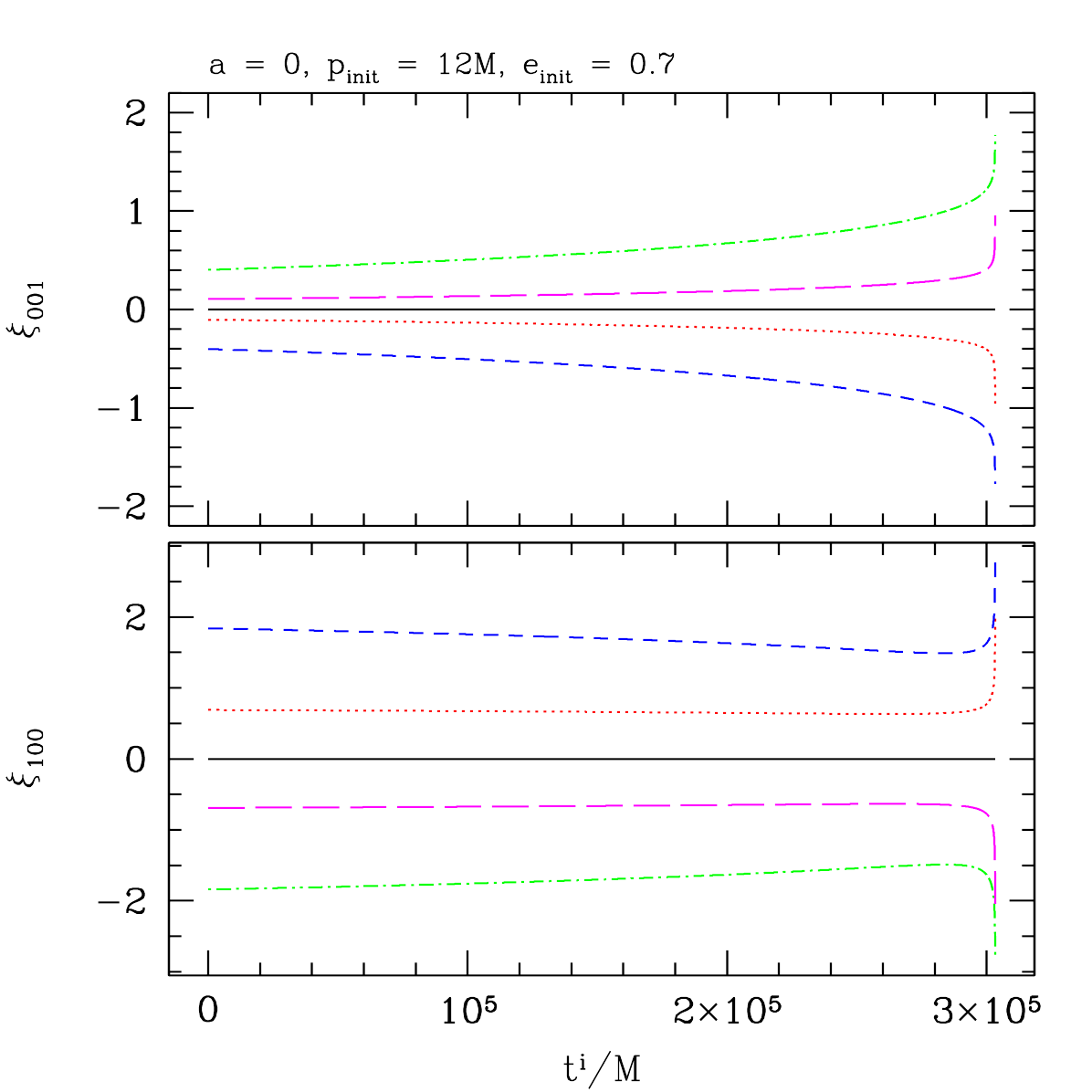}
\caption{The same as Fig.\ \ref{fig:a0.0_smallecc_xi}, but for an inspiral with $p_{\rm init} = 12M$, $e_{\rm init} = 0.7$.  These phases show more variation in this case than when $e_{\rm init} = 0.2$, although the curves still show only gentle variations until the system approaches the end of inspiral.}
\label{fig:a0.0_largeecc_xi}
\end{figure}

Figures \ref{fig:a0.0_smallecc_xi} and \ref{fig:a0.0_largeecc_xi} show how the phases $\xi_{001}$ (top panels) and $\xi_{100}$ (bottom panels) evolve over these inspirals.  We show these phases for initial anomaly angles $\chi_{r0} = 0$ (solid [black] curves), $\pi/6$ (dotted [red]), $\pi/2$ (short-dashed [blue]), $3\pi/2$ (dot-dashed [green]), and $11\pi/6$ (long-dashed [magenta]).  For the small eccentricity case, both $\xi_{001}$ and $\xi_{100}$ are nearly flat over the inspiral, though they show significant variation in the very last moments.  The variation is larger in the higher eccentricity case for $\xi_{001}$, changing by almost a radian over the inspiral for $\chi_{r0} = \pi/2$ and $3\pi/2$ even before reaching the large change at the very end.  In all cases, $\xi_{100}$ and $\xi_{001}$ are smooth and well behaved.  They are also relatively simple to calculate, only requiring information about the geodesic with parameters $p$ and $e$.  Since computing $A_{lm0n}$ is an expensive operation, one should only compute the fiducial amplitudes $\check A_{lm0n}$ and use the phase $\xi_{m0n} = m\xi_{100} + n\xi_{001}$ to convert.

To calibrate how well the phases $\xi_{m0n}$ allow us to account for initial conditions, we compare the waveform assembled voice-by-voice with one computed independently using a time-domain Teukolsky equation solver.  For the comparison waveform, we compute the worldline followed by an inspiraling body, use it to build the source for the time-domain Teukolsky equation as described in Ref.\ \cite{Sundararajan:2008zm}, and then compute the waveform using the techniques developed in Refs.\ \cite{Sundararajan:2008zm,Zenginoglu:2011zz}.  The time-domain solver projects its output onto spherical $(l,m)$ modes of spin-weight $-2$; we focus our comparison on voices with $l = 2$, $m = 2$.

\begin{figure}[htp]
\includegraphics[width=0.48\textwidth]{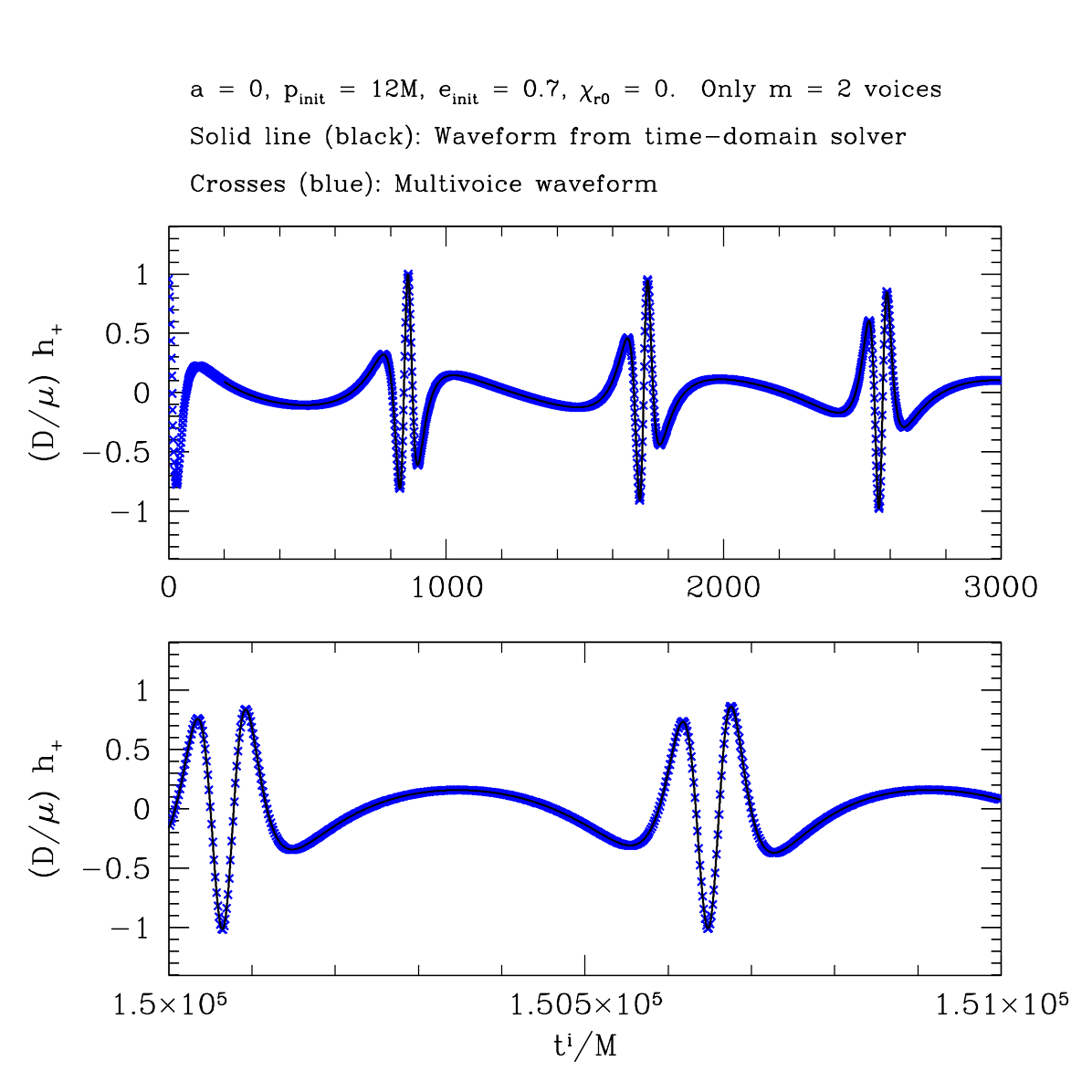}
\caption{The waveform for inspiral into a Schwarzschild black hole with $p_{\rm init} = 12M$, $e_{\rm init} = 0.7$, $\chi_{r0} = 0$ at mass ratio $\varepsilon = 10^{-3}$ computed voice-by-voice using the methods described here (blue crosses), and computed using a time-domain Teukolsky equation solver for the worldline of an inspiral with these initial conditions.  Though only voices with $l = 2$, $m = 2$ are included, the multivoice data are otherwise identical to the $\chi_{r0} = 0$ data shown in Fig.\ \ref{fig:a0.0_hp}.  Top panel shows a stretch $\Delta t^{\rm i} = 3000M$ near the beginning of inspiral; bottom shows a stretch $\Delta t^{\rm i} = 1000M$ near $t^{\rm i} = 1.5\times10^5M$, roughly the middle of this inspiral.  The two calculations agree perfectly in these two snapshots; we in fact find that they hold this agreement all the way to the end at $t^{\rm i} \simeq 3 \times 10^5M$.}
\label{fig:a0.0_compare000}
\end{figure}

\begin{figure}[htp]
\includegraphics[width=0.48\textwidth]{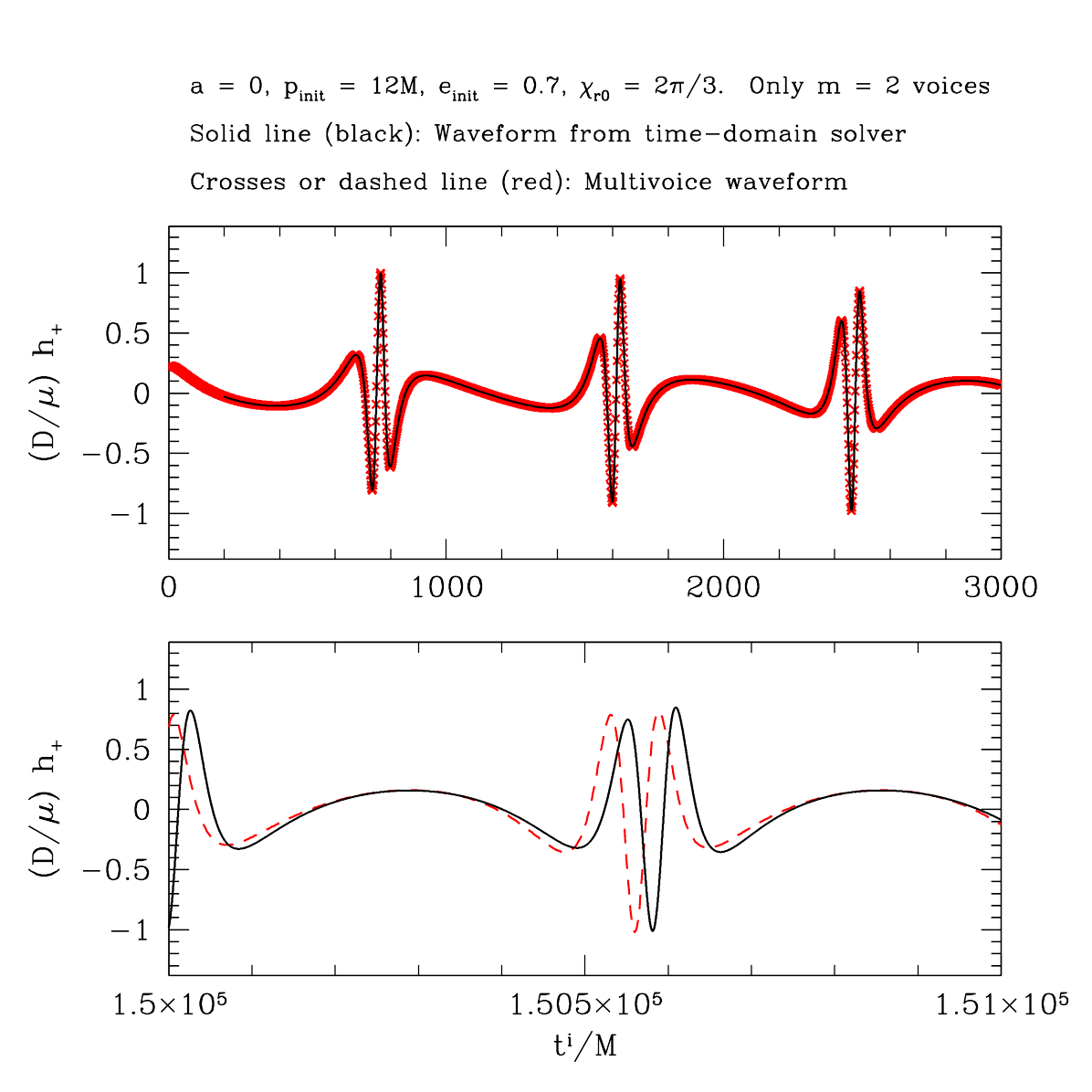}
\caption{Same as Fig.\ \ref{fig:a0.0_compare000}, but for $\chi_{r0} = 2\pi/3$; the multivoice data shown here are identical to the $l = 2$, $m = 2$ $\chi_{r0} = 2\pi/3$ data shown in Fig.\ \ref{fig:a0.0_hp}.  The two waveforms agree very well at early times, but a secular offset accumulates in this case; as shown in the lower panel, the two waveforms are roughly 2 radians out of phase by $t^{\rm i} \simeq 1.5\times 10^5M$.  This grows to about 4 or 5 radians by the end of inspiral.  Our hypothesis is that the time-domain integrator includes slow-time contributions (i.e., contributions from source terms that evolve on the longer inspiral timescale $T_{\rm i}$) which are left out of the adiabatic waveform.}
\label{fig:a0.0_compare120}
\end{figure}

\begin{figure}[htp]
\includegraphics[width=0.48\textwidth]{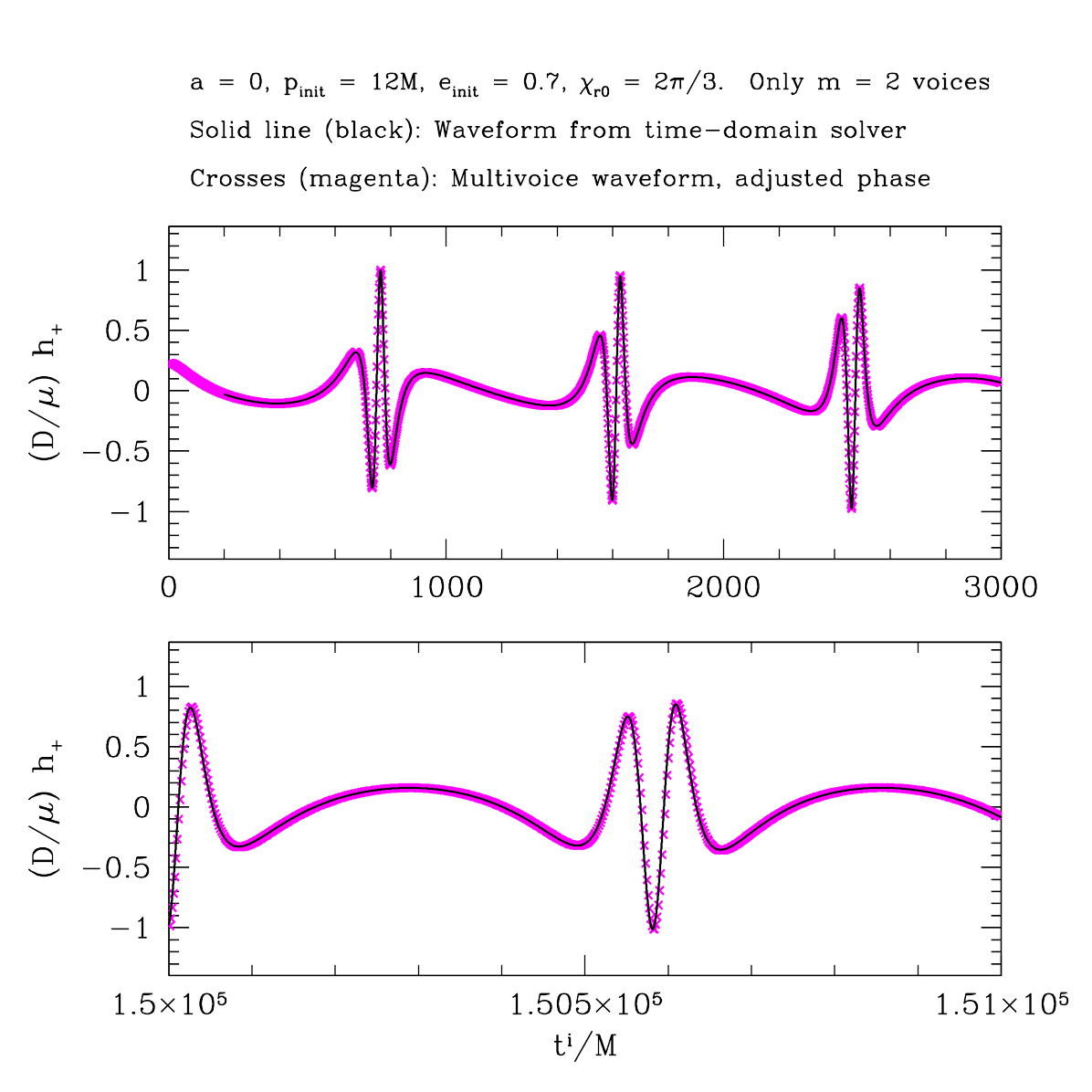}
\caption{Same as Fig.\ \ref{fig:a0.0_compare120}, but using the {\it ad hoc} phase modification introduced in Eq.\ (\ref{eq:PhiPAmn}).  This correction adjusts the orbital phase with a term that depends on the inspiral rate.  Though this correction has not been rigorously computed, it nonetheless substantially improves the agreement between the two waveforms, at least over this domain of $t^{\rm i}$.  (As described in the text, the waveforms drift from one another as inspiral continues.)  This supports the hypothesis that this drift arises due to slow-time variations neglected in the adiabatic approximation.}
\label{fig:a0.0_compare120_efix}
\end{figure}

Figures \ref{fig:a0.0_compare000}, \ref{fig:a0.0_compare120}, and \ref{fig:a0.0_compare120_efix} summarize the results that we find for Schwarzschild inspiral with $p_{\rm init} = 12M$, $e_{\rm init} = 0.7$.  Figure \ref{fig:a0.0_compare000} shows what we find when the initial anomaly angle $\chi_{r0} = 0$.  In this case, we find that the waveform assembled voice-by-voice and the time-domain comparison remain in phase for the entire inspiral.  In the figure, we compare the two waveforms for a stretch of duration $\Delta t^{\rm i} = 3000M$ at the beginning of inspiral as well as a stretch of duration $\Delta t^{\rm i} = 1000M$ in the middle (near $t^{\rm i} = 1.5\times 10^5 M$).  The two waveforms lie on top of each other in both cases; for this choice of $\chi_{r0}$ we find an excellent match all the way to the end at $t^{\rm i} \simeq 3\times10^5M$.

Figure \ref{fig:a0.0_compare120} shows the inspiral waveforms when we put $\chi_{r0} = 2\pi/3$.  In this case, we find a secular drift which accumulates as inspiral proceeds.  The top panel is again a stretch $\Delta t^{\rm i} = 3000M$ from the beginning of inspiral; as in Fig.\ \ref{fig:a0.0_compare000}, the two computed waveforms lie on top of one another.  However, by $t^{\rm i} = 1.5 \times 10^5M$, the two waveforms are about 2 radians out of phase, as can be seen in the lower panel of this figure.  This mismatch grows to about 4 or 5 radians by the end of the inspiral.

As discussed in Sec.\ \ref{sec:diss_evolve}, our solution neglects the impact of ``slow-time'' derivatives on EMRI evolution, leaving out time derivatives of terms which vary on the inspiral timescale $T_{\rm i}$.  As such, our solution only solves the Teukolsky equation (\ref{eq:teuk}) up to errors of $O(\varepsilon)$.  The time-domain solver by contrast finds a solution which, up to numerical discretization, solves Eq.\ (\ref{eq:teuk}) at all orders in $T_{\rm o}/T_{\rm i}$.  Our hypothesis is that this phase offset is because the time-domain solver captures at least some of the ``slow-time'' derivatives which are missed by the adiabatic construction.  Interestingly, the magnitude of the offset depends strongly on $\chi_{r0}$.  By examining multiple values of $\chi_{r0}$, we find that the effect varies at least approximately as $1 - \cos\chi_{r0}$.  This suggests that a slow-time variation in $\xi_{m0n}$ may play a particularly important role in this secular drift.

To test the hypothesis that the offset is due to overlooked slow-time terms, we replaced the accumulated phase, Eq.\ (\ref{eq:Phimkn}), with the following {\it ad hoc} modification:
\begin{eqnarray}
\Phi^{\rm mod}_{m0n}(t^{\rm i}) &=& \int_{t_0}^{t^{\rm i}}\biggl[m\Omega_\phi\left(1 + (1 - \cos\chi_{r0})\frac{3}{2}\frac{dp/dt}{p\Omega_\phi}\right)
\nonumber\\
&+&n\Omega_r\left(1 + (1 - \cos\chi_{r0})\frac{3}{2}\frac{dp/dt}{p\Omega_r}\right)\biggr]dt\;.
\label{eq:PhiPAmn}
\end{eqnarray}
The factor of $(1 - \cos\chi_{r0})$ in this modification accounts for the empirical dependence on $\chi_{r0}$ that we found; the factor of $dp/dt$ connects this phase to the inspiral, and the factors $1/p$ and $1/\Omega_{r,\phi}$ provide dimensional consistency.  The numerical factor $3/2$ was determined empirically.  It is interesting to note that a slow-time evolution in the Newtonian limit yields 
\begin{equation}
\frac{d\Omega}{dt} = -\frac{3}{2}\frac{dp}{dt}\frac{\Omega}{p} - 3e\frac{de}{dt}\frac{\Omega}{1 - e^2}\;.
\end{equation}
Our empirical phase modification appears to be consistent with a weak-field correction associated with the rate at which $p$ changes due to inspiral.

We strongly emphasize that Eq.\ (\ref{eq:PhiPAmn}) is completely {\it ad hoc}, and has not been justified by any careful calculation.  However, we find that it does surprisingly well improving the match between the two calculations.  Figure \ref{fig:a0.0_compare120_efix} is the equivalent of Fig.\ \ref{fig:a0.0_compare120}, but with Eq.\ (\ref{eq:PhiPAmn}) used to compute the phase rather than Eq.\ (\ref{eq:Phimkn}).  Notice that the two waveforms lie on top of one another, at least over the domain shown here.  As inspiral proceeds, our {\it ad hoc} fix becomes less accurate: we find a roughly 1 radian offset between the two waveforms when $t^{\rm i} \simeq 2.5 \times 10^5M$, growing to several radians by the end of inspiral.  We find nearly identically improved matches examining inspirals with different values of $\chi_{r0}$, and for different choices of the mass ratio $\varepsilon$.  Interestingly, including terms in $de/dt$ inspired by the weak-field rate of change of $\Omega$, does not help, suggesting that the similarity to the weak-field formula may be a coincidence.

This analysis indicates that the phase offset we find is consistent with a post-adiabatic effect, and therefore is missed by construction when making adiabatic waveforms.  The surprising effectiveness of our {\it ad hoc} fix suggests it may not be too difficult to analytically model this behavior and improve these waveforms.

\subsection{Waveform voices}
\label{sec:schw_voices}

The waveforms shown and discussed in Sec.\ \ref{sec:schw_h_TD} are fairly complicated, especially for the high eccentricity case.  By contrast, the individual voices which contribute to these waveforms are very simple, evolving smoothly and simply on the much longer inspiral timescale.

\begin{figure*}[ht]
\includegraphics[width=0.48\textwidth]{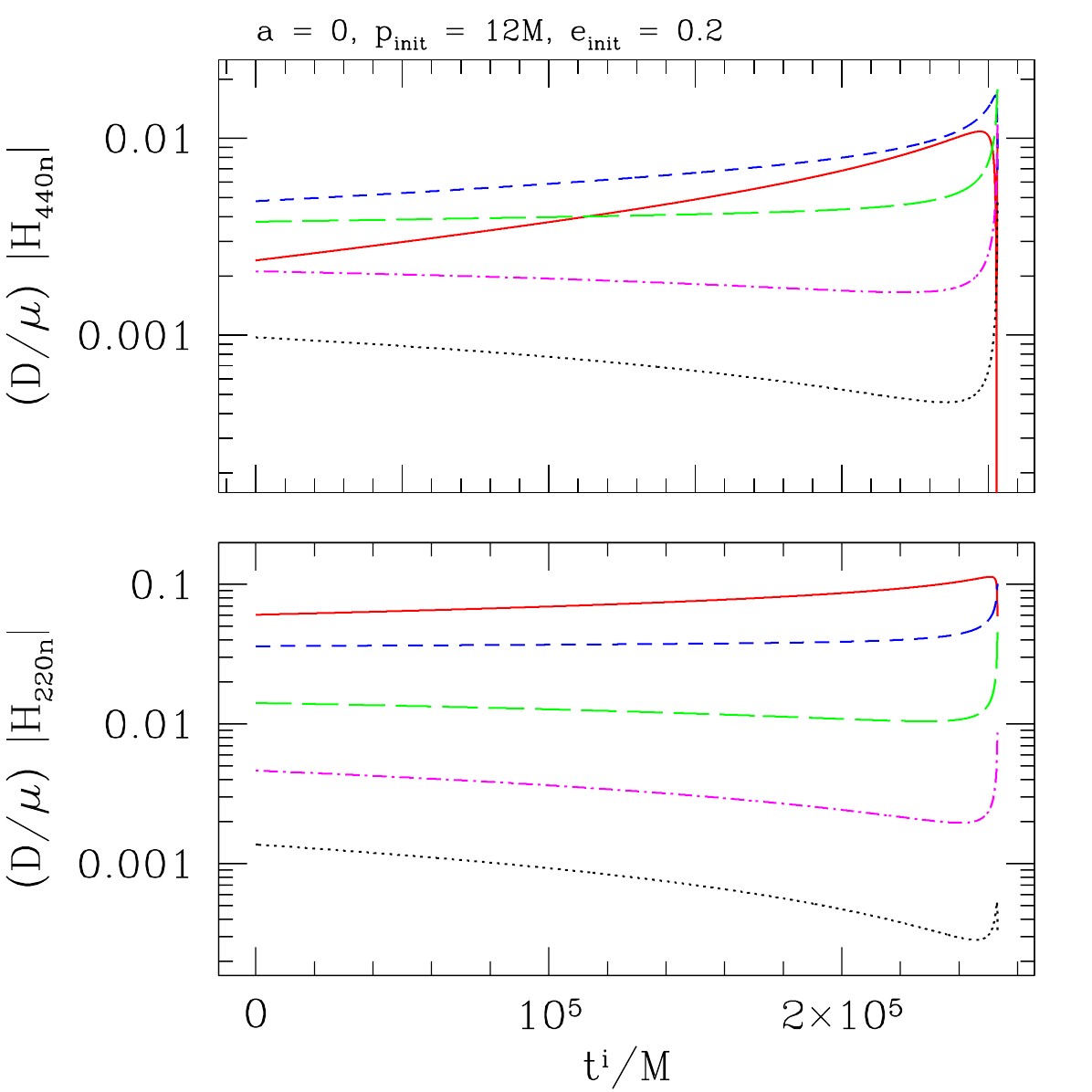}
\includegraphics[width=0.48\textwidth]{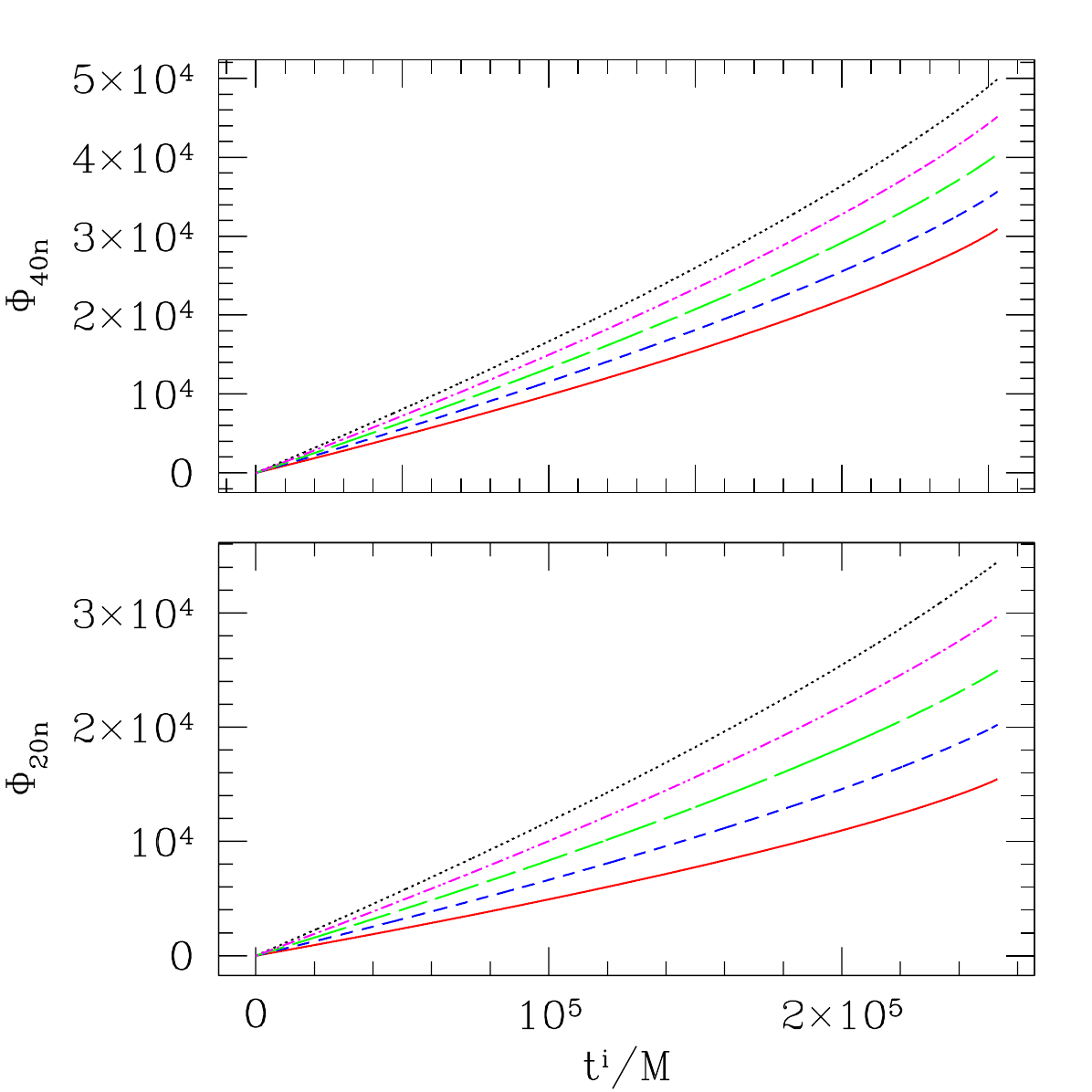}
\caption{Some of the voices which contribute to $h_+$ for the case $(p_{\rm init}, e_{\rm init}) = (12M,0.2)$, as shown in the left panels of Fig.\ {\ref{fig:a0.0_hp}}.  The magnitude of the amplitude $|H_{lm0n}(t^{\rm i})|$ is on the left; the phase $\Phi_{m0n}(t^{\rm i})$ is on the right.  Top panels show voices with $l = 4$ and $m = 4$, bottom panels show $l = 2$ and $m = 2$.  Solid (red) curves are data with $n = 0$, short-dashed (blue) blue curves are $n = 1$, long-dashed (green) are $n = 2$, dot-dashed (magenta) are $n = 3$, and dotted (black) are $n = 4$.  In all cases both the amplitude and the phase evolve smoothly and simply.  As functions of time, these voices can be sampled much less densely than $h_+$ must be sampled in order to accurately track its behavior.}
\label{fig:a0.0_smallecc_posvoices}
\end{figure*}

\begin{figure}[htp]
\includegraphics[width=0.48\textwidth]{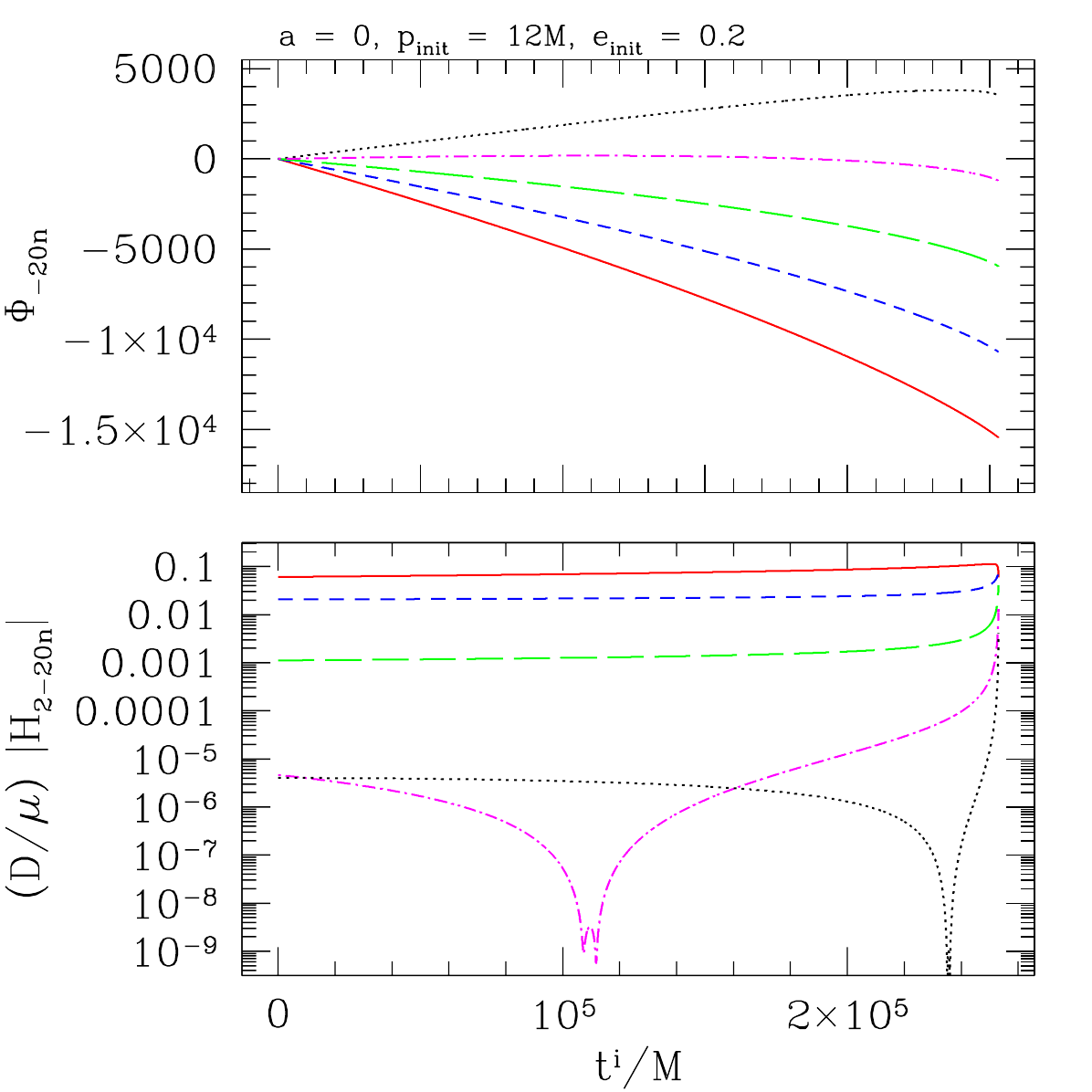}
\caption{Some voices with $l = 2$ and $m = -2$ which contribute to $h_+$ for the small eccentricity case shown in Fig.\ \ref{fig:a0.0_hp}.  Colors and definitions are exactly as in Fig.\ \ref{fig:a0.0_smallecc_posvoices}.  The behavior of modes with $n = 3$ and $n = 4$ are interesting: During the inspiral, there are moments at which the phase evolution reverses direction, corresponding to the voice's instantaneous frequency changing sign.  The amplitudes corresponding to these voices pass through zero at times very close to these moments.  The zero passage of these amplitudes leads to the sharp appearance that we see here.  The real and imaginary parts of $H_{2-23}$ and $H_{2-24}$ are perfectly smooth.}
\label{fig:a0.0_smallecc_negvoices}
\end{figure}

Figures \ref{fig:a0.0_smallecc_posvoices} and \ref{fig:a0.0_smallecc_negvoices} show individual voices that contribute for the case with $e_{\rm init} = 0.2$.  A handful of the voices we show look jagged in these figures due to how we have presented the data: the amplitude passes through zero in some cases, so $|H_{lm0n}|$ appears spiky on a log-linear plot.  These zero passings correspond to moments when the instantaneous frequency $F_{m0n} =(1/2\pi)d\Phi_{m0n}/dt$ changes sign.  In this low eccentricity case, the voices with $l = 2$, $m = \pm2$, $n = 0$ have the largest amplitudes.

\begin{figure}[htbp]
\includegraphics[width=0.48\textwidth]{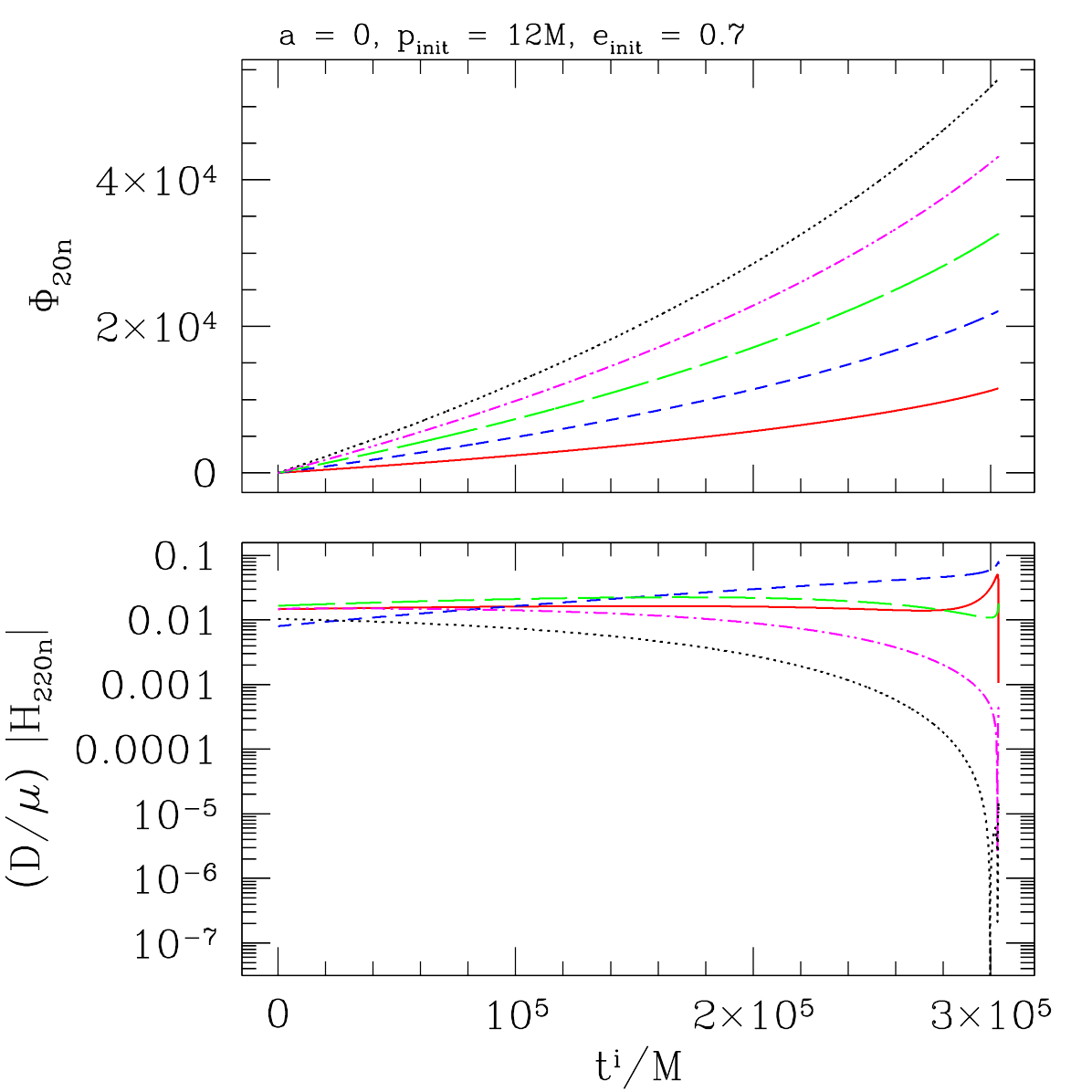}
\caption{Some voices with $l = 2$, $m = 2$ which contribute to the large eccentricity waveform shown in Fig.\ \ref{fig:a0.0_hp}.  Top panel shows the voices' phase, lower panel their amplitudes.  Solid (red) curves are data with $n = 0$, short-dashed (blue) curves are $n = 3$, long-dashed (green) are $n = 6$, dot-dashed (magenta) are $n = 9$, and dotted (black) are $n = 12$.  In contrast to the small eccentricity case, the voice with $n = 0$ does not dominate over most of the inspiral.  In fact, of the voices shown, the one which starts weakest ($n = 3$) evolves to be become the loudest by the end of inspiral.  Nonetheless, all amplitudes and phases evolve in a smooth and simple way, just as in the low eccentricity case.}
\label{fig:a0.0_largeecc_posvoices}
\end{figure}

Figure \ref{fig:a0.0_largeecc_posvoices} shows some of the voices which contribute for the case with $e_{\rm init} = 0.7$.  Again we see that the voices' amplitudes and phases are smooth and well behaved.  In contrast to the small eccentricity case, modes with $n = 0$ do not dominate here.  Of the voices we plot, $n = 6$ dominates at early times, though it falls below several other voices as the inspiral ends.  The voice with $n = 3$ starts out weakest, but becomes strongest roughly halfway through this inspiral.

\begin{figure*}[htbp]
\includegraphics[width=0.48\textwidth]{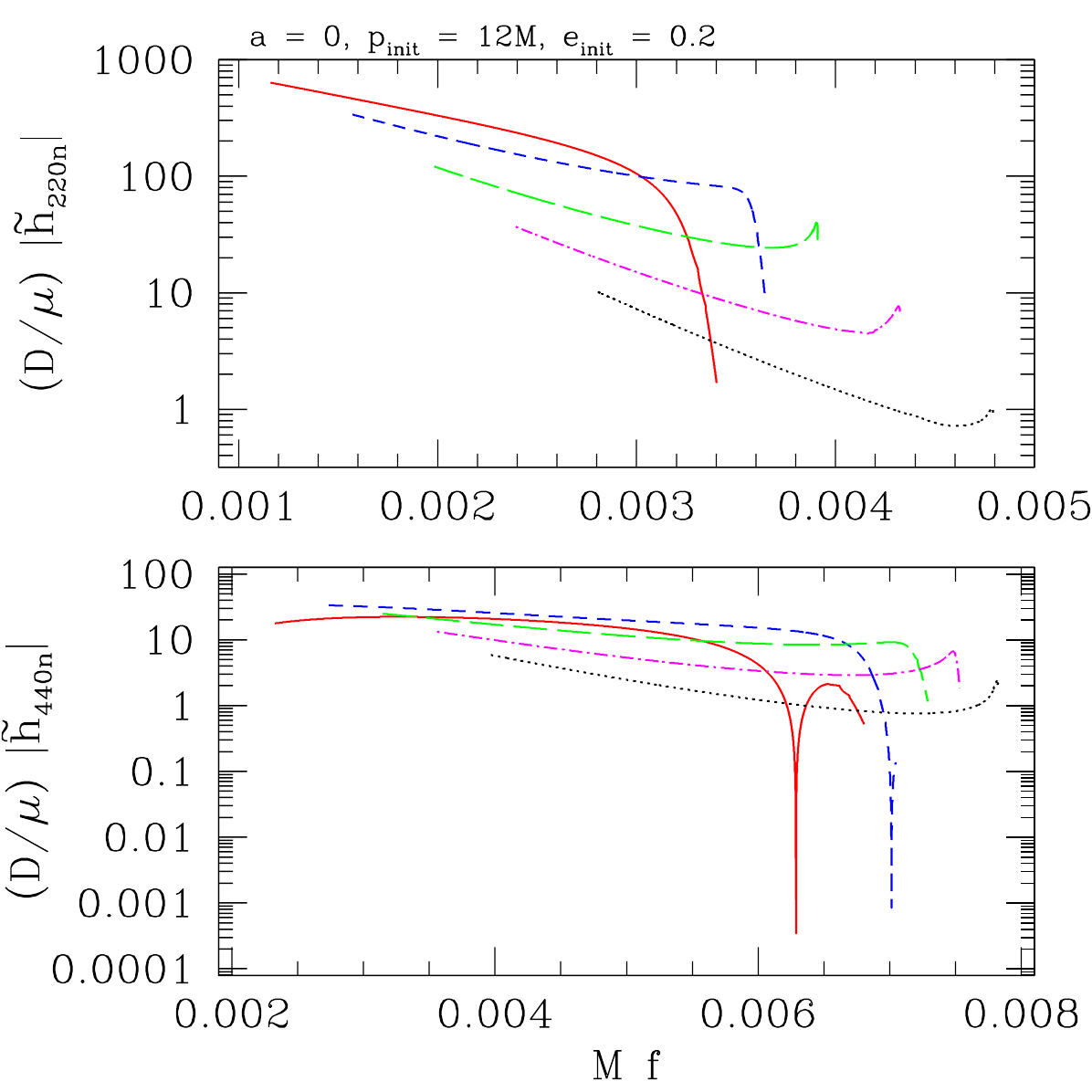}
\includegraphics[width=0.48\textwidth]{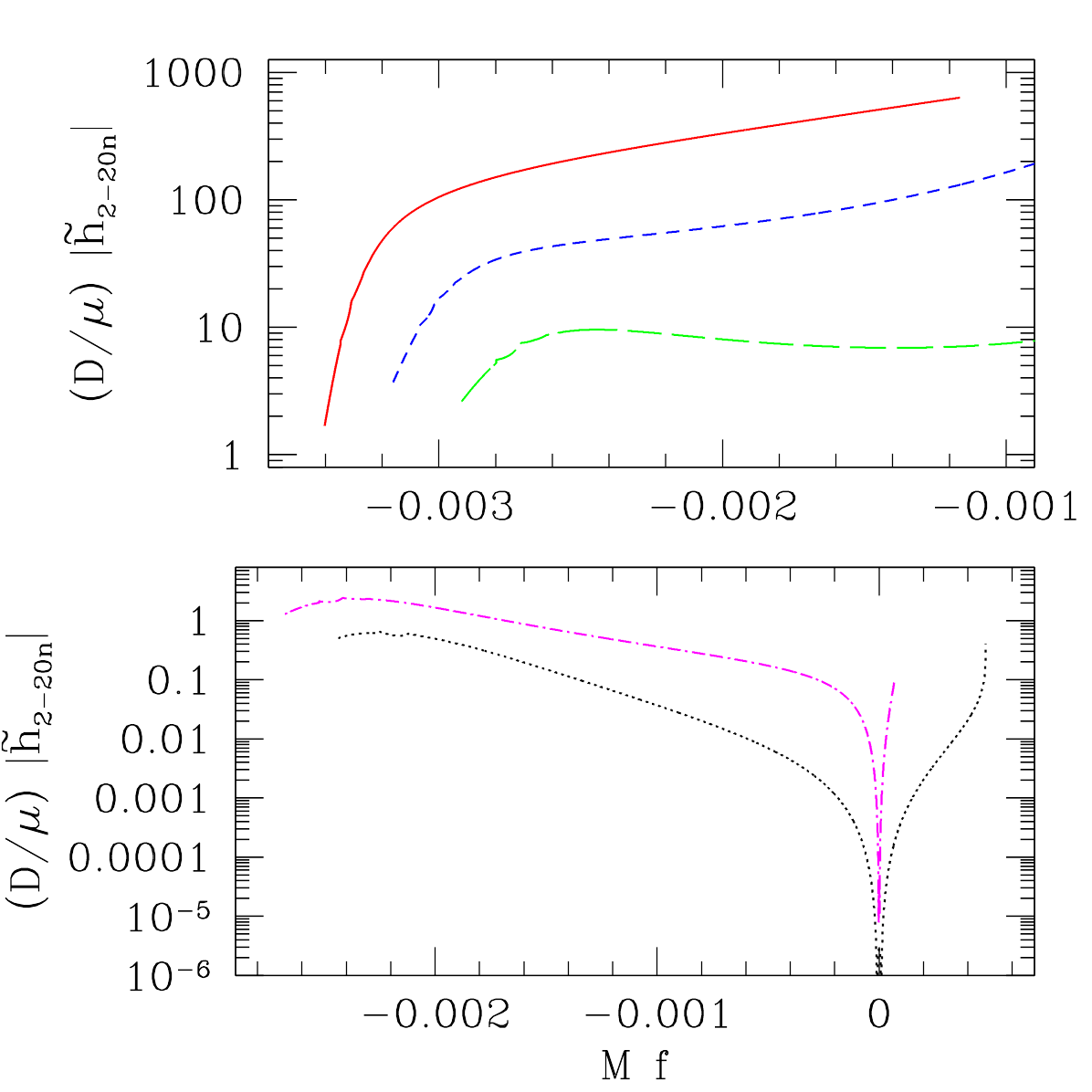}
\caption{Frequency domain representation of the voices shown in Figures \ref{fig:a0.0_smallecc_posvoices} and \ref{fig:a0.0_smallecc_negvoices}.  Top left shows voices with $l = m = 2$; bottom left shows voices with $l = m = 4$; both panels on the right show voices with $l = 2$, $m = -2$.  In all cases, the solid (red) curves are the frequency-domain amplitudes with $n = 0$; short-dashed curves (blue) are for $n = 1$; long-dashed (green) are $n = 2$; dot-dashed (magenta) are $n = 3$; and dotted (black) are $n = 4$.  (We use two panels for the voices with $m = -2$ to showcase the range in amplitude of this case.)  Because we generate the waveform in the time domain and use the stationary phase Fourier transform, each voice spans a slightly different range of frequency.}
\label{fig:a00_smallecc_fd}
\end{figure*}

\begin{figure}[hbp]
\includegraphics[width=0.48\textwidth]{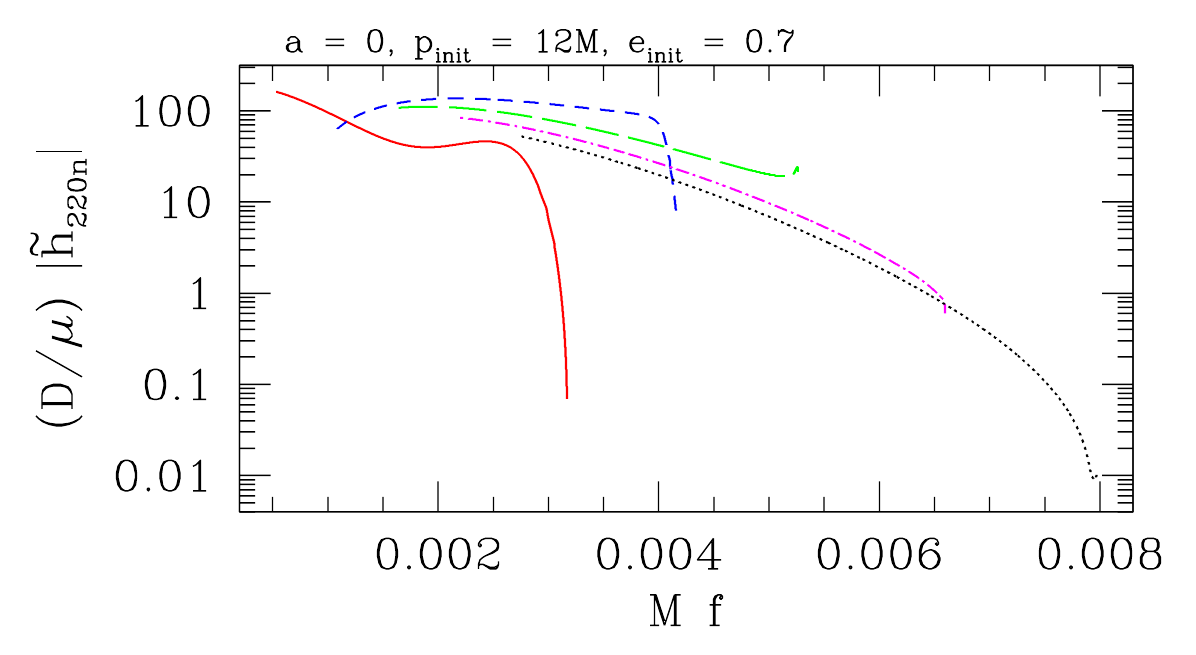}
\caption{Frequency domain representation of the voices shown in Fig.\ \ref{fig:a0.0_largeecc_posvoices}.  All curves show voices with $l = m = 2$.  The solid (red) curve is the frequency-domain amplitude with $n = 0$; short-dashed (blue) is for $n = 3$; long-dashed (green) is $n = 6$; dot-dashed (magenta) for $n = 9$; and dotted (black) for $n = 12$.}
\label{fig:a00_largeecc_fd}
\end{figure}

All of the voices we examine have this behavior: both the amplitudes and phases of individual voices evolve smoothly on the inspiral timescale $T_{\rm i}$.  This property is shared by the voices' frequency-domain behavior.  Following Sec.\ \ref{sec:fourier}, we compute the frequency-domain representation of the voices examined here; Figs.\ \ref{fig:a00_smallecc_fd} and \ref{fig:a00_largeecc_fd} show our results.  Again, we see that all the voices evolve smoothly over the inspiral.  The apparent spikiness in some cases (for example, the voice with $l = m = 4$, $n = 4$ for $e_{\rm init} = 0.2$) is because this voice's amplitude passes through zero, and we show its magnitude on a log scale.  It's also worth noting that the frequency range of different voices varies.  This is because our analysis begins in the time domain, and then transforms to the frequency domain using the SPA.  Different voices thus start at different frequencies, and reach different frequencies at the end of inspiral.

Both the time-domain and the frequency-domain representation of these voices can be computed quickly and efficiently.  Future work, particularly data-analysis-focused applications which compare EMRI waves to detector noise, could benefit substantially by focusing upon the waveform in voice-by-voice fashion, studying which voices are most relevant for detection and measurement as a function of source parameters.

\section{Results II: Example Kerr EMRI waveforms and their voices}
\label{sec:results_kerr}

We next consider a few examples of inspiral into Kerr black holes.  Although certain important details differ from the results discussed in Sec.\ \ref{sec:results_schw}, much of what we find for Kerr inspiral waveforms is qualitatively quite similar to those we find for the Schwarzschild case.  As such, we keep this discussion brief, focusing on the most important highlights of this analysis.  Future work will explore these waveforms and their properties in depth.

We begin in Sec.\ \ref{sec:kerr_2D} with a discussion of two constrained cases: one that initially has zero eccentricity, but is inclined with respect to the hole's equatorial plane; and a second case that is equatorial, but starts with large eccentricity.  We then discuss one example of fully generic (inclined and eccentric) inspiral in Sec.\ \ref{sec:kerr_3D}.

\subsection{Constrained orbital geometry: Spherical and equatorial inspirals}
\label{sec:kerr_2D}

We begin our Kerr study with two cases of inspiral into black holes with spin $a = 0.9M$.  In the first case, we examine an orbit that is spherical, with large inclination to the black hole's equatorial plane.  We take $(r_{\rm init}, x_{I,{\rm init}}) = (10M, 0.5)$.  This inspiral reaches the last stable orbit at $r = 3.820M$, $x_I = 0.483$.  The orbit's inclination is nearly constant during inspiral, with $x_I$ decreasing very slightly.  A similarly slight decrease of $x_I$ is seen in all of the cases we have examined.

\begin{figure*}
\includegraphics[width=0.48\textwidth]{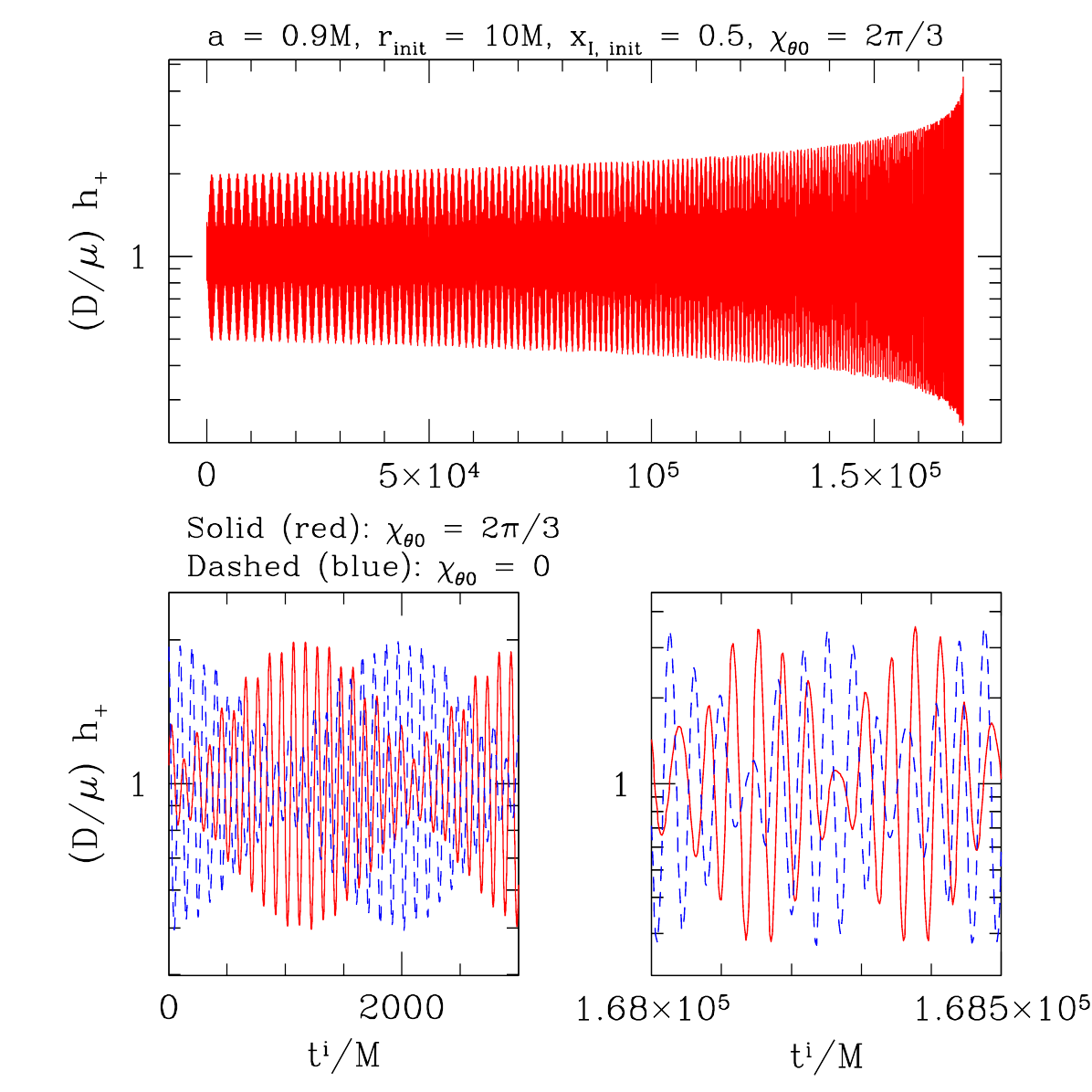}
\includegraphics[width=0.48\textwidth]{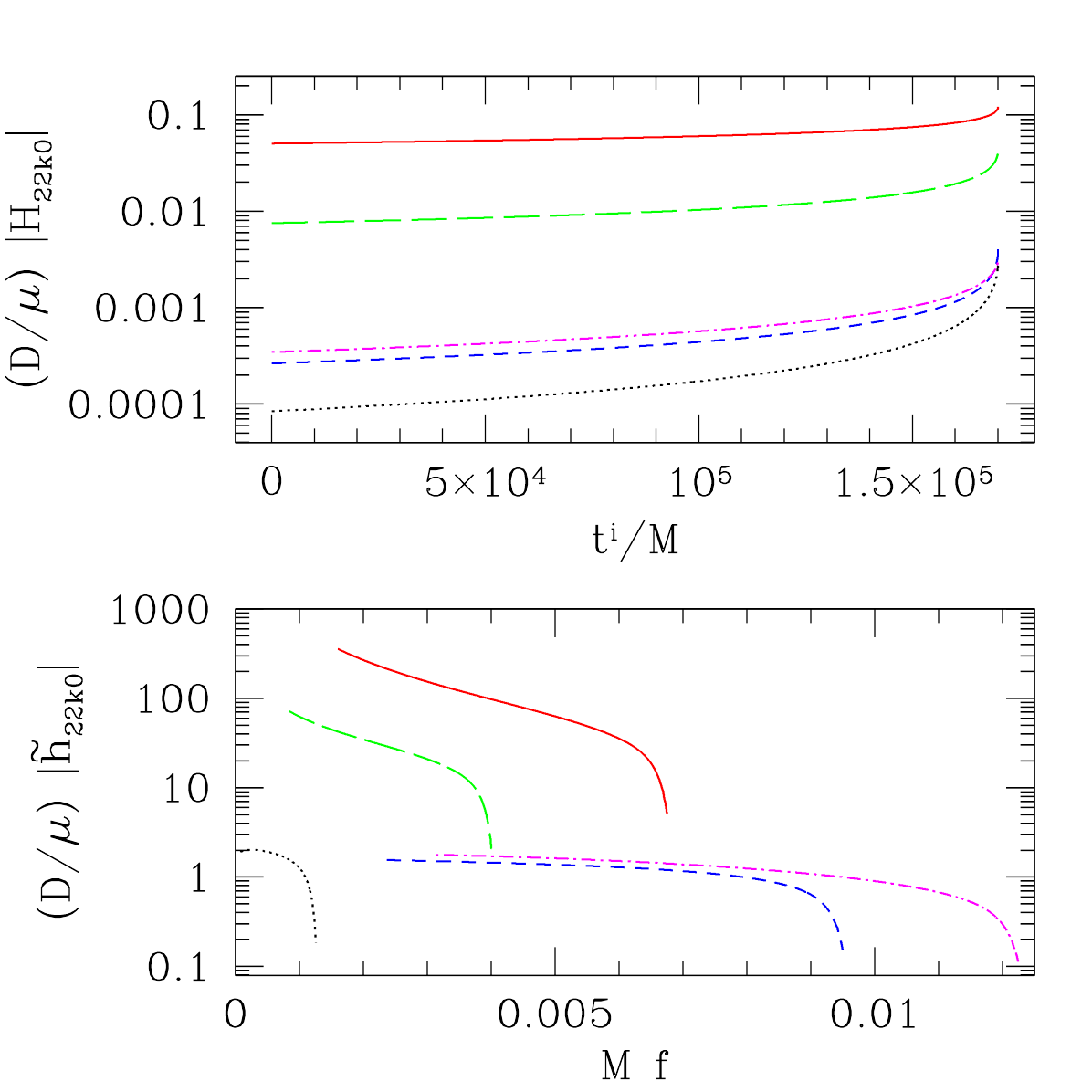}
\caption{Left panels: the waveform $h_+$ for an example of spherical inspiral into a black hole with $a = 0.9M$.  This example has $(r_{\rm init}, x_{I,{\rm init}}) = (10M, 0.5)$, is at mass ratio $\varepsilon = 10^{-3}$, and is observed in the black hole's equatorial plane.  During inspiral, the orbit's radius steadily shrinks while $x_I$ decreases very slightly (see text for details).  The polar anomaly angle $\chi_{\theta0} = 2\pi/3$.  Top left shows the waveform over the entire inspiral; bottom left panels zoom onto early and late times.  The lower panels also illustrate the influence of the anomaly angle, comparing $\chi_{\theta0} = 2\pi/3$ (solid [red] curves) with inspiral on the fiducial geodesic (dash [blue] curves).  We again see that this angle has a large effect on the waveform, in this case changing the initial orientation so that the frame-dragging induced modulation begins at a different phase.  Right panels: some of the $l = m = 2$ voices which contribute to this waveform, both time domain (top) and frequency domain (bottom).  Each line is a different $k$ index: solid (red) is $k = 0$, short-dashed (blue) is $k = 1$, long-dashed (green) is $k = -1$, dot-dashed (magenta) is $k = 2$, and dotted (black) is $k = -2$.  Across the examples of zero eccentricity inspiral we have examined, the tendency is that voices with $|k + m| = l$ are the strongest, and fall off as $|k + m|$ moves away from this peak.}
\label{fig:a0.9_circ}
\end{figure*}

The left-hand panels of Fig.\ \ref{fig:a0.9_circ} show the gravitational waveform we find in this case.  For the time-domain waveform, we plot contributions from all modes with $l \in [2,3,4]$, $m \in [-l, \ldots, l]$, $k \in [0, \ldots, 10]$, plus modes that are simply related by symmetry.  (For inspirals with zero eccentricity, only voices with $n = 0$ contribute.)  The strong influence of spin-orbit modulation can be seen in the lower left-hand panels, which zoom in on early and late times.  These lower panels also illustrate the role of the initial polar anomaly angle $\chi_{\theta0}$, contrasting the waveform with $\chi_{\theta0} = 2\pi/3$ versus the one with $\chi_{\theta0} = 0$.  The two waveforms are similar in shape but shifted, consistent with the fact that $\chi_{\theta0}$ controls the system's initial conditions: when $\chi_{\theta0} = 0$, the orbit is at $\theta = \theta_{\rm min}$ when $t = 0$; for $\chi_{\theta0}$, the orbit starts at a value of $\theta$ roughly midway between the equator and $\theta_{\rm max} = \pi - \theta_{\rm min}$.

The right-hand panels of Fig.\ \ref{fig:a0.9_circ} show several of the voices with $l = m = 2$ which contribute to this waveform.  The trend we have found across all the cases we examine is that voices with $|k + m| = l$ are loudest, and fall off as $|k + m|$ moves away from this peak.  This trend can be seen in the cases shown here: the strongest voice has $k = 0$, followed by $k = -1$.  In the time domain, the voices with $k = 2$ and $k = 1$ have roughly the same amplitude; the voice with $k = -2$ is the weakest of those shown here.  Interestingly, the voices with $k = 2$, $k = 1$, and $k = -2$ have similar magnitude in the frequency domain.  The factor $1/\sqrt{\dot F}$ which enters the frequency-domain magnitude compensates for the fact that this voice's time-domain amplitude is smaller by a factor of 2 or 3.  Such differences have important implications for the measurability of these signals.

The second case we examine is an equatorial eccentric inspiral, with $(p_{\rm init}, e_{\rm init}) = (12M, 0.7)$.  These are the same initial conditions we used for the high-eccentricity inspiral into Schwarzschild we examined in Sec.\ \ref{sec:results_schw}.  For Kerr with spin $a = 0.9M$, these initial conditions lead to a much longer inspiral that goes very deep into the strong field, becoming nearly circular before plunge: inspiral lasts for $\Delta t^{\rm i} \simeq 6.1 \times 10^5M$, roughly twice the duration of the high-eccentricity Schwarzschild inspiral, and ends with $(p_{\rm final}, e_{\rm final}) = (2.40M, 0.057)$.

\begin{figure*}
\includegraphics[width=0.48\textwidth]{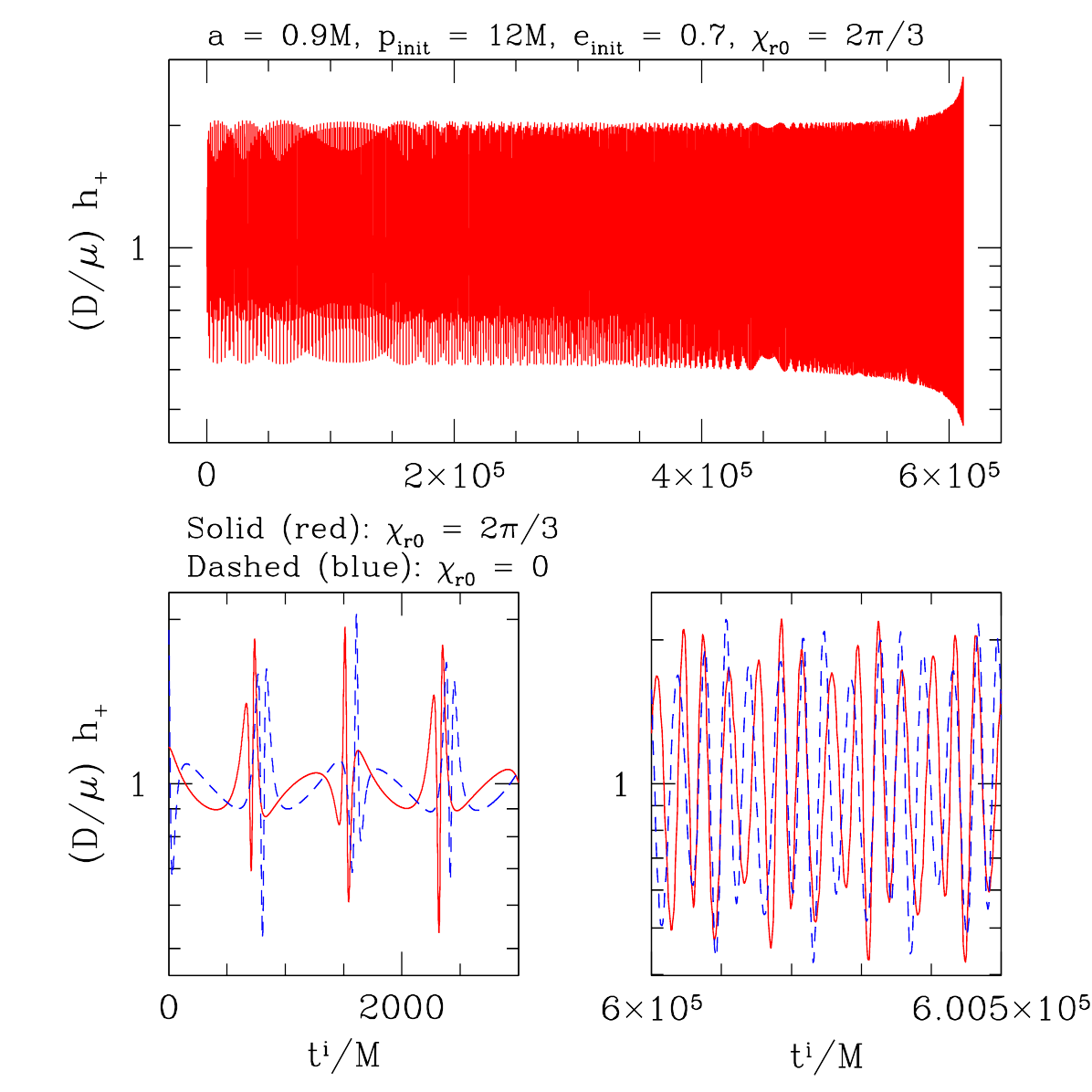}
\includegraphics[width=0.48\textwidth]{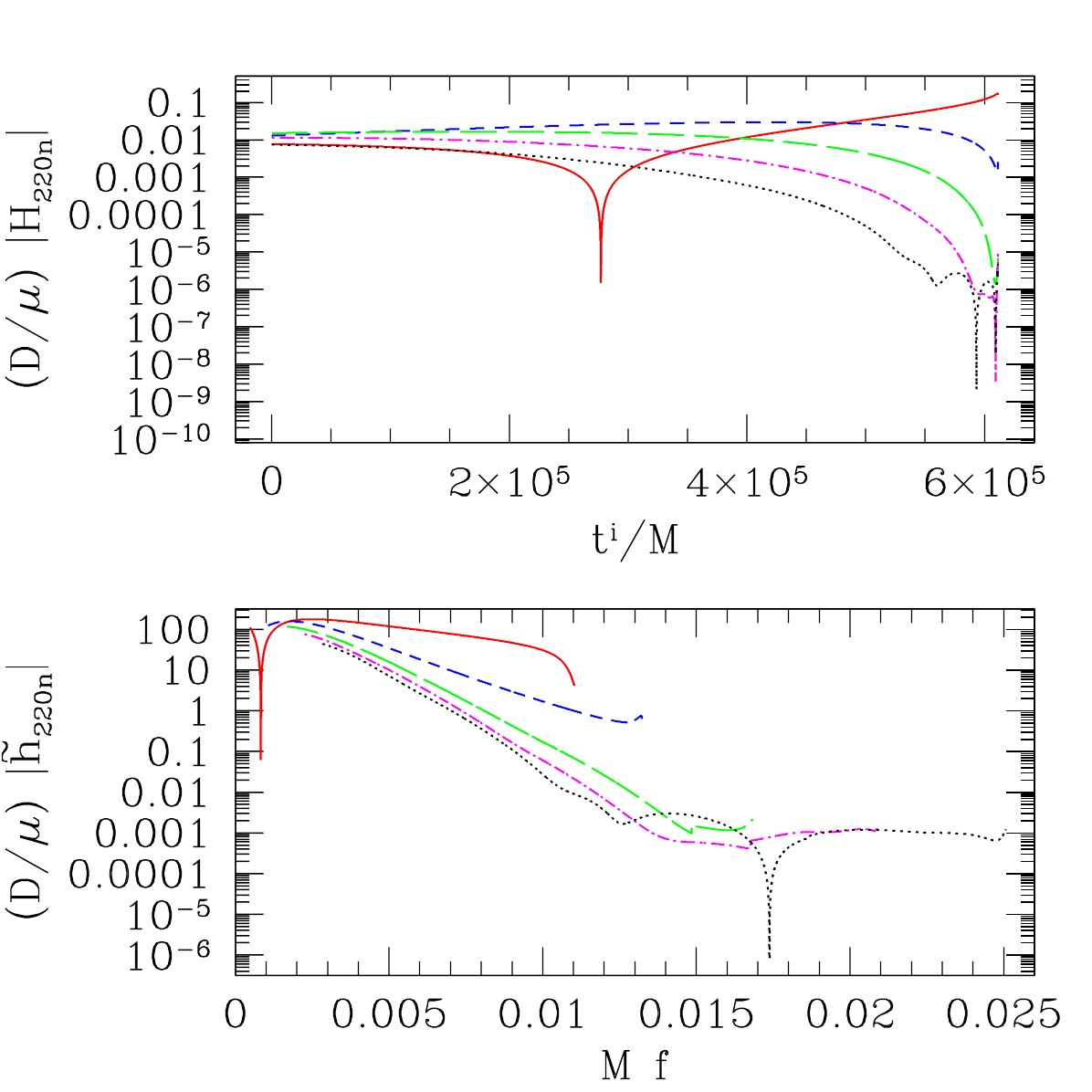}
\caption{Left panels: the waveform $h_+$ for an example of eccentric equatorial inspiral into a black hole with $a = 0.9M$.  This example has $(p_{\rm init}, e_{\rm init}) = (12M, 0.7)$, is at mass ratio $\varepsilon = 10^{-3}$, and is observed in the black hole's equatorial plane.  The initial conditions are similar to those used to make the large eccentricity case shown in Fig.\ \ref{fig:a0.0_hp}, and the two waveforms are similar at early times.   However, at this rapid spin, inspiral goes very deep into the strong field, becoming nearly circular before plunge; we find $(p_{\rm final}, e_{\rm final}) = (2.40M, 0.057)$.  The late waveform is consistent with low eccentricity late in the inspiral.  Top left shows the waveform for $\chi_{r0} = 2/3$ over the inspiral; in addition to zooming onto early and late times, bottom left illustrates the influence of the anomaly angle, comparing $\chi_{r0} = 2\pi/3$ (solid [red] curves) with inspiral on the fiducial geodesic (dashed [blue] curves).  Right panels: some of the $l = m = 2$ voices which contribute to this waveform, both time domain (top) and frequency domain (bottom).  Solid (red) curves show the voices with $n = 0$; short-dashed (blue) are for $n = 3$; long-dashed (green) are $n = 6$; dot-dashed (magenta) are $n = 9$; and dotted (black) curves show $n = 12$.}
\label{fig:a0.9_ecc}
\end{figure*}

The left-hand panels of Fig.\ \ref{fig:a0.9_ecc} show the gravitational waveform for this inspiral.  For the time-domain waveform, we include contributions from modes with $l \in [2,3,4]$, $m \in [-l, \ldots, l]$, $n \in [0, \ldots, 40]$, plus modes that are simply related by symmetry; as in the Schwarzschild cases we examined, only voices with $k = 0$ contribute since there is no $\theta$ motion.  The early waveform is qualitatively quite similar to the early large eccentricity waveform we found for Schwarzschild (Fig.\ \ref{fig:a0.0_hp}).  The late waveform, by contrast, is quite different, reflecting the fact that the orbit has nearly circularized as it approaches the final plunge.  As in the Schwarzschild case, we see that the initial radial anomaly angle $\chi_{r0}$ has a large impact on the waveform.

The right-hand panels of Fig.\ \ref{fig:a0.9_ecc} show some of the $l = m = 2$ voices which contribute to this waveform.  An interesting trend we see is that the importance of different voices changes dramatically during inspiral.  The evolution of the $n = 0$ voice is especially dramatic: it is fairly weak at early times (and in fact passes through zero during the inspiral), but dominates the waveform at late times.  This makes sense --- at late times the system's geometry is nearly circular, and voices corresponding to radial harmonics play a substantially less important role.

\subsection{Generic Kerr inspiral}
\label{sec:kerr_3D}

We conclude our discussion of results by looking at a Kerr inspiral that is both inclined from the equatorial plane and eccentric.  As discussed in Sec.\ \ref{sec:implement}, we have not yet generated dense data sets covering a wide range of such orbits.  The example shown here demonstrates that the techniques we have developed to build adiabatic EMRI waveforms have no difficulty with such cases.  Although generic EMRI waveforms have been developed using ``kludges,'' to our knowledge this is the first generic example that uses strong-field backreaction and strong-field wave generation for the entire calculation. 

\begin{figure*}
\begin{center}
\includegraphics[width=0.75\textwidth]{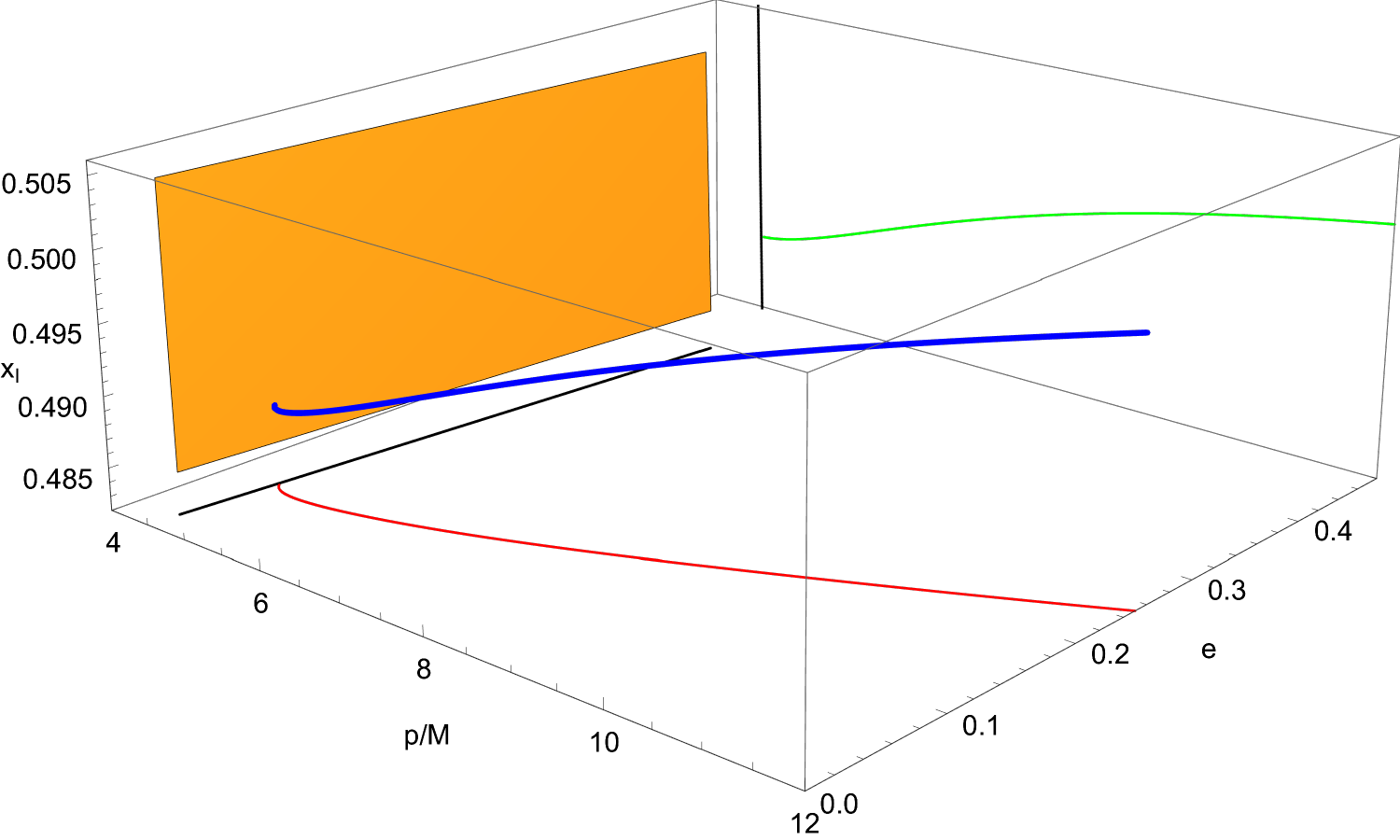}
\end{center}
\caption{The inclined and eccentric trajectory in $(p, e, x_I)$ for the generic inspiral we examine.  This trajectory, illustrated by the blue curve, begins at $(p_{\rm init}, e_{\rm init}, x_{I,{\rm init}}) = (12M, 0.25, 0.5)$ and proceeds until the smaller body encounters the LSO (a section of which is illustrated by the orange plane) at $(p_{\rm final}, e_{\rm final}, x_{I, {\rm final}})= (4.64M, 0.084, 0.488)$.  A projection of this trajectory into the $(p,e)$ plane is illustrated by the red curve, along with the projection of the LSO at the final value of $x_I$ (black line in lower plane); a projection of this trajectory into the $(p, x_I)$ plane is illustrated by the green curve, along with the projection of the LSO at the final value of $e$ (black line on back ``wall'' of the box).}
\label{fig:a0.7_gentraj}
\end{figure*}

\begin{figure*}
\includegraphics[width=0.48\textwidth]{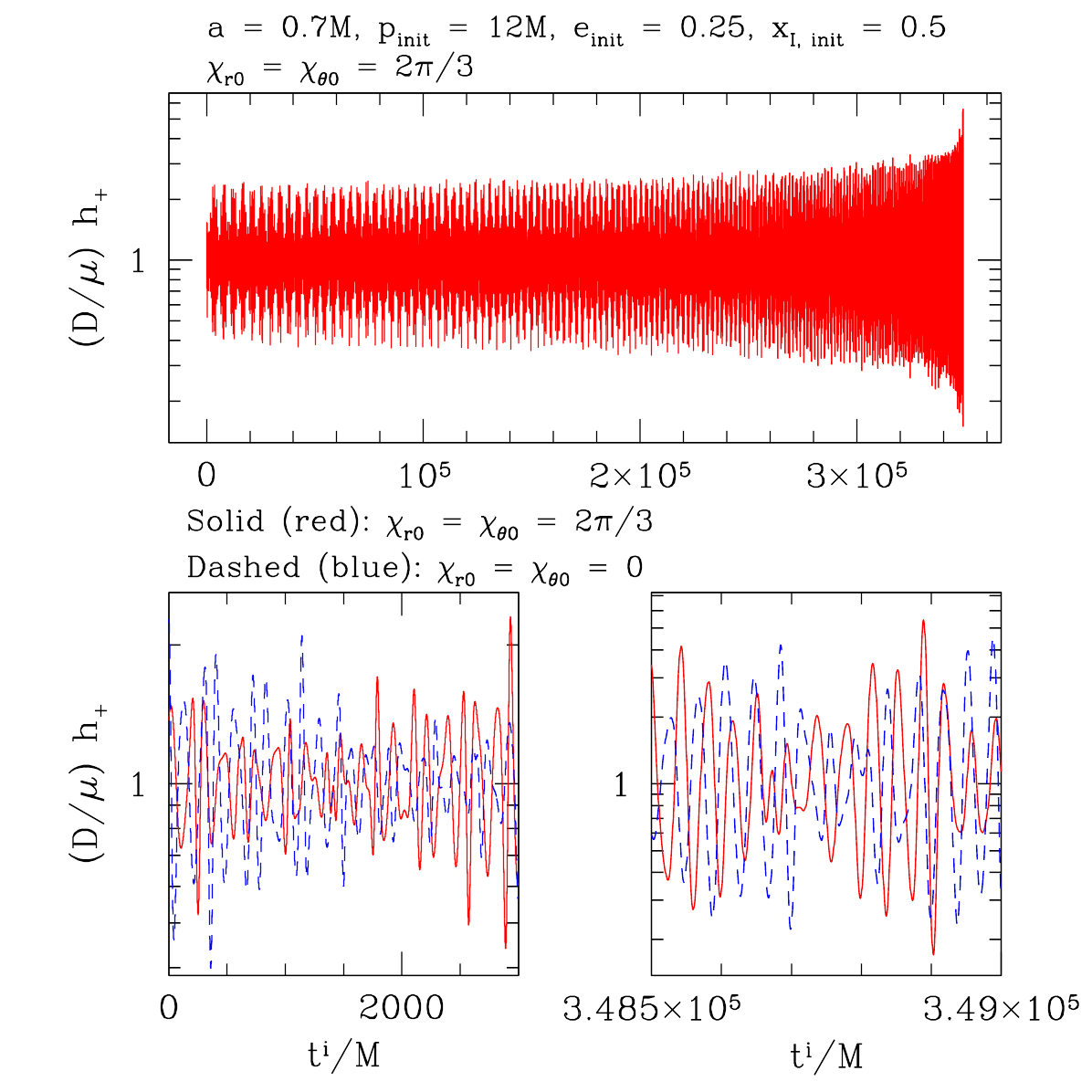}
\includegraphics[width=0.48\textwidth]{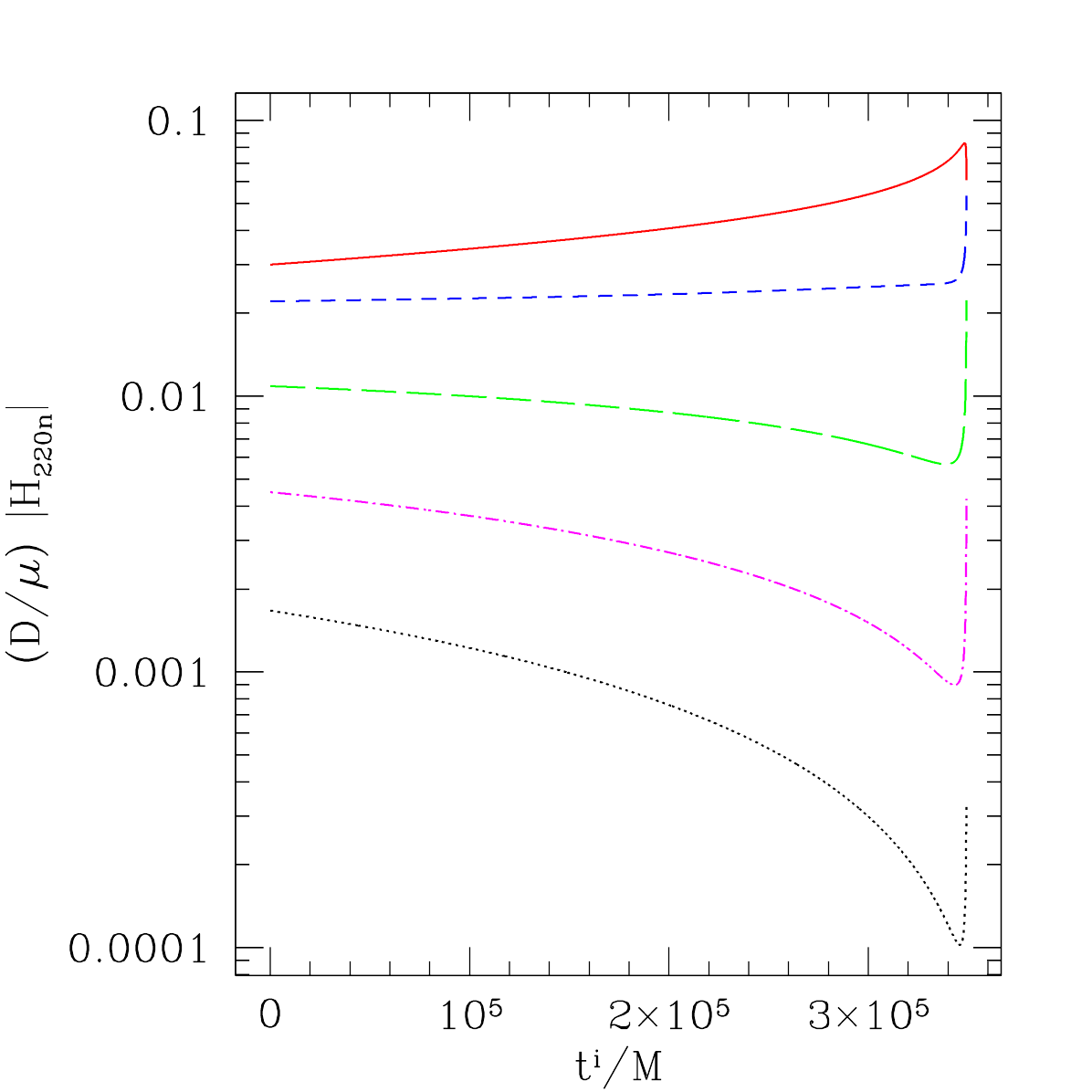}
\caption{Left panel: the waveform $h_+$ for an example of generic inspiral into a black hole with $a = 0.7M$.  This example corresponds to $(p_{\rm init}, e_{\rm init}, x_{I,{\rm init}}) = (12M, 0.25,0.5)$, and has a mass ratio $\varepsilon = 10^{-3}$; the small body inspirals until it encounters the LSO at $(p_{\rm final}, e_{\rm final}, x_{I,{\rm final}}) = (4.64M, 0.084, 0.488)$.  We show the waveform as observed in the black hole's equatorial plane.  Upper left panel shows the waveform over the inspiral; lower left panels zoom onto early and late times.  The lower left panels also illustrate the influence of the anomaly angle, comparing $\chi_{r0} = \chi_{\theta0} = 2\pi/3$ (solid [red] curves) with inspiral on the fiducial geodesic (dashed [blue] curves).  Right panel: Some of the voices that contribute to this waveform.  The voices shown here are for $l = m = 2$, $k = 0$; solid (red) curve is $n = 0$, short-dashed (blue) is $n = 1$, long-dashed (green) is $n = 2$, dot-dashed (magenta) is $n = 3$, and dotted (black) is $n = 4$.  The behavior of these voices is quite simple, with $n = 0$ strongest, and the voices falling away as $n$ increases.}
\label{fig:a0.7_gen1}
\end{figure*}

\begin{figure*}
\includegraphics[width=0.48\textwidth]{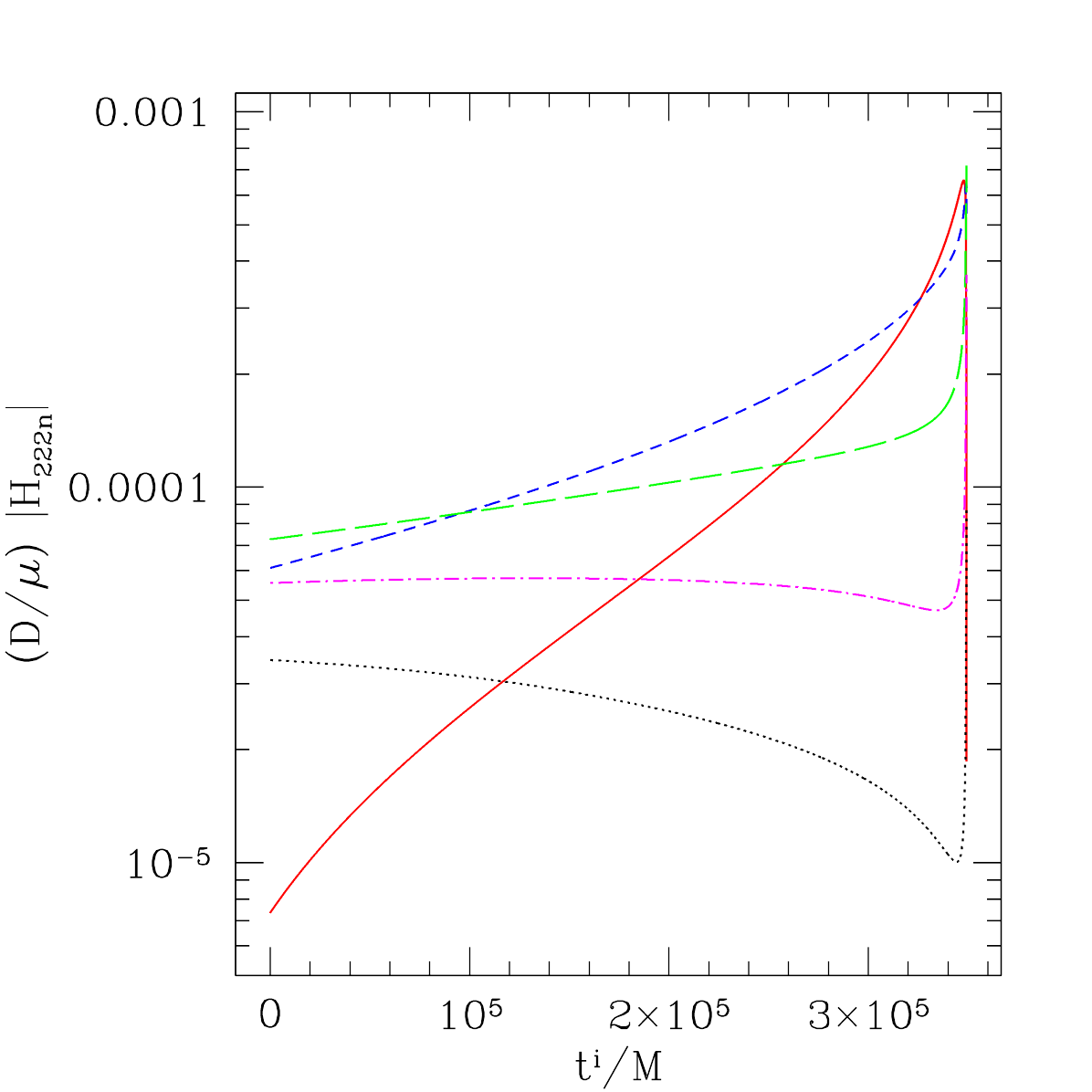}
\includegraphics[width=0.48\textwidth]{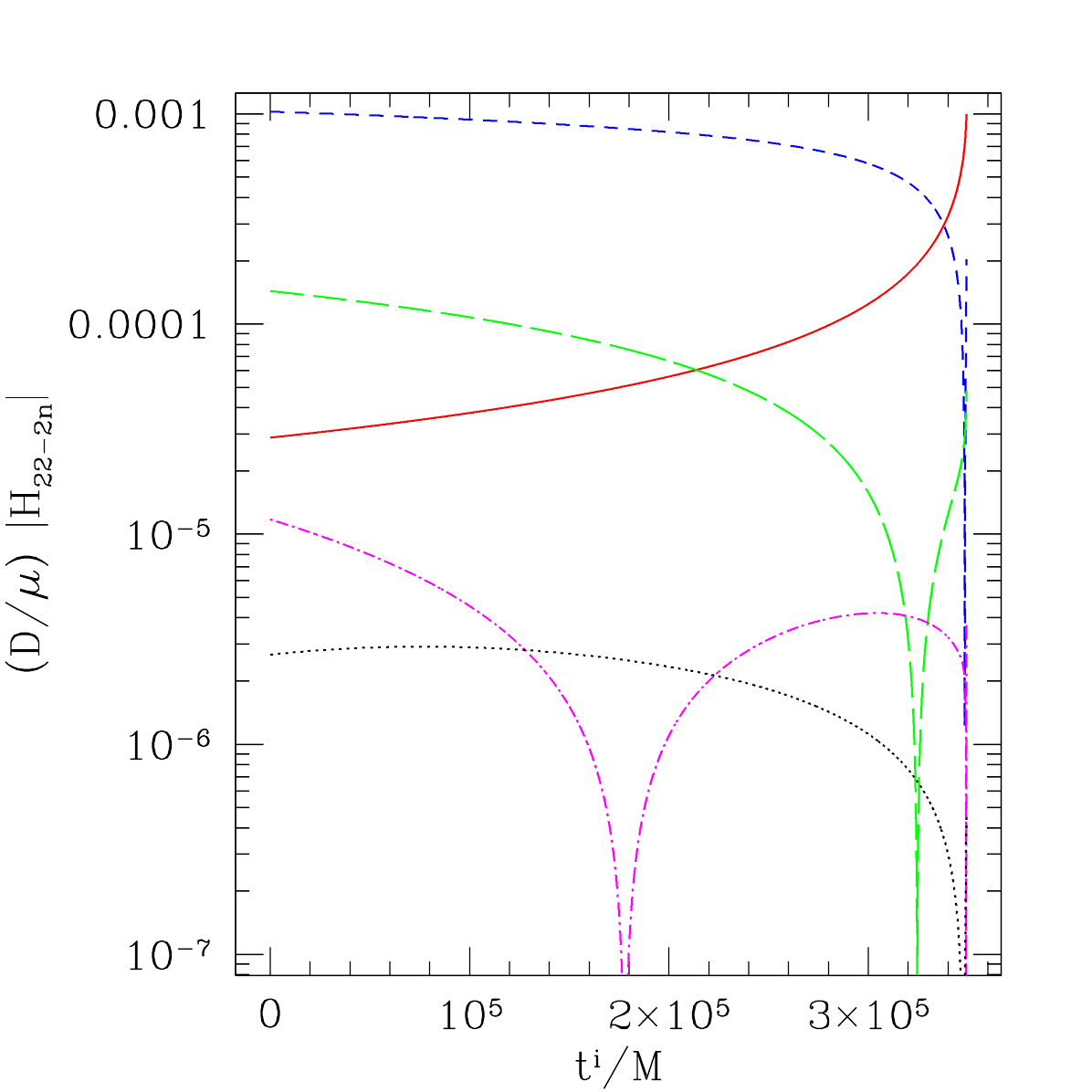}
\caption{Additional voices that contribute to the waveform shown in Fig.\ \ref{fig:a0.7_gen1}.  Left panel shows voices with $k = 2$; right panel shows voices with $k = -2$.  As in the right-hand panel of Fig.\ \ref{fig:a0.7_gen1}, the solid (red) curves are $n = 0$, short-dashed (blue) are $n = 1$, long-dashed (green) are $n = 2$, dot-dashed (magenta) are $n = 3$, and dotted (black) are $n = 4$.  The behavior is somewhat more complicated for these voices; although larger values of $n$ tend to be weaker, the evolution for small values of $n$ evolves rather differently than in the $k = 0$ case.  For both the cases shown here, the $n = 0$ voice in particular evolves in importance.}
\label{fig:a0.7_gen2}
\end{figure*}

The case we examine begins at $(p_{\rm init}, e_{\rm init},x_{I, {\rm init}}) = (12M, 0.25, 0.5)$.  At mass ratio $\varepsilon = 10^{-3}$, inspiral lasts for $\Delta t^{\rm i} \simeq 3.5 \times 10^5M$, at which time the smaller body encounters the LSO at $(p_{\rm final}, e_{\rm final}, x_{I, {\rm final}})= (4.64M, 0.084, 0.488)$.  As in the spherical cases, notice that the total change in inclination is very small: the change $\delta x_I = 0.012$ corresponds to the inclination angle $I$ increasing by about $0.79^\circ$.  Figure \ref{fig:a0.7_gentraj} shows the trajectory that the smaller body follows in $(p,e,x_I)$.  The left-hand panels of Fig.\ \ref{fig:a0.7_gen1} shows the time-domain $+$-polarization of the waveform that we find in this case, including all modes with $l \in [2,3,4]$, $m \in [-l, \ldots, l]$, $k \in [0, \ldots, 10]$, $n \in [0, \ldots, 25]$ (as well as modes simply related to this modes by symmetry).  The right-hand panel of Fig.\ \ref{fig:a0.7_gen1} and both panels of Fig.\ \ref{fig:a0.7_gen2} show some of the voices that contribute to this waveform ($l = m = 2$, $k = 0$ voices in Fig.\ \ref{fig:a0.7_gen1}; $l = m = 2$, $k = \pm 2$ voices in Fig.\ \ref{fig:a0.7_gen2}).

Perhaps not surprisingly, the picture that emerges for generic orbits is a blend of features seen in the spherical and equatorial limits.  The time domain waveform has large peaks corresponding to periapsis passage, separated by lower amplitude troughs as the orbit moves through apoapsis; the relatively small contrast between the peaks and troughs is consistent with this case's fairly modest initial eccentricity.  Superposed on this structure is a more rapid ``whirling'' associated with frame-dragging.  This is especially clear late in the waveform when the orbit has moved to small radius, and resembles behavior seen in the waveform for the spherical case at late times.

For $k = 0$, the time-domain structure of voices that contribute to this waveform is similar to what we saw in the low-eccentricity Schwarzschild case shown in Fig.\ \ref{fig:a0.0_smallecc_posvoices}: $n = 0$ is the strongest voice, with contributions steadily decreasing as $n$ increases.  For $k = \pm 2$, we see more variation: $n = 0$ often starts out relatively weak, but becomes much stronger late in the inspiral after eccentricity is significantly decreased.  The frequency domain structure of the voices is consistent with this picture; given the already large number of plots in this paper, we do not include figures showing this.

\section{Conclusion}
\label{sec:conclude}

In this paper, we have shown how to to use precomputed frequency-domain Teukolsky equation solutions to make EMRI waveforms.  In this way, one can compute and store in advance the most computationally expensive aspects of EMRI analysis, and then use the methods we describe to rapidly assemble waveforms using the stored data.  We particularly emphasize the usefulness of the multivoice waveform structure, which facilitates identifying particularly important waveform multipoles and harmonics, both in the time and frequency domains.

The framework and techniques we describe here have not been optimized, so there is much scope for efficiency gains.  In a companion analysis \cite{Chua:2020stf}, some of the present authors examined several algorithmic improvements, showing that we can in fact construct analysis-length time-domain EMRI waveforms in the Schwarzschild limit in under a second.  There are two ways that the results of the present work can be incorporated into the framework of \cite{Chua:2020stf}.  The first is to extend to Kerr inspirals.  The main challenges here are due to the higher dimensional parameter space, and the need to sum over more modes, since waveforms depend on an additional harmonic frequency.  Inspirals also extend deeper into the strong field, so more modes are likely to make strong contributions to the waveform.  The framework in \cite{Chua:2020stf} is designed to overcome these challenges.  Its neural network interpolation is a promising technique for dealing with the increased dimensionality, and its use of GPU-based hardware acceleration alleviates the computational burden of summing over thousands of additional $(l,m,k,n)$-modes.

The second extension is to incorporate frequency-domain waveforms.  For the Schwarzschild limit, Ref.\ \cite{Chua:2020stf} already demonstrates how to efficiently interpolate waveform amplitudes; the remaining new challenge is the efficient calculation of the stationary time as a function of frequency.  This can likely be achieved by sparsely evaluating $t(f)$ at a few key frequencies and interpolating to compute additional values.  Knowing $t(f)$ is in addition likely to be useful for connecting the frequency-domain waveforms we describe here to a detector response function.  The LISA detector's response changes with time as the antenna orbits the Sun, introducing time- and frequency-dependent modulations, and changing the sensitivity to the two gravitational-wave polarizations as the orientation of the antenna relative to a source varies over the orbit.  Though both these extensions will complicate the development of fast waveforms, they do not change the fundamental message of Ref.\ \cite{Chua:2020stf}.  We are confident that adapting those methods with the data sets and framework described here will be quite effective.

Because of the high cost of generating the waveform amplitude data, once computed, the data should be widely shared.  The data sets we have computed, described in Sec.\ \ref{sec:implement}, have been released through the Black Hole Perturbation Toolkit \cite{BHPToolkit} (hereafter ``the Toolkit'').  These data were all computed using a fairly small (roughly 1000 core) in-house cluster at the MIT Kavli Institute.  This scale of cluster is adequate for making inspiral data for 2-D orbits (spherical or equatorial), but is too small to be useful for fully generic 3-D (inclined and eccentric Kerr) data sets.  We have ported our frequency-domain Teukolsky equation solver, GREMLIN, to the US National Science Foundation's XSEDE \cite{XSEDE} environment, and plan to develop generic data sets there.  It should be noted, however, that there is a lot to be learned from 2-D cases about how best to lay out orbits on the grid (for example, the work in progress mentioned in Sec.\ \ref{sec:implement} that shows the importance of dense coverage at small eccentricity).  Data sets released through the Toolkit are likely to be revised with some regularity as we learn more about the best way to lay out the data grids.

Plans are in place to release the GREMLIN code which was used to generate these data sets, and which includes tools for post-processing analysis and generating EMRI waveforms.  A reduced functionality version of GREMLIN (specialized to circular, equatorial orbits) has already been released \cite{BHPToolkit}.  The generic version will be released after it is cleaned of proprietary libraries and fully comports with Open Source licensing requirements.

In terms of analyses of the source physics, there are several natural places to extend what we have done here.  One clear step is to consider how physics beyond the adiabatic approximation affects waveforms.  As described in Sec.\ \ref{sec:results_schw}, we see evidence from comparison with waveforms computed by a time-domain Teukolsky solver that neglecting terms which vary on the long inspiral timescale leads to an initial-condition-dependent secular drift in the waveform.  Our hypothesis is that this arises due to a slow evolution in the $\xi_{mkn}$ phases described in Sec.\ \ref{sec:amps_and_evolve}.

In a similar vein, we expect it will not be difficult to incorporate the orbit-averaged first-order conservative self force.  In brief, on average the conservative self force changes the rate at which the orbit precesses, and can be modeled as a slow ``anomalous'' change to the rate of periastron advance and the advance of the line of nodes.  These anomalous changes can in turn be incorporated into an osculating geodesic framework by allowing the parameters $\chi_{r0}$, $\chi_{\theta0}$, and $\phi_0$ to slowly evolve under the influence of this force.  See, for example, Ref.\ \cite{Pound:2007th} for further discussion.  A similar effect due to the orbit-averaged coupling of the smaller body's spin to the background curvature probably can also be modeled in such a way.  It should also be possible to include many {\it non}-orbit-averaged self forces and spin-curvature couplings \cite{Osburn:2015duj,Warburton:2017sxk,Akcay:2019bvk,Piovano:2020zin,Skoupy:2021iwb} by using a near-identify transformation \cite{vandeMeent:2018rms}.  Such non-orbit averaged analyses will be needed in order to model the impact of resonances \cite{Flanagan:2012kg,Bonga:2019ycj}, for example.  It is likely that an orbit-averaged-based analysis accurately describes systems away from resonances, but one will need to do a more complicated (and expensive) analysis in the vicinity of each resonance to model how the system evolves through each resonance crossing, and practical schemes for efficiently combining the two within template models must be devised. Regardless, extensions of this sort may make it possible for this framework to incorporate the most important post-adiabatic effects quite easily, significantly improving the ability of these models to serve as templates for EMRI measurements.

Finally, we note that recent phenomenological frequency-domain gravitational wave models are being calibrated in the large mass-ratio limit with Teukolsky waveforms \cite{Pratten:2020fqn}. The directly constructed frequency-domain waveforms presented in this work could be useful in this effort.

\section*{Acknowledgments}

An early presentation of some of the results and methods described here was given at the 22nd Capra Meeting on Radiation Reaction in General Relativity, hosted by the Centro Brasiliero de Pequisas F\'isicas in Rio de Janeiro, Brazil in June 2019; we thank the participants of that meeting for useful discussions, and particularly Marc Casals for organizing the meeting.  The time-domain waveform computations were performed on the MIT/IBM {\it Satori} GPU supercomputer supported by the Massachusetts Green High Performance Computing Center (MGHPCC) and the {\it CARNiE} cluster at UMass Dartmouth.  SAH thanks the MIT Kavli Institute for providing computing resources and support; L.\ S.\ Finn for (long ago) suggesting the multivoice framework for examining EMRI waveforms; Stanislav Babak for emphasizing the importance of understanding EMRI waveform structure in the frequency domain; Steve Drasco and William T.\ Throwe for past contributions to GREMLIN development; Gustavo Velez for insight into the Jacobian between $(dE/dt, dL_z/dt, dQ/dt)$ and $(dp/dt,de/dt,dx_I/dt)$; Talya Klinger for work porting GREMLIN to the XSEDE computing environment; and both Talya Klinger and Sayak Datta for discussions of accuracy and systematic error in EMRI data sets.  We also thank Leo Stein for raising the issue of the pathology discussed in Ref.\ \cite{Klein:2014bua}, which was important to consider as a contrast to a similar issue we examine.  SAH's work on this problem was supported by NASA ATP Grant 80NSSC18K1091 and NSF Grant PHY-1707549. NW  acknowledges support from a Royal Society--Science Foundation Ireland Research Fellowship. This publication has emanated from research conducted with the financial support of Science Foundation Ireland under Grant number 16/RS-URF/3428. GK acknowledges support from NSF grants PHY-2106755 and DMS-1912716. AJKC acknowledges support from the NASA grant 18-LPS18-0027.  MLK acknowledges support from NSF grant DGE-0948017.

\appendix

\section{Formulas for $E$, $L_z$, and $Q$}
\label{app:geodesicconstants}

In this appendix, we list formulas for the geodesic constants of the motion $E$, $L_z$, and $Q$ as functions of our preferred parameters $p$, $e$, and $x_I \equiv \cos I$.   Formulas for these constants were first worked out by Schmidt \cite{Schmidt:2002qk}, and are provided in particularly clean form for generic orbits by van de Meent \cite{vandeMeent:2019cam}.  We list them here using our preferred parameterization, and also include formulas for spherical orbits.

The three geodesic constants are given by
\begin{eqnarray}
E &=& \sqrt{\frac{\kappa\rho + 2\varpi\sigma - 2{\rm sgn}(x_I)\sqrt{\sigma(\sigma\varpi^2 + \rho\varpi\kappa - \eta\kappa^2)}}{\rho^2 + 4\eta\sigma}}\;,
\nonumber\\
\label{eq:Eformula}\\
L_z &=& -\frac{g(r_{\rm a})E - \sqrt{g(r_{\rm a})^2 + h(r_{\rm a})f(r_{\rm a})E^2 - h(r_{\rm a})d(r_{\rm a})}}{h(r_{\rm a})}\;,
\nonumber\\
\label{eq:Lzformula}\\
Q &=& (1 - x_I^2)\left[a^2(1 - E^2) + \frac{L_z^2}{x_I^2}\right]\;,
\label{eq:Qformula}
\end{eqnarray}
where $r_{\rm a} = p/(1 - e)$ is the coordinate radius of the orbit's apoapsis.  For generic orbits, the quantities appearing here are given by
\begin{eqnarray}
\kappa &=& d(r_{\rm a})h(r_{\rm p}) - d(r_{\rm p})h(r_{\rm a})\;,
\\
\varpi &=& d(r_{\rm a})g(r_{\rm p}) - d(r_{\rm p})g(r_{\rm a})\;,
\\
\rho &=& f(r_{\rm a})h(r_{\rm p}) - f(r_{\rm p})h(r_{\rm a})\;,
\\
\eta &=& f(r_{\rm a})g(r_{\rm p}) - f(r_{\rm p})g(r_{\rm a})\;,
\\
\sigma &=& g(r_{\rm a})h(r_{\rm p}) - g(r_{\rm p})h(r_{\rm a})\;,
\end{eqnarray}
where $r_{\rm p} = p/(1 + e)$ is the coordinate radius of the orbit's periapsis, and where
\begin{eqnarray}
d(r) &=& \Delta(r)[r^2 + a^2(1 - x_I^2)]\;,
\\
f(r) &=& r^4 + a^2[r(r + 2M) + (1 - x_I^2)\Delta(r)]\;,
\\
g(r) &=& 2aMr\;,
\\
h(r) &=& r(r - 2M) + \frac{(1 - x_I^2)\Delta(r)}{x_I^2}\;,
\end{eqnarray}
Note that we have switched notation slightly versus Refs.\ \cite{Schmidt:2002qk} and \cite{vandeMeent:2019cam}: we use $\varpi$ in these lists rather than $\epsilon$ to avoid notational collision with $\epsilon = \sqrt{M^2 - a^2}/2Mr_+$, as well as with the very similar $\varepsilon \equiv \mu/M$.

For spherical orbits with $e = 0$, $r_{\rm a} = r_{\rm p} \equiv r_{\rm o}$.  In this limit, we use a different form of these quantities:
\begin{eqnarray}
\kappa &=& d(r_{\rm o})h'(r_{\rm o}) - d'(r_{\rm o})h(r_{\rm o})\;,
\\
\varpi &=& d(r_{\rm o})g'(r_{\rm o}) - d'(r_{\rm o})g(r_{\rm o})\;,
\\
\rho &=& f(r_{\rm o})h'(r_{\rm o}) - f'(r_{\rm o})h(r_{\rm o})\;,
\\
\eta &=& f(r_{\rm o})g'(r_{\rm o}) - f'(r_{\rm o})g(r_{\rm o})\;,
\\
\sigma &=& g(r_{\rm o})h'(r_{\rm o}) - g'(r_{\rm o})h(r_{\rm o})\;.
\end{eqnarray}
In these formulas, $'$ denotes $\partial/\partial r$.

\section{Jacobian from\\$(dE/dt, dL_z/dt, dQ/dt)$ to $(dp/dt, de/dt, dx_I/dt)$}
\label{app:jacobian}

Once the rates of change $dE/dt$, $dL_z/dt$, and $dQ/dt$ have been computed, we use them to compute how a system evolves from one geodesic to another in the adiabatic limit.  As part of this, we would like to know how the geodesic geometry parameters $p$, $e$, and $x_I$ change due to this backreaction.  In this appendix, we write out the details of this procedure.

A generic Kerr orbit has turning points in its radial motion at apoapsis, $r_{\rm a} = p/(1 - e)$, and at periapsis, $r_{\rm p} = p/(1 + e)$.  It also has a turning point in its polar motion at $\theta_m$.  (The second polar turning point at $\pi - \theta_m$ yields no new information because of reflection symmetry about $\theta = \pi/2$.)  Recall that $\theta_m$ is related to our inclination angle $I$ by Eq.\ (\ref{eq:Idef}).   The radial turning points mean that $R(r_{\rm a}) = 0$ and $R(r_{\rm p}) = 0$, where $R(r)$ is defined in Eq.\ (\ref{eq:rdot}); the polar turning point means that $\Theta(\theta_m) = 0$, where $\Theta(\theta)$ is defined in Eq.\ (\ref{eq:thdot}).  We require that these conditions hold as the system evolves due to backreaction, which means that we require
\begin{eqnarray}
\frac{d}{dt}R(r_{\rm a}) = 0\;,\quad\frac{d}{dt}R(r_{\rm p}) = 0\;,\quad\frac{d}{dt}\Theta(\theta_m) = 0\;.
\end{eqnarray}
Expanding these total time derivatives, we find that the results can be written as a matrix equation,
\begin{equation}
\begin{pmatrix}
J_{E r_{\rm a}} & J_{L_z r_{\rm a}} & J_{Q r_{\rm a}} \\
J_{E r_{\rm p}} & J_{L_z r_{\rm p}} & J_{Q r_{\rm p}} \\
J_{E x_I} & J_{L_z x_I} & J_{Q x_I}
\end{pmatrix}
\cdot
\begin{pmatrix}
dE/dt \\ dL_z/dt \\ dQ/dt
\end{pmatrix}
=
\begin{pmatrix}
dr_{\rm a}/dt \\ dr_{\rm p}/dt \\ dx_I/dt
\end{pmatrix}
\end{equation}
Computing the various matrix elements $J_{ab}$, we find simple expressions for $dr_{\rm a,p}/dt$ and $dx_I/dt$.  Examine the change to the inclination first:
\begin{widetext}
\begin{eqnarray}
\frac{dx_I}{dt} &=& (1 - x_I^2)\frac{2\sqrt{1 - x_I^2}\sqrt{Q - a^2(1 - E^2)(1 - x_I^2)}(dL_z/dt) - x_I(dQ/dt) - 2x_I(1 - x_I^2)a^2E(dE/dt)}{2[Q - a^2(1 - E^2)(1 - x_I^2)^2]}
\label{eq:dxIdt1}
\\
&=& \frac{2x_I(1 - x_I^2)L_z(dL_z/dt) - x_I^3[(dQ/dt) + 2(1 - x_I^2)a^2E(dE/dt)]}{2[L_z^2 + x_I^4a^2(1 - E^2)]}\;.
\label{eq:dxIdt2}
\end{eqnarray}
\end{widetext}
Notice that Eq.\ (\ref{eq:dxIdt1}) is singular when $Q \to 0$ (which coincides with $|x_I| \to 1$), and Eq.\ (\ref{eq:dxIdt2}) is singular when $L_z \to 0$ ($|x_I| \to 0$).  We use Eq.\ (\ref{eq:dxIdt1}) when $|x_I| \le 0.5$, and (\ref{eq:dxIdt2}) when $|x_I| > 0.5$.  The two expressions yield identical results except right at their singular points.

To present our results for $dr_{\rm a,p}/dt$, we first define
\begin{eqnarray}
{\cal D}(r) &\equiv& 2M[Q + (L_z - aE)^2] - 2r[L_z^2 + Q + a^2(1 - E^2)]
\nonumber\\
& &+ 6Mr^2 - 4r^3(1 - E^2)\;.
\end{eqnarray}
Using this, we have
\begin{eqnarray}
 J_{Er_{\rm a,p}} &\equiv& \frac{4aM(L_z - aE)r_{\rm a,p} - 2Er_{\rm a,p}^2(a^2 + r_{\rm a,p}^2)}{{\cal D}(r_{\rm a,p})}\;,
\nonumber\\
J_{L_zr_{\rm a,p}} &\equiv& \frac{4M(aE - L_z)r_{\rm a,p} + 2L_zr_{\rm a,p}^2}{{\cal D}(r_{\rm a,p})}\;,
\nonumber\\
J_{Qr_{\rm a,p}} &\equiv& \frac{r_{\rm a,p}^2 - 2Mr_{\rm a,p} + a^2}{{\cal D}(r_{\rm a,p})}\;.
\end{eqnarray}
We then find
\begin{equation}
\frac{dr_{\rm a,p}}{dt} = J_{Er_{\rm a,p}}\frac{dE}{dt} + J_{L_zr_{\rm a,p}}\frac{dL_z}{dt} + J_{Qr_{\rm a,p}}\frac{dQ}{dt}\;,
\end{equation}
Once $dr_{\rm a,p}/dt$ are known, it is simple to compute $dp/dt$ and $de/dt$:
\begin{eqnarray}
\frac{dp}{dt} &=& \frac{(1 - e)^2}{2}\frac{dr_{\rm a}}{dt} + \frac{(1 + e)^2}{2}\frac{dr_{\rm p}}{dt}\;,
\label{eq:dpdt}\\
\frac{de}{dt} &=& \frac{(1 - e^2)}{2p}\left[(1 - e)\frac{dr_{\rm a}}{dt} - (1 + e)\frac{dr_{\rm p}}{dt}\right]\;.
\end{eqnarray}

In the spherical limit $e = 0$, $r_{\rm a} = r_{\rm p} \equiv r_{\rm o}$, and the conditions $R(r_{\rm a}) = 0$, $R(r_{\rm p}) = 0$ are redundant.  The equations we have derived here do not work in that limit.  Spherical orbits are instead governed by the conditions $R(r_{\rm o}) = 0$, $R'(r_{\rm o}) = 0$, where $R' \equiv \partial R/\partial r$.  Past work \cite{Kennefick:1995za,Ryan:1995xi} long ago proved that adiabatic dissipative self interaction evolves a spherical orbit into a new spherical orbit; the first derivation of $dQ/dt$ \cite{Sago:2005fn} proved that this result respected the ``spherical goes to spherical'' constraint.  Given $(dE/dt,dL_z/dt, dQ/dt)$ from a spherical orbit, one can infer $dr_{\rm o}/dt$ by enforcing the condition
\begin{equation}
\frac{d}{dt}R'(r_{\rm o}) = 0\;.
\label{eq:dtRprime}
\end{equation}
(Alternately, one can enforce the conditions $dR(r_{\rm o})/dt = 0$, $dR'(r_{\rm o})/dt = 0$.  Doing so yields solutions for $dr_{\rm o}/dt$ and $dQ/dt$ given $dE/dt$ and $dL_z/dt$.  This is how backreaction on spherical orbits was computed \cite{Hughes:1999bq} before $dQ/dt$ was fully understood \cite{Sago:2005fn}.)  Implementing Eq.\ (\ref{eq:dtRprime}), we find
\begin{equation}
\frac{dr_{\rm o}}{dt} = J^{\rm c}_{Er_{\rm o}}\frac{dE}{dt} + J^{\rm c}_{L_zr_{\rm o}}\frac{dL_z}{dt} + J^{\rm c}_{Qr_{\rm o}}\frac{dQ}{dt}\;,
\end{equation}
where the Jacobian elements for spherical orbits are
\begin{eqnarray}
J^{\rm c}_{Er} &=& \frac{2aM(L_z - aE) - 2a^2Er - 4Er^3}{{\cal D}^{\rm c}(r)}\;,
\\
J^{\rm c}_{L_zr} &=& \frac{-2M(L_z - aE) + 2L_z r}{{\cal D}^{\rm c}(r)}\;,
\\
J^{\rm c}_{Qr} &=& \frac{r - M}{{\cal D}^{\rm c}(r)}\;,
\end{eqnarray}
with
\begin{equation}
{\cal D}^{\rm c}(r) = -[a^2(1 - E^2) + L_z^2 + Q] + 6Mr - 6(1 - E^2)r^2\;.
\end{equation}

\bibliography{refs}

\end{document}